\documentclass{aa}  
\usepackage{orcidlink}
\usepackage{graphicx}
\usepackage{txfonts}
\usepackage{lipsum}
\usepackage{subcaption}

\usepackage{lscape}            
\usepackage{placeins}           
\usepackage{tablefootnote}
\begin{document}

   \title{Probing the influence of the protocluster environment on galaxy morphology at $z$ = 2.23}

   \author{Golden-Marx, Emmet\orcidlink{0000-0001-5160-6713}\thanks{\email{emmet.goldenmarx@inaf.it}}\inst{\ref{aff1}\ref{aff2}}, Cai, Z.\orcidlink{0000-0001-8467-6478}\inst{\ref{aff2}}, Shi, D.\orcidlink{0000-0002-4314-5686}\inst{\ref{affil3}}, Wang, X.\orcidlink{0000-0002-9373-3865}\inst{\ref{affil4}\ref{affil5}\ref{affil6}},  Lemaux, B.C.\orcidlink{0000-0002-1428-7036}\inst{\ref{affil7}\ref{affil8}}, Vulcani, B.\orcidlink{0000-0003-0980-1499}\inst{\ref{aff1}}, Häußler, B.\orcidlink{0000-0002-1857-2088}\inst{\ref{affil9}}, Renard, P.\orcidlink{0000-0002-5953-4491}\inst{\ref{aff2}}, Shen, L.\orcidlink{0000-0001-9495-7759}\inst{\ref{affil10}\ref{affil11}}, Giddings, F.\orcidlink{0009-0003-2158-1246}\inst{\ref{affil12}} 
   }

   \institute{INAF - Osservatorio di Padova, Vicolo Osservatorio 5, 35122 Padova, Italy\label{aff1}
   \and
   Department of Astronomy, Tsinghua University, Beijing 100084, China\label{aff2}
   \and 
   Center for Fundamental Physics, School of Mechanics and Optoelectronic Physics, Anhui University of Science and Technology, Huainan 232001, China \label{affil3}
   \and
   School of Astronomy and Space Science, University of Chinese Academy of Sciences (UCAS), Beijing 100049, China\label{affil4}
   \and
   National Astronomical Observatories, Chinese Academy of Sciences, Beijing 100101, China\label{affil5}
   \and
   Institute for Frontiers in Astronomy and Astrophysics, Beijing Normal University, Beijing 102206, China\label{affil6}
   \and 
   Gemini Observatory, NSF NOIRLab, 670 N. A'ohoku Place, Hilo, Hawai'i, 96720, USA\label{affil7}
   \and
   Department of Physics and Astronomy, University of California, Davis, One Shields Ave., Davis, CA 95616, USA\label{affil8}
   \and
   European Southern Observatory, Alonso de Córdova 3107, Vitacura, Santiago de Chile, Chile\label{affil9}
   \and
   Department of Physics and Astronomy, Texas A\&M University, College Station, TX, 77843-4242 USA\label{affil10}
   \and 
   George P.\ and Cynthia Woods Mitchell Institute for Fundamental Physics and Astronomy, Texas A\&M University, College Station, TX, 77843-4242 USA\label{affil11}
   \and
   Institute for Astronomy, University of Hawai‘i, 2680 Woodlawn Drive, Honolulu, HI 96822, USA\label{affil12} 
   }

   \date{Received March 3, 2025}
   \titlerunning{Protocluster galaxy morphology}
   \authorrunning{Golden-Marx et al. 2025}

  \abstract{

   As galaxies evolve in dense cluster and protocluster environments, they interact and quench their star formation, which gradually transforms the dominant galaxy population from star-forming galaxies to quiescent, red galaxies.  This transformation is identifiable by observing galaxy colors and can also be seen in the morphological transformation of late-type galaxies into early-type galaxies, which creates the morphology-density relation observed when comparing populations in clusters to field galaxies at a given epoch.  However, high-$z$ ($z$ $>$ 2) galaxy morphology studies are hindered by the high angular resolution necessary to characterize morphology.}
   {We present a study of HST WFC3 F160W observations of protoclusters from the MAMMOTH survey (BOSS1244 and BOSS1542) at $z$ $\sim$ 2.23 with populations of previously identified H$\alpha$ emitters.}  
   {By measuring the S\'ersic index of 151 H$\alpha$ emitters, we look for the early morphological transformation of high-$z$ star-forming galaxies in these well-studied, large, non-virialized protoclusters, which we believe to be the precursors of present-day galaxy clusters.} 
   {We find that the morphology of the populations of star-forming galaxies in protoclusters does not differ from their field counterparts.  However, we also identify a population of clumpy, potentially merging galaxies, which could lead to an increase in the population of early-type galaxies within these structures.  Additionally, in BOSS1244, which has two previously identified massive quiescent galaxies including a brightest cluster galaxy (BCG), we find an abundance of early-type galaxies near both the BCG and two co-eval high-$z$ quasars.}  
   {Although we find a strong similarity between the morphology of field and protocluster galaxies, the population of early-type star-forming galaxies surrounding the spectroscopically confirmed quiescent BCG in BOSS1244, something not seen in BOSS1542, may point to differences in the evolutionary state of these two co-eval protoclusters and be a sign of an early forming cluster core in BOSS1244.}


   \maketitle
   \nolinenumbers

\section{Introduction}\label{sect:intro}
Galaxy clusters are gravitationally bound cosmic megacities that host large populations of galaxies and provide a unique laboratory for exploring galaxy evolution.  These structures are evolving with galaxy populations that mature dramatically.  At low redshift, clusters are characterized by populations of red, early-type, elliptical galaxies and a densely clustered central region \citep[e.g.,][]{Miller2005,Rykoff2014}.  However, at $z$ $>$ 2, most extended structures are better identified as protoclusters, the early stage in galaxy cluster formation, which are characterized by an overdensity of galaxies that are predicted to gravitationally collapse into a cluster at $z$ $=$ 0 \citep[e.g.,][]{Muldrew2015,Overzier2016,Chiang2017,Alberts2022}.  These systems typically include large populations of star-forming and dusty galaxies \citep[e.g.,][]{Kurk2004,Casey2015,Hill2020,Edward2024}, making them distinct from their local counterparts. Unlike low-$z$ clusters, which are more commonly dynamically relaxed, protoclusters are not virialized, extend well beyond the typical Mpc size scale of low-$z$ clusters \citep[e.g.,][]{Cucciati2014,Forrest2023,Staab2024,Shah2024}, and can include clumpier substructures with multiple density peaks \citep[e.g.,][]{Cucciati2014,Hatch2017}.

While large samples of statistically robust clusters exist out to $z$ $\sim$ 1.5 with well-studied galaxy populations \citep[e.g.,][]{Gladders2000,Miller2005,Lemaux2012,Wing2011,Rykoff2014,Gonzalez2019,Golden-Marx2019,Balogh2021} and well-constrained properties linking star formation quenching and the build-up of red sequence populations to denser cluster environments \citep[e.g.,][]{Lemaux2019,Tomczak2019,Werner2022}, the impact of the protocluster environment on galaxy populations is less well defined.  In simulations at 2 $<$ $z$ $<$ 5, protoclusters account for $\approx$ 20$\%$ of the cosmic star formation rate density \citep{Chiang2017} and the build-up of stellar mass differs from the field due to a top-heavy population of massive galaxies \citep[e.g.,][]{Muldrew2018}.  Furthermore, simulations find that protocluster galaxies at $z$ $<$ 3 quench and build up their stellar mass faster than their field counterparts, particularly among infalling satellite galaxies \citep[e.g.,][]{Muldrew2015,Contini2016}.  Observationally, protoclusters include large populations of star-forming galaxies \citep[e.g.,][]{Edward2024}, although, at cosmic noon, star-forming galaxies also appear to be the dominant population in the field \citep{Muzzin2013A}, particularly among massive galaxies \citep[e.g.,][]{Marchesini2014,Vulcani2016}.  

Importantly, the demographics of these protocluster galaxies and any properties identified are highly dependent on the choice of tracer \citep[][]{Overzier2016,Alberts2022}.  Protoclusters have been detected via searches for overdensities of line-emitting galaxies including Lyman-$\alpha$ emitters (LAEs) and H-$\alpha$ emitters (HAEs) \citep[e.g.,][]{Kurk2004,Cai2017,Umehata2019,Zheng2021,Naufal2023}, by examining the environments of radio-loud active galactic nuclei (AGNs; \citealp[e.g.,][]{Hatch2011,Wylezalek2013,Shimakawa2018,Shen2021}), and by searching for overdensities of dusty, star-forming galaxies \citep[e.g.,][]{Casey2015,Hill2020}.  Additionally, robust photometric and spectroscopic redshift surveys can be used to probe well-studied survey fields to identify protocluster populations \citep[e.g.,][]{Cucciati2014,Lemaux2022,Forrest2023,Forrest2024,Staab2024,Shah2024}.  Although most galaxies identified via these methods are star-forming, recent James Webb Space Telescope (JWST) observations have found additional populations of high-$z$ quiescent galaxies in protocluster environments at 2.0 $<$ $z$ $<$ 6.0 \citep[e.g.,][]{McConachie2022,Naufal2024,Tanaka2024,Kiyota2025,McConachie2025}.

Fundamental to protocluster and cluster science is understanding how galaxy populations evolve.  At $z$ $<$ 1.5, galaxy populations are typically quantified by color-magnitude diagrams, color-color diagrams, and star formation rate (SFR) measurements to estimate the quiescent fraction as a function of environment \citep[e.g.,][]{Rudnick2012,Marchesini2014,Cooke2015,Cooke2016,Cerulo2016,Nantais2016,Tomczak2019,Lemaux2019,Golden-Marx2019,Werner2022,Euclid-Cleland2025}.  However, star formation quenching in cluster galaxies is not the only evolutionary transformation occurring between cosmic noon (2 $<$ $z$ $<$ 4) and the present day.  Galaxies at cosmic noon have primarily late-type, clumpy, and/or irregular morphologies \citep[e.g.,][]{Mortlock2013, Shibuya2016,Chen2022,Kartaltepe2023,Treu2023,Ferreira2023,Jacobs2023,Shen2024,Smethurst2025}, while low-$z$ galaxies, particularly those in clusters, are populated by bulge-dominated, early-type galaxies.  This is fundamentally defined in the morphology-density relation, where denser structures host larger populations of early-type elliptical and S0/lenticular galaxies \citep[e.g.,][]{Dressler1980,Postman2005,Vulcani2011a,Vulcani2011b,Cappellari2016,Vulcani+2023,Euclid-Cleland2025,Euclid-Qulley2025}.  Although star formation quenching and morphological transformations are linked to the build-up of present day cluster populations, these two processes act in different ways. 

Crucially, the populations of irregular and late-type galaxies seen at cosmic noon must quench and morphologically transform into their low-$z$, early-type counterparts by $z$ $\sim$ 1.0 \citep[e.g.,][]{Dressler1980,Gladders2000,Postman2005,Rudnick2012,Cooke2016,Cerulo2016,Cerulo2017,Nantais2017,Lemaux2019,Tomczak2019}.  As such, it is important to understand under what conditions galaxies in dense environments begin to quench and/or transform their morphology.  Particularly, does quenching occur before or after the morphological transformation begins?  Interestingly, the time scales of these processes appear to differ.  In clusters, quenching occurs primarily among lower mass galaxies between 1.0 $<$ $z$ $<$ 1.5, as seen in the similarity in the quenched populations and the persistence of the color- and SFR-density relations \citep[e.g.,][]{Rudnick2012,Lee-Brown2017,Lemaux2019,Tomczak2019}, implying that massive cluster galaxies quench first.  Quenching occurs through a combination of processes including strangulation, ram pressure stripping, harassment, and starvation, each of which act on different timescales \citep[e.g.,][]{Gunn1972,Larson1980,Balogh2000,Quilis2000}.  Thus, the quenching timescale evolves with the cluster dynamical timescale (1 - 3 Gyr) \citep{Foltz2018,Roberts2019}, although it may be better traced by the longer cluster infall timescale \citep[e.g.,][]{Werner2022,Kim2023}. 

However, the morphological transformation of cluster galaxies is not directly tied to the same processes, although they are linked \citep{Correa2019}.  At low redshift, \citet{Sampaio2024} find that the quenching timescale for low-mass galaxies is approximately double the morphological transformation timescale (2.5\,Gyr vs 1.2\,Gyr), implying they transform first.  However, given the dominance of the low-$z$ cluster environment, this may not hold among massive protocluster galaxies.  At cosmic noon, the morphological transformation may be tied to compaction, where mergers or other instabilities bring existing stars and gas towards the galaxy center, resulting in dense, star-forming, spheroidal galaxies (quenched by AGNs) \citep[e.g.,][]{Barro2013,Barro2017}.  Morphologically, this would result in early-type galaxies that are highly star-forming.  In simulations, compaction begins at cosmic noon, with timescales of 0.5 - 1\,Gyr, while quenching can occur over multiple Gyrs \citep{Zolotov2015}.  However, given the importance of galaxy mergers in compaction \citep[e.g.,][]{Barro2013,Zolotov2015,Barro2017} and in the formation of massive, quenched, early-type galaxies \citep[e.g.,][]{Toomre1977,Naab2006,Delucia2011,Cappellari2016,Rodriguez-Gomez2017}, the merger timescale may be a better tracer.  However, merger rates depends on protocluster mass, so these timescales are variable \citep[e.g.,][]{Liu2025}.  The differences in the timescales for quenching and morphological transformation allow for quenching to occur both before and/or after early morphological transformations, pointing to the importance of characterizing galaxy morphology in the early universe.

Characterizing galaxy morphology requires high-resolution imaging.  At high redshift, this can be done with space-based telescopes including JWST, the Hubble Space Telescope (HST), or \textit{Euclid}, as well as ground-based telescopes with adaptive optics.  Regardless of the redshift, morphology studies can be done by modeling the light profile of galaxies using parametric models, including the S\'ersic model \citep{Sersic1963}, a generalized version of the deVaucouleurs profile \citep{deVaucouleurs1948}, where the S\'ersic index, n, $=$ 1 describes a classical spiral galaxy and n $=$ 4 describes a classical elliptical galaxy. Alternatively, non-parametric models of galaxy morphology measurements have been used to describe galaxy light profiles in terms of concentration, asymmetry, and clumpiness \citep[e.g.,][]{Lotz2004,Lotz2006,Sazonova2020,Naufal2023}. 

The differences between cluster and protocluster populations make galaxy morphology a powerful tool for studying the evolution of these cosmic structures.  However, the required high angular resolution needed to characterize galaxy morphology limits the number of these studies at high redshift.  As such, there is currently no consensus for the morphology-density relation in protoclusters and high-$z$ clusters ($z$ $>$ 1.0).  \citet{Chan2021} used the sample of 11 GOGREEN clusters \citep{Balogh2021} with masses from 8 $\times$ 10$^{12}$\,M$_{\odot}$ $<$ M$_{200}$ $<$ 8 $\times$ 10$^{14}$\,M$_{\odot}$ at 1.0 $<$ $z$ $<$ 1.4 to study 832 galaxies and found little differences between the morphology of star-forming cluster and field galaxies, but did detect an excess of oblate, flattened quiescent cluster galaxies. \citet{Sazonova2020} analyzed four massive clusters (2 $\times$ 10$^{14}$\,M$_{\odot}$ $<$ M$_{200}$ $<$ 11 $\times$ 10$^{14}$\,M$_{\odot}$) at 1.2 $<$ $z$ $<$ 1.8 and found that two clusters show strong evidence of richer environments hosting more bulge-dominated galaxies.  Similarly, \citet{Strazzullo2023} used five massive (M$_{200}$ $>$ 4 $\times$ 10$^{14}$\,M$_{\odot}$) Sunyaev-Zel'dovich clusters at 1.4 $<$ $z$ $<$ 1.7 and found that the majority of their quiescent galaxies are bulge-dominated and in the densest environments (and similarly that the majority of their star-forming galaxies are disky).  However, \citet{Strazzullo2023} also found that the morphological parameters that characterize field and cluster quiescent galaxies are remarkably similar, implying that the causes of quenching and the morphological transformation, at least among massive quiescent galaxies, may be similar between the protocluster environment and the field, just more pervasive in protoclusters.  \citet{Noordeh2021} used a well-studied $z$ $\sim$ 2.0 cluster (M$_{500}$ $\sim$ 6.3 $\pm$ 1.5 $\times$ 10$^{13}$\,M$_{\odot}$; \citealp{Mantz2018,Willis2020}) and found a strong division based on galaxy color and morphology among cluster galaxies.  Additionally, \citet{Strazzullo2013} found a dense proto-red sequence core of bulge-dominated galaxies in a $z$ $\sim$ 2.0 cluster (M$_{200}$ $\sim$ 5.3 $\times$ 10$^{13}$\,M$_{\odot}$). Furthermore, \citet{Mei2023} and \citet{Afanasiev2023}, both of which used the same spectroscopically confirmed clusters/protoclusters (10$^{13.5}$\,M$_{\odot}$ $<$ M$_{halo}$ $<$ 10$^{14.5}$\,M$_{\odot}$) from the Clusters Around Radio Loud AGN (CARLA) Survey
found an abundance of "bulge'' galaxies in all of their systems, including a $z$ $\approx$ 2.8 protocluster \citep{Mei2023}, and an abundance of small protocluster galaxies relative to the field \citep{Afanasiev2023}.  In contrast, \citet{Peter2007} found no evidence of any trends between morphology and environment in their study of a protocluster at $z$ $\sim$ 2.3 with HST. In total, these studies include $\approx$ 40 spectroscopically-confirmed galaxy clusters/protoclusters, which makes any result subject to cosmic variance.  Importantly, only two of these systems are at $z$ $>$ 2.1 \citep[e.g.,][]{Peter2007,Mei2023} and except for \citet{Peter2007}, each probe galaxy populations including potentially quenched galaxies.  As such, the transformation of cluster galaxy morphology cannot be decoupled from quenching. Importantly, the majority of systems are massive, so it is impossible to determine the role of cluster mass in the evolution of the morphology-density relation without a more expansive sample.

In characterizing galaxy evolution using morphology, particularly at high redshift, it is important to account for galaxy mergers, which may impact the build-up of the morphology-density relation in high-$z$ protoclusters due to the role of galaxy mergers in massive early-type galaxy formation.  Although no consensus exists on the merger rates within protoclusters \citep[][]{Delahaye2017,Coogan2018,Watson2019,Monson2021,Liu2023,Liu2025,Giddings2025}, the high density of galaxies in protoclusters allows for the possibility of mergers.  Because the velocity dispersion of protocluster galaxies has been measured as $\sim$ 300\,km/s in simulations \citep{Cucciati2014} and $<$ 500\,km/s in observations \citep[e.g.,][]{Toshikawa2020,Liu2025}, these properties further enhance the likelihood of galaxy interactions and indicate mergers play a key role in protocluster galaxy evolution.  Observationally, it is important to note that protoclusters not only foster galaxy mergers, but also host clumpy galaxies, the abundance of which peak at cosmic noon \citep[e.g.,][]{Shibuya2016,Umehata2025}.  As shown in \citet{Ribeiro2017}, these clumpy galaxies can also result from major mergers.  However, because it is difficult to differentiate clumpy galaxies that are not merging from merging systems without accurate spatially resolved spectroscopy, some fraction of visually identified mergers may instead be non-merging clumpy galaxies.  While different, both populations can result in eventual galaxy transformations and understanding what fraction of protocluster galaxies are merging and/or clumpy is an important question in tracing the evolution of these cluster galaxies.

Because the timescales and the processes associated with quenching and the morphological transformation of protocluster galaxies differ, it is vital to determine if star-forming protocluster galaxies experience early processing that would begin to transform their morphology before quenching?  While JWST has proven invaluable for galaxy evolution and morphology studies, particularly at $z$ $>$ 2 \citep[e.g.,][]{Kartaltepe2023,Treu2023,Shen2024,Lee2024,vanderWel2024,Kawinwanichaki2025,Gozaliasl2025}, results related to protocluster systems (e.g., COSMOS-Web; \citealp{Shuntov2025,Shuntov2025b}) are still forthcoming.  

To answer these questions, we present new results from the MApping the Most Massive Overdensities through Hydrogen (MAMMOTH) survey \citep[e.g.,][]{Cai2016,Cai2017}.  Specifically, we use HST observations of two protoclusters at $z$ $\sim$ 2.23 (P.I. Zheng Cai).  Although we only include two protoclusters, we are tracing star-forming galaxies in protoclusters, allowing us to examine a very narrow, yet important window into the evolution of protocluster galaxies.  By examining the morphology of non-quenched, massive galaxies, we can better determine if the morphological transformation of star-forming galaxies is influenced by the protocluster environment and if this transformation occurs prior to quenching, or if instead the protocluster and co-eval field galaxies appear morphologically similar at cosmic noon, pointing towards the morphological transformation of protocluster galaxies occurring after quenching has begun.   

The remainder of this paper is as follows.  We introduce the MAMMOTH survey in Section~\ref{sect:data-m} and our analysis methods in Section~\ref{sect:Galapagos}.  We present and contextualize our results in Sections~\ref{sect:results-M} and \ref{Sect:Discussion}.  We adopt a flat $\Lambda$CDM cosmology, using H$_{0}$ = 70km\,s$^{-1}$\,Mpc$^{-1}$, $\Omega_{m}$ = 0.3, and $\Omega_{\Lambda}$=0.7.  Unless noted, all distances are given as proper distances and all magnitudes are given as AB magnitudes. 

\section{Data}\label{sect:data}
\subsection{MAMMOTH protoclusters}\label{sect:data-m}

\begin{figure*}
\begin{center}
\includegraphics[scale=0.8,trim={0.0in 0.0in 0.0in 0.0in},clip=true]{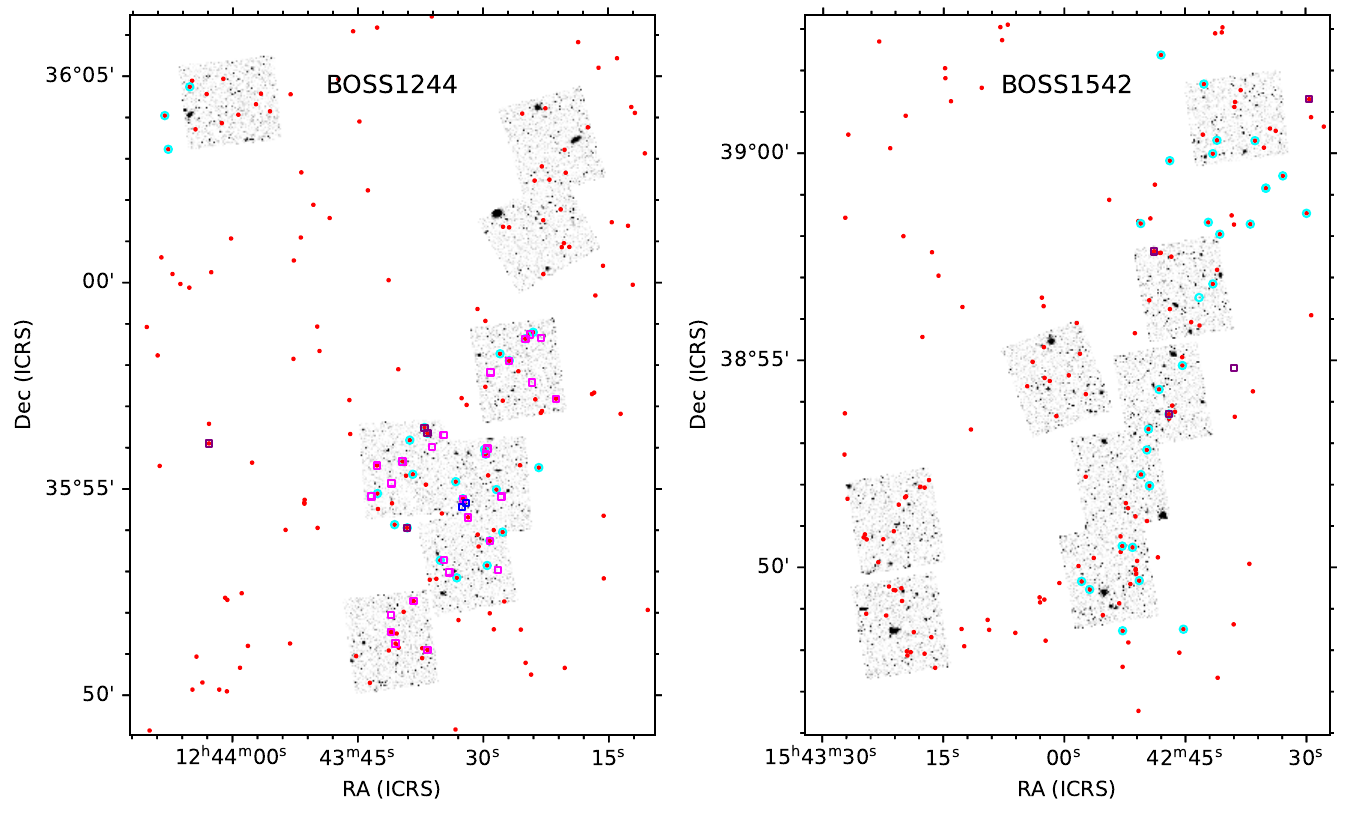}

\caption{The HST coverage of the MAMMOTH protoclusters in our sample.  In each panel, we present our HST WFC3 F160W images.  The overlaid red circular regions denote the HAEs.  The overlaid cyan circular regions denote HAEs that were spectroscopically confirmed in \citet{Shi2021} and fall within the larger HST footprint (a small number of them fall outside the coverage region shown).  In the left panel, the galaxies that were spectroscopically confirmed via HST grism observations are overlaid in magenta squares.  Similarly, the quiescent galaxies identified in \citet{Shi2024} are shown in blue squares.  In both BOSS1244 and BOSS1542, the SDSS-identified co-eval quasars in each field are shown in purple squares.  Some of these quasars were previously identified as HAEs.  In both BOSS1244 and BOSS1542, the densest regions are found in the lower right quadrant with HST coverage.  The total area shown for BOSS1244 and BOSS1542 is $\sim$ 12$\arcmin$ $\times$ 17$\arcmin$.}

\label{Fig:protoclusters}
\end{center}
\end{figure*}

To study the impact of the protocluster environment on galaxy morphology, we need a statistically robust sample of protocluster galaxies at $z$ $>$ 2.0.  To that end, we use well-studied protoclusters from the MAMMOTH survey \citep[e.g.,][]{Cai2016,Cai2017,Zheng2021,Shi2021,Zhang2022,Wang2022,Liu2023,Shi2024,Liu2025,Zhou2025}.  MAMMOTH systems were identified via strong Lyman-$\alpha$ absorption features in SDSS BOSS spectroscopy \citep[e.g.,][]{Cai2016,Cai2017}, indicating the existence of large-scale cool gas, which suggests that these structures are dynamically young (see BOSS1244 and BOSS1542 in Figure~\ref{Fig:protoclusters} for the HST Wide Field Camera 3 [WFC3] F160W coverage of these protoclusters).  Each MAMMOTH system discussed here has been shown to be a massive protocluster structure hosting large populations of high-$z$ HAE galaxies \citep[e.g.,][]{Zheng2021,Shi2021}.

Because our MAMMOTH protocluster populations are selected via HAEs, it is important to understand any biases among this population.  While robust spectroscopic surveys allow for a more uniform identification of protocluster populations down to lower masses, these surveys can be observationally expensive, particularly given the large angular extent of protoclusters, and are currently limited to well-studied survey fields \citep[e.g.,][]{Morishita2023,Staab2024,Forrest2024,Shah2024,Forrest2025,Watson2025}.  As such, HAEs offer a unique alternative to expensive spectroscopy, particularly given that photometric redshift estimates based on narrow-band selected emitting galaxies are accurate \citep[e.g.,][]{Shi2021} and these observations require far less telescope time.  In comparing samples of HAEs at $z$ $\sim$ 2 to co-eval star-forming galaxies, \citet{Oteo2015} found that the distribution of stellar masses and SFRs of HAEs resemble typical star-forming galaxies.  Although the median HAE stellar mass and SFR is slightly above normal star-forming galaxies, HAEs only exclude the bluest and least massive typical star-forming galaxies.  Additionally, \citet{Oteo2015} found that HAE estimated SFRs agree within 0.3\,dex of dust-corrected UV SFRs and that these galaxies lie on the star formation main sequence down to M$_{*}$ $\sim$ 10$^{9.5}$\,M$_{\odot}$.  As such, HAEs appear to be a strong proxy for the total protocluster population, allowing us to trace a representative population of these structures. 

The first protocluster in our sample is BOSS1244.  Originally reported in \citet{Zheng2021}, BOSS1244 was characterized by a rich population of HAEs corresponding to an overdensity of 5.6 $\pm$ 0.3$\sigma$ over a 54 $\times$ 32 $\times$ 32 cMpc$^{3}$ region and includes 244 HAEs \citep{Zheng2021}.  Each HAE was detected via deep K$_{s}$ broadband and H$_{2}$S$_{1}$ narrowband imaging taken with the wide-field infrared camera on the Canada-France-Hawaii Telescope (see \citealp{Zheng2021} for a full description of the image reduction and HAE detection in these protoclusters).  To be classified as an HAE, \citet{Zheng2021} required a minimum H$\alpha$ equivalent width of 45~\AA, which corresponds to $\sim$ 5.1M$_{\odot}$/yr at $z$ $\sim$ 2.24, and should allow for the inclusion of the fainter end of the luminosity function for HAEs as seen in \citet{Lee2012}.  However, as seen in Figure~\ref{Fig:backgroundsample}, the majority of the SFRs for the HAEs in both protoclusters are above 20\,M$_{\odot}$/yr.  We note that this bias in our sample limits our analysis of the morphology of protocluster galaxies to primarily bright, star-forming galaxies. 

BOSS1244 was spectroscopically confirmed and shown to consist of two substructures at $z$ = 2.230 and $z$ = 2.246 (\citealp{Shi2021}; despite the separate redshift peaks, we treat BOSS1244 as one structure in this analysis).  \citet{Shi2021} used the Multiple Mirror Telescope/Magellan Infrared Spectrograph (MMT/MIRS) to observe 46 target HAEs and the Large Binocular Telescope/LUCI (LBT/LUCI) near-infrared spectrograph to observe an additional 15 HAEs (as well as 3 overlapping targets).  In total, 46 HAEs were confirmed to be at the target redshift, including 5 HAEs that are quasars (see Figure~\ref{Fig:protoclusters} for the distribution of spectroscopically confirmed galaxies; see \citealp{Shi2021} for a complete description of the spectroscopic sample of HAEs in BOSS1244).  From these spectroscopic observations, \citet{Shi2021} measure a line-of-sight velocity dispersion of $\sim$ 400\,km\,s$^{-1}$ in each component and cluster masses of M$_{200}$ $=$ 2.8 $\pm$ 2.2 $\times$ 10$^{13}$\,M$_{\odot}$ and M$_{200}$ $=$ 3.0 $\pm$ 2.0 $\times$ 10$^{13}\,$M$_{\odot}$. 

Furthermore, additional protocluster member galaxies were identified as part of the analysis in \citet{Shi2024}, which used HST WFC3 grism slitless spectroscopy to identify two spectroscopically-confirmed quiescent galaxies, including a BCG. Beyond the two spectroscopically-confirmed quiescent galaxies, \citet{Shi2024} report 14 additional spectroscopic members, which were not previously identified narrow-band selected HAEs (for a complete description of these HST grism observations, see \citealp{Shi2024} and \citealp{Wang2022}).  Although we focus on the sample of HAEs, we do include an additional discussion of the spectroscopic sample in Section~\ref{sect:1244}. 

Based on the detection of these HAEs, \citet{Zheng2021} constructed overdensity contours to trace the distribution of emitters.  As seen in both \citet{Zheng2021} and \citet{Shi2021}, BOSS1244 shows two large density peaks, which were confirmed via the velocity distributions measured in \citet{Shi2021}.  Each of these density peaks shows a more concentrated distribution of galaxies, akin to the distribution of galaxies expected in a low-$z$ relaxed galaxy cluster.  Based on the velocity distribution of these HAEs and their estimated masses, \citet{Shi2021} predict that these two components will collapse into a single cluster structure by $z$ = 0.

For this morphology analysis, MAMMOTH also undertook a large HST campaign (P.I. Zheng Cai, Proposal ID = 15266) using the WFC3 camera in the F160W band ($\lambda$$_{o}$=1536.9\,nm, $\Delta$$\lambda$=268.3\,nm) to characterize the galaxy populations in BOSS1244, BOSS1542, and BOSS1441 (see Appendix~\ref{sect:1441} for our analysis of BOSS1441).  At $z$ $\sim$ 2.23, the WFC3 F160W filter is ideal for characterizing the morphology of galaxies as the majority of the optical stellar emission will be redshifted into the IR.  Additionally, the full-width-half-maximum (FWHM) of the PSF in the F160W band is $\sim$ 0$\farcs$151 ($\sim$ 1.25\,kpc at $z$ = 2.23), which allows us to characterize the morphology of high-$z$ galaxies.  The final science images were produced using \textsc{ASTRODRIZZLE} with a scale of 0$\farcs$06/pixel. The limiting magnitude of these HST WFC3 F160W observations is $\sim$ 24.97\,magnitudes for both fields.  Because protocluster structures are much larger than the field-of-view (FOV) of HST WFC3 (2.1 $\times$ 2.3 arcmin$^{2}$), the observations were designed to observe the densest regions of each protocluster, allowing us to cover the most possible HAEs (see the left panel of Figure~\ref{Fig:protoclusters}).  For BOSS1244, this results in 8 WFC3 F160W pointings each with an exposure time of 2614.684\,s covering 91 HAEs (see Table~\ref{tb:emitters} for the breakdown of the number of emitters in each protocluster and how many are observed with HST).  

The second protocluster is BOSS1542, which was spectroscopically confirmed at $z$ = 2.241\citep{Shi2021}.  BOSS1542 has an overdensity of 4.9 $\pm$ 0.3$\sigma$ over the same area as BOSS1244 and includes 223 HAEs that were identified in the same manner as those in BOSS1244 \citep{Zheng2021}.  BOSS1542 was similarly spectroscopically confirmed using MMT/MIRS spectroscopy to observe 23 HAEs and LBT/LUCI was used to confirm an additional 36 galaxies \citep{Shi2021}.  In total, 35 galaxies were found at the target redshift, including three quasars previously identified as HAEs.  From these observations, \citet{Shi2021} estimates a velocity dispersion of $\sim$ 250\,km\,s$^{-1}$ and a mass of M$_{200}$ $=$ 0.79 $\pm$ 0.31 $\times$ 10$^{13}$\,M$_{\odot}$.  Similar to BOSS1244, only the densest regions of the protocluster were observed with HST WFC3 F160W (see the right panel of Figure~\ref{Fig:protoclusters}), resulting in the 8 HST pointings each with an exposure time of 2614.684\,s covering 92 HAEs.    

As discussed in both \citet{Zheng2021} and \citet{Shi2021} and explicitly seen in Figure 1 of \citet{Shi2021}, despite having similar observations and populations of HAEs, the distribution of HAEs in BOSS1542 is more elongated and filamentary, especially when compared to that of BOSS1244.  Despite similar total masses, this difference in the overall protocluster morphology may hint at differences in the dynamical state of BOSS1244 and BOSS1542.  That these protoclusters may be co-eval, but observed at different points in their evolution could add additional uncertainty to our measurement of the morphology of protocluster galaxies as a function of environment as we are only probing two systems.

\begin{table*}[h!]
\caption{Protocluster sample selection statistics}                 
\label{tb:emitters}   
\centering                      
\begin{tabular}{ c c c}      
\hline\hline             

 & BOSS1244 & BOSS1542 \\        
\hline                     
   Total number of emitters & 244 & 223 \\   
   Number of emitters observed with HST & 91& 92 \\
   Number of emitters with successful \textsc{Galapagos} fittings (reduced-$\chi^{2}$ $<$ 2)  & 80  & 71 \\
   Number of n $\geq$ 2 galaxies & 25 & 24 \\ 
\hline                                
\end{tabular}
\end{table*}

\subsection{Estimating galaxy morphology}\label{sect:Galapagos}
We estimate the morphology of protocluster galaxies using a S\'ersic model \citep{Sersic1963} to characterize the light profile of each galaxy.  Given that the MAMMOTH protoclusters are at $z$ $\sim$ 2.23, where the classical model of a spiral galaxy (n = 1) and elliptical galaxy (n = 4) does not necessarily characterize the majority of galaxies \citep[e.g.,][]{Ferreira2023}, we classify our galaxies as either early-type (n $\geq$ 2), or late-type (n $<$ 2) following a similar criteria presented in literature \citep[e.g.,][]{Strazzullo2013,Noordeh2021}.  We also include an additional subset of early-type galaxies, which we refer to as ``strongly'' bulge-dominated early-type galaxies (n $\geq$ 3) to account for the most early-type looking galaxies.    

To measure the morphology of each protocluster galaxy, we use \textsc{Galfit} \citep{Peng2002,Peng2010}, a 2D data analysis tool that directly fits a galaxy's light profile and measures its morphology.  \textsc{Galfit} works by creating a model for each galaxy, subtracting the model from the original image, and creating a residual to estimate the goodness of the fit.   For this analysis, we use a PSF made from stacking non-saturated stars observed in the regions surrounding BOSS1244 and then use both a science image cutout and a sigma image to create a model of a single component S\'ersic galaxy, chosen so as to not overfit the data, and characterize each galaxy in terms of its radius (r), S\'ersic index, axis ratio, ellipticity, and magnitude.  Beyond any initial conditions regarding values of S\'ersic index and radius, \textsc{Galfit} simultaneously fits all components of a given galaxy, taking into account both the detected object and the sky background \citep{Peng2002,Peng2010}, as well as convolving all modeled profiles based on the provided PSF to simulate telescope optics.

To increase the computational efficiency of our analysis, we use \textsc{Galapagos-2/MEGAMORPH} \citep{Barden2012,Haussler2013}, an IDL-based \textsc{Galfit} wrapper that combines \textsc{Galfit} with \textsc{Source Extractor} \citep{Bertin1996}.  Instead of providing a single postage stamp image of each target galaxy, \textsc{Galapagos} samples the entire mosaic and does two runs of \textsc{Source Extractor} in a ``hot'' and ``cold'' detection mode to create multiple detection maps to separate and identify all galaxies in a given field, which is particularly useful in dense protocluster fields.  From these detections, we can use \textsc{Galapagos} to create both image cutouts and fit all of the parameters via \textsc{Galfit}.  When running \textsc{Galapagos}, we follow the same input parameters as specified in \citet{vanderWel2012} and \citet{Afanasiev2023} (e.g., 0.2 $\leq$ n $\leq$ 8, 0$\farcs$3 $\leq$ r, and use a tophat\_9.0\_9x9 filter for the cold \textsc{Source Extractor} run and a gauss\_4.0\_7x7 filter for the hot \textsc{Source Extractor} run), each of which analyzed the morphology of high-$z$ ($z$ $\sim$ 2) galaxies observed with HST.  Importantly, given the necessary precision for high-$z$ morphology measurements, we limit our analysis to galaxies brighter than m$_{F160W}$ $<$ 24.5 magnitudes, following \citet{vanderWel2012}, which is approximately half a magnitude above our detection limit and includes $\sim$ 97$\%$ of our HAEs.  Although the limiting magnitudes differ between this sample and the CANDELS co-eval field sample (see Section \ref{sect:background}), the 24.5 magnitude analysis limit is brighter than both detection limits, allowing us to compare our analysis to results from \citealp{vanderWel2012}).  Although the S\'ersic index measurements for galaxies fainter than this limit have similar error bars to their brighter counterparts, we find that the for a small sample of these faint galaxies, the radii tend to be overestimated, possibly due to confusion with the background, which impacts the goodness of the overall fit.  

\begin{figure*}
\begin{center}
\includegraphics[scale=0.35,trim={0.0in 0.0in 0.0in 0.0in},clip=true]{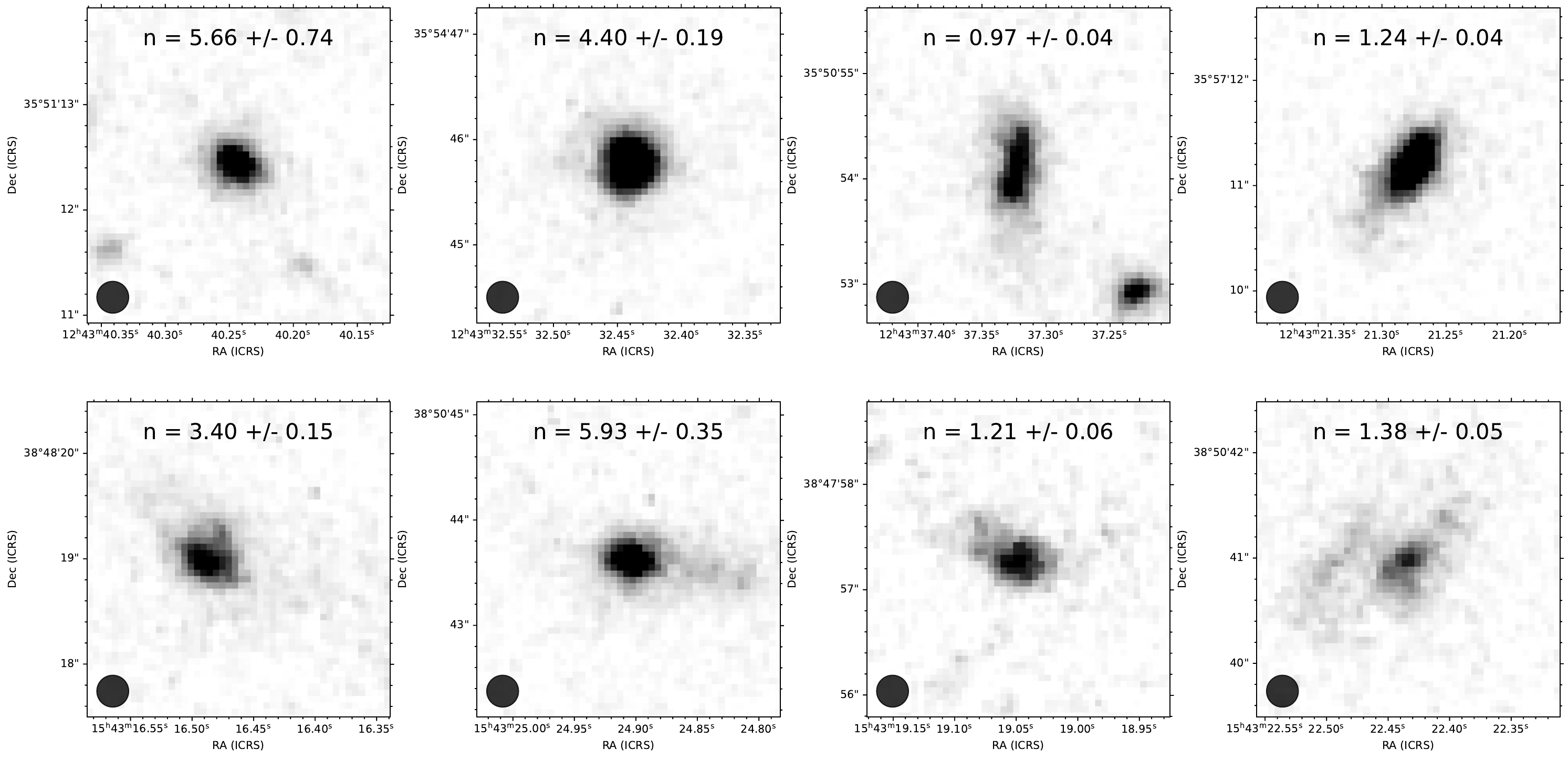}

\caption{Examples of image cutouts of HST WFC3 F160W detected HAEs in BOSS1244 (top row) and BOSS1542 (bottom row).  Early-type galaxies are shown in the left two columns and late-type galaxies are shown in the right two columns.  The S\'ersic index of each galaxy is included at the top of each cutout.  Each cutout is 2$\farcs$88 $\times$ 2$\farcs$88.  The size of the FWHM of the HST WFC3 F160W PSF is shown as the black circle in the bottom of each cutout.}

\label{Fig:Examples}
\end{center}
\end{figure*}

After running \textsc{Galapagos}, we visually inspect each target, model, and residual to ensure correct identification and check for any galaxies where the fit residual was particularly poor.  We find that \textsc{Galapagos} fails for $\sim$ 5$\%$ of our targets (e.g., no resulting fit is given, despite a galaxy being included as a target). Typically, these galaxies are the brightest, unresolved galaxies in our sample, some of which are among the sample of HAEs that were identified as quasars in \citet{Shi2021}, where the size of the galaxy is consistent with that of our observed PSF.  Additionally, we further inspect the sample and remove any galaxy where the reduced-$\chi^{2}$ of the fit was greater than 2, to only keep the best fit galaxies in our sample (see Appendix~\ref{sect:GalapagosFits} and Figure~\ref{Fig:Residuals} for examples of the residuals of our S\'ersic index fits).  This leaves us with 80 galaxies in BOSS1244 and 71 galaxies in BOSS1542 (see Figure~\ref{Fig:Examples} for examples of HAEs).  Additionally, while \textsc{Galapagos} identifies the target galaxy nearest to our emitting galaxy input coordinates, because the angular resolution of the K$_s$-band images is much worse than our HST images (the seeing was $\sim$0$\farcs$65 - 0$\farcs$8 for the K$_s$-band images), we have multiple examples of previously blended systems.  For our analysis, if we have two distinct galaxies in HST, we choose the galaxy closest to the coordinates of the HAE (which is typically the brighter galaxy).  However, for five pairs, the two galaxies are truly blended in the K$_s$ band, with the central coordinate of the HAEs falling squarely between the two galaxies in the HST imaging.  In those instances, we include both galaxies. 

\begin{figure*}
\begin{center}
\includegraphics[scale=0.35,trim={0.0in 0.0in 0.0in 0.0in},clip=true]{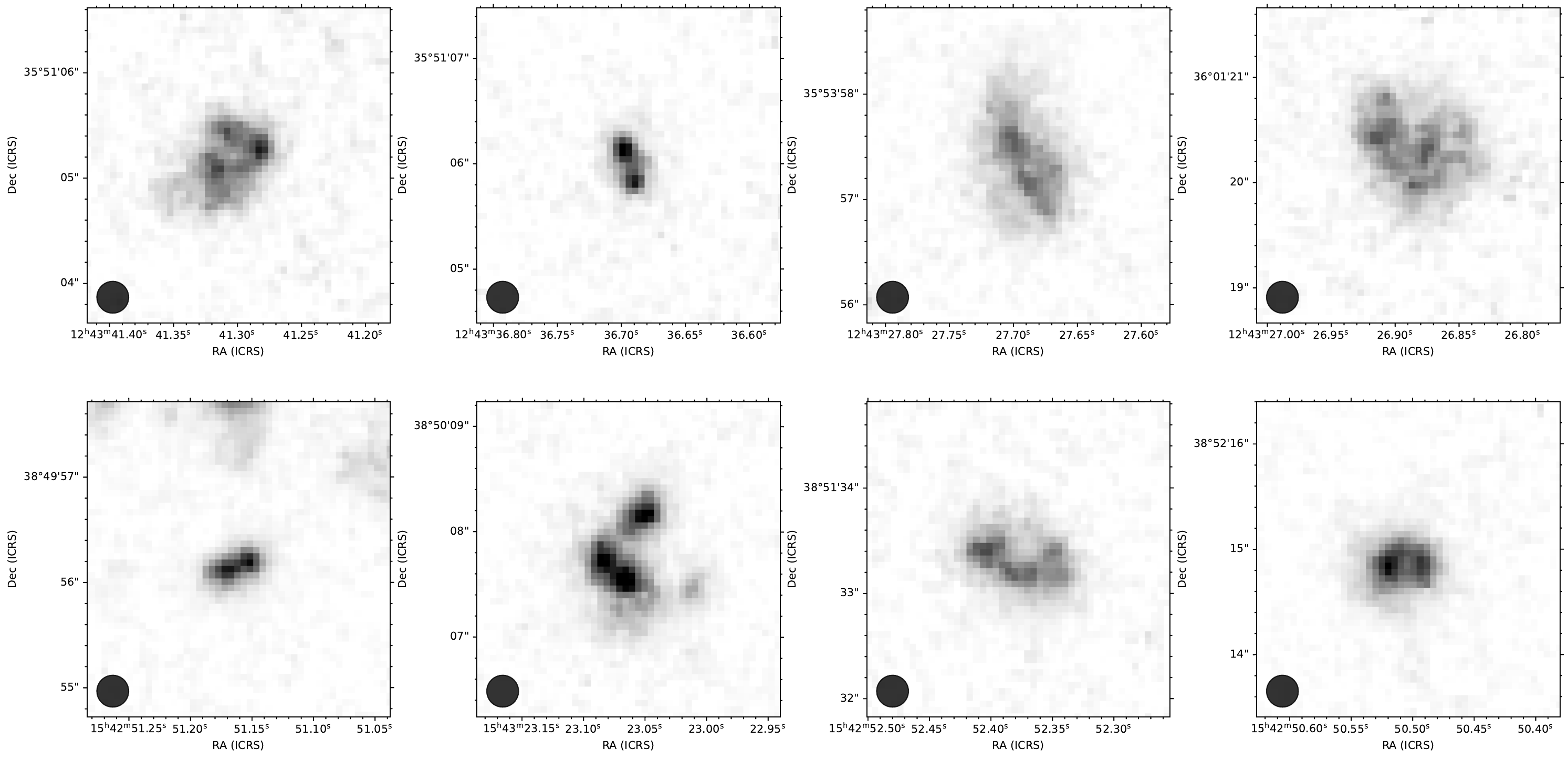}

\caption{Examples of image cutouts of HST WFC3 F160W detected multi-peak galaxies identified in BOSS1244 (top row) and BOSS1542 (bottom row).  Each cutout is 2$\farcs$88 $\times$ 2$\farcs$88.  The size of the FWHM of the HST WFC3 F160W PSF is shown as the black circle in the bottom of each cutout.  These galaxies were detected as a singular galaxy with \textsc{Source Extractor/Galapagos}, but show multiple bright peaks, potentially indicative of a merging and/or clumpy  galaxy.  }

\label{Fig:Mergers}
\end{center}
\end{figure*}

Given the proclivity of mergers in high-$z$ protoclusters \citep[e.g.,][]{Watson2019,Liu2023,Liu2025,Giddings2025}, as well as clumpy galaxies \citep[e.g.,][]{Shibuya2016,Umehata2025,Kalita}, it is possible that some of these galaxies are merging or clumpy systems.  As shown in \citet{Liu2023}, both BOSS1244 and BOSS1542 host a number of potentially merging galaxy pairs.  Specifically, \citet{Liu2023} perform an analysis of potential ``close pairs," galaxies, which are near each other in physical space, but do not show obvious signs of mergers \citep[e.g.,][]{Kartaltepe2007,Lotz2011,Snyder2017,Giddings2025}.  Here, we adopt a more cautious approach for identifying merging systems.  For the close pairs identified in \citet{Liu2023}, we only report on the pair members that are HAEs (we do identify one close pair of HAEs in BOSS1244 and two close pairs of HAEs in BOSS1542).  However, potentially merging galaxies are not limited to close pairs in protoclusters or in the field.  Following our initial visual inspection of the \textsc{Galapagos} postage stamp and the model residual image, we inspect any merger candidates using \citet{Kartaltepe2015} as a guide to classify potentially isolated, merging (or clumpy), and interacting galaxies (see Figure~\ref{Fig:Mergers}). Using this approach, we look for the existence of tidal tails/elongated galaxy structures between nearby galaxies or the appearance of merging systems.  Although we identify potentially interacting systems, our interpretations are limited by our lack of spectroscopy for many of the multi-galaxy interacting systems.  As such, we focus on mergers, where we visually identify a single galaxy (based on \textsc{Galapagos} and \textsc{Source Extractor} detections) with a ``multi-peak'', seen visually as multiple bright peaks in the same galaxy (see Figure~\ref{Fig:Mergers} for examples of multi-peak systems; see Section~\ref{sect:mergers-M} for further discussion of these systems).  Although some of these may be edge-on spiral galaxies or clumpy star-forming galaxies, we find some show additional signs of mergers (e.g., diffuse tidal tails). 

\subsection{Measuring the local density}
Since the HST coverage of our protocluster candidates does not cover the entire structure, we cannot create a robust measurement of the protocluster density relative to the co-spatial field at all points in the protocluster.  Instead, we focus on measuring the local density of every HAE.  Although all galaxies in our sample are emitters, we only characterize the morphology of HST detected galaxies brighter than m$_{F160W}$ $<$ 24.5\,mag \citep{vanderWel2012}.  While there is not a one-to-one relation between the K$_{s}$ band magnitude and the HST WFC3 F160W magnitude, given that 97$\%$ of our galaxies are above the detection threshold, we can estimate the density without adding additional sources of error.  For our analysis, we characterize the local density by measuring the number of emitting galaxies within 300\,kpc\footnote{Although we use 300\,kpc in this work, we measure the difference in environment using regions between 200\,kpc and 800\,kpc and ultimately chose 300\,kpc because of the clearest measurement of HAEs in low and high-density regions that allowed for two similarly-sized density bins.} of each HAE.  

As our density measurements trace a population of star-forming and relatively massive HAEs, it is important to note that there may be some biases regarding how HAEs map to the density of the entire protocluster population.  Specifically, if either protocluster contains a large population of fainter star-forming or quiescent galaxies that are not co-spatial to the denser regions of the HAEs, then some of our lower density regions may actually have higher densities than our HAE estimate.  In fact, in simulations, \citet{Baxter2025} traced the location of the highest density peak in protoclusters at 1 $<$ $z$ $<$ 5 and find that the location of the highest density peak in protoclusters at $z$ $>$ 2 does depend on the stellar mass or SFR cuts, suggesting a degree of variability among density measurements.  Using our co-eval background field (see Section~\ref{sect:background}), we tested our density measurements with different stellar mass and star formation rate cuts.  Doing so, we find the most scatter among the lowest density regions.  However, these lower density measurements are generally below the lower density bins in the protocluster sample.  This suggests that the majority of the density measurements in the protocluster sample should not be impacted greatly by changes in our selection criteria.  Although these density fluctuations are likely a smaller effect in the densest regions, some degree of error in our morphology-density measurements (see Section~\ref{sect:results-M}) may be due to the density measurements.

\subsection{The comparison field samples}\label{sect:background}
To determine whether the morphology of protocluster galaxies is influenced by the protocluster environment, we need a co-eval field sample. For comparison, we use the well-analyzed CANDELS \citep[e.g.,][]{Grogin2011,Koekemoer2011,Kodra2023} survey fields because \citet{vanderWel2012} previously characterized the morphology of CANDELS galaxies and measured the S\'ersic index of the galaxies using the HST WFC3 F160W band.  Since our HST imaging was done in the same F160W band and both the protocluster sample and a co-eval field sample have limiting magnitude depths fainter than the 24.5 magnitude analysis threshold and used identical \textsc{Galapagos} analysis tools, we can compare our observations to these values to constrain the impact of the protocluster environment on galaxy morphology.  However, because our HAEs have a very narrow redshift range, based on our narrow-band emission (2.246 $\pm$ 0.02, in agreement with the measurements from \citealp{Shi2021}), we have to statistically create our co-eval field galaxy sample.  We first treat all galaxies with m$_{F160W}$ $\leq$ 24.5\,magnitudes (the magnitude limit of our morphology analysis) in the CANDELS fields as potential high-$z$ galaxies. Using robustly measured photometric redshifts \citep{Kodra2023}, we account for the likelihood that each galaxy lies at the redshift of our HAEs.  To do this, we do a first-order estimate of the redshift probability distribution function for each galaxy assuming that each galaxy's photometric redshift follows a Gaussian distribution centered on the reported value with the width estimated from the error in the photometric redshift.  We then integrate the area under the curve corresponding to our narrow redshift window (2.246 $\pm$ 0.02).  Because the median photometric redshift uncertainty for CANDELS high-$z$ galaxies is $\Delta$$z$ $\sim$ 0.2, the statistical likelihood that these galaxies fall within our redshift range is low, typically $\sim$ 40$\%$ at most.  Thus, to capture the statistical co-eval field sample, we sum the total weight of the likelihood of each galaxy being in our redshift range assuming Gaussian error.  To measure the density of the galaxies, we use the same 300\,kpc (at $z$ $\sim$ 2.23) region as our target fields (noting that we exclude galaxies near the edges of the CANDELS fields to avoid undersampling the density) and measure the weighted likelihood that all of the galaxies in the 2D projection are at our target redshift.  As a result, we can estimate the density and morphology of typical non-protocluster galaxies at cosmic noon.  

\begin{figure}
\begin{center}
\includegraphics[scale=0.65,trim={0.15in 0.1in 0.1in 0.1in},clip=true]{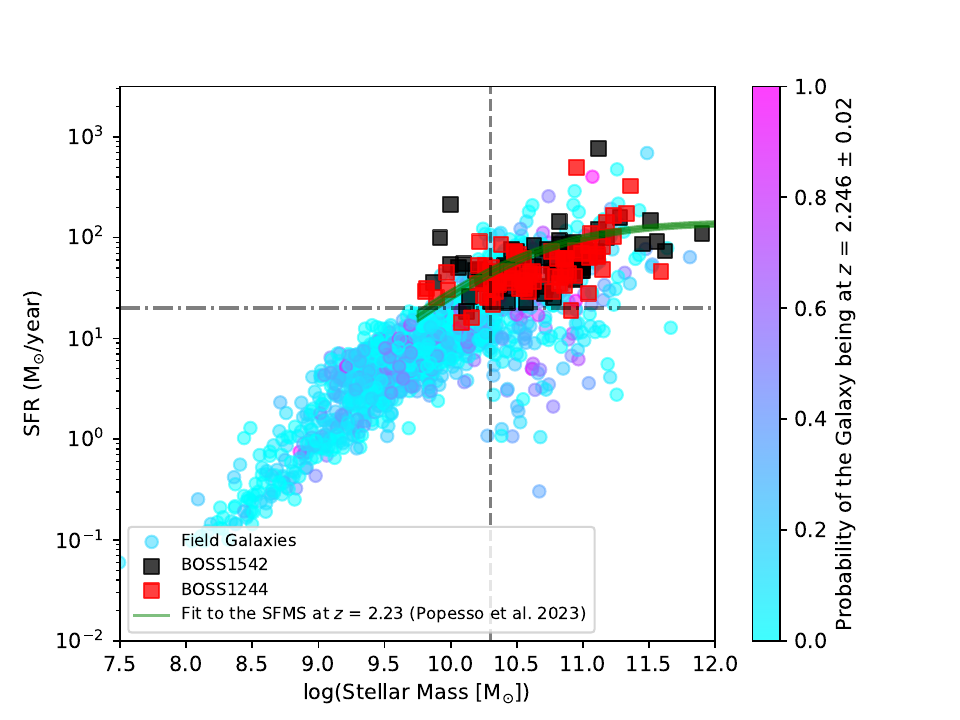}

\caption{The SFR as a function of stellar mass for our samples of HAEs from BOSS1244 and BOSS1542 overplotted against the co-eval sample of field galaxies with SFRs and stellar masses from \citet{Osborne2024} in the CANDELS fields.  The color of the co-eval field sample galaxy represents the likelihood that a given galaxy is at $z$ = 2.246 $\pm$ 0.02 based on the photometric redshifts from \citet{Kodra2023}.  We overplot an estimate of the star formation main sequence at $z$ = 2.23 based on the analysis in \citet{Popesso2023} in green.  Because the star formation main sequence is calibrated using a Kroupa IMF, we convert the star formation rates and stellar masses to reflect the Chabrier IMF used in \citet{Osborne2024}.  Although we measure the likelihood that all galaxies, regardless of the median redshift estimate are at $z$ = 2.246, in the above plot, we only include those galaxies with a median photometric redshift (or spectroscopic redshift) of 2 $<$ $z$ $<$ 3. The vertical line represents the 90$\%$ stellar mass completeness limit for BOSS1244 and BOSS1542 (10$^{10.3}$\,M$_{\odot}$) based on the analysis in \citet{Liu2023}.  The horizontal line shows our estimate of the SFR completeness limit (20\,M$_{\odot}$/yr).}

\label{Fig:backgroundsample}
\end{center}
\end{figure}

Although less common, some fraction of the high-$z$ galaxies from the CANDELS catalogs from \citet{Kodra2023} are quiescent galaxies \citep[e.g.,][]{Merlin2019}.  As our sample of protocluster HAEs does not include quiescent galaxies\footnote{Although not included in the initial sample, as mentioned in Section~\ref{sect:data}, there are two spectroscopically-confirmed quiescent galaxies in BOSS1244.  We include these galaxies in our analysis in Section~\ref{sect:1244}}, we create an additional co-eval field consisting only of those galaxies in CANDELS with measured SFRs and stellar masses from \citet{Osborne2024}.  \citet{Osborne2024} estimate their SFRs using UV, optical, and mid-IR photometry for 63,266 galaxies at 0.7 $<$ $z$ $<$ 2.3 in the GOODS-S, UDS, COSMOS, and EGS fields.  With these additional parameters, we create an additional co-eval field sample.  This CANDELS co-eval field subsample (labeled ``CANDELS M$_{*}$ \& SFR lim'' in Figures~\ref{Fig:Morph-density} and \ref{Fig:Morph-density-v2}) consists of all galaxies with m$_{F160W}$ < 24.5 with a stellar mass greater than 10$^{10.3}$M$_{\sun}$ and a SFR > 20 M$_{\odot}$/yr.  The stellar mass limit and the stellar masses of the galaxy sample in BOSS1244 and BOSS1542 are presented in \citet{Liu2023}.  Although the detection limit for identifying an HAE in our sample was reported to yield a SFR of 5.1\,M$\sun$/yr in \citet{Zheng2021}, as shown in Figure~\ref{Fig:backgroundsample}, the minimum dust-corrected HAE star-formation rate, calculated following the description for HAEs in \citet{Zheng2021}, is well above this.  As such, for a better statistical sample, we use a limiting SFR of 20\,M$_{\odot}$/yr.  For the co-eval field subsample, we again measure the density of galaxies following the same prescription as we did for the CANDELS field sample, where we weight the likelihood that each galaxy is within our narrow redshift range (as seen based on the color of the co-eval field galaxies in Figure~\ref{Fig:backgroundsample}) and then measure the weighted number of galaxies within 300\,kpc of each targeted galaxy (see Table~\ref{tb:field} for the weighted number of galaxies in each sample).

We acknowledge that although we apply similar stellar mass and SFR cuts between our two samples, the protocluster population is entirely HAEs, while the background sample is star-forming galaxies.  As highlighted in \citet{Oteo2015} and as seen in Figure~\ref{Fig:backgroundsample}, HAEs should probe a similar parameter space to typical star-forming galaxies at $z$ $\sim$ 2.0 and only differ among the least massive systems. Since our magnitude limit is $\sim$ 90$\%$ complete down to masses of 10$^{10.3}$\,M$_{\odot}$, we are not probing low-mass galaxies.  As such, we do see strong agreement between the parameters of these galaxies in terms of SFR and stellar mass (see Appendix~\ref{sect:Field} and Figure~\ref{Fig:StatsComp} for a comparison of the distribution of S\'ersic index, stellar mass, and SFR between the protocluster and co-eval field samples).  Additionally, we note that all of our HAEs, as well as the co-eval field sample fall within 1\,dex of the star formation main sequence at $z$ = 2.23 from \citet{Popesso2023} (see Figure~\ref{Fig:backgroundsample}).

\begin{table*}[h!]
\caption{Co-eval field samples}                 
\label{tb:field}   
\centering                        
\begin{tabular}{l c c }      
\hline\hline              
 & CANDELS & CANDELS M$_{*}$ \& SFR lim \\        
\hline                    
   Total Weighted Number of Galaxies & 138.04 & 39.57 \\    
   Weighted Number of Bright Galaxies & 21.22 & 16.33  \\
   
\hline
\footnotetext{}
                           
\end{tabular}

\end{table*}

\section{Results}\label{sect:results-M}
\subsection{The morphology-density relation}
After measuring the morphology of each HAE with \textsc{Galapagos} and the density of the surrounding galaxies, we plot the morphology as a function of local density for the star-forming HAEs in BOSS1244 and BOSS1542 (see Figure~\ref{Fig:Morph-density}; for the analysis of BOSS1441, see Appendix~\ref{sect:1441} and Figure~\ref{Fig:Morph-density-ALL3}). 
To search for environmental trends, we separate our galaxies into a low- and high-density sample, aiming for two approximately equal size density bins.  To quantify the error in these measurements, we use a bootstrap technique where we run our analysis 1000 times, treating the initial error in the S\'ersic index from \textsc{Galapagos} as Gaussian.  We estimate what fraction of galaxies are early-type in each bin and the error in this measurement, which accounts for galaxies that are just below/very near our S\'ersic index division and are borderline cases.  We then add our error in quadrature with the error estimated from the binomial confidence intervals to get the final error bars shown in Figures~\ref{Fig:Morph-density} and \ref{Fig:Morph-density-v2}.  For the density measurements, we estimate the error by taking the total number of emitters within the K$_{s}$-band FOV and randomly placing them in the field and then measuring the number of galaxies within 300\,kpc of each galaxy.  We similarly bootstrap this density and estimate the median number of nearby neighbors as the error in the density.    

\begin{figure}[hbt!]
\begin{center}
\includegraphics[scale=0.7,trim={0.0in 0.6in 0.0in 1.0in},clip=true]{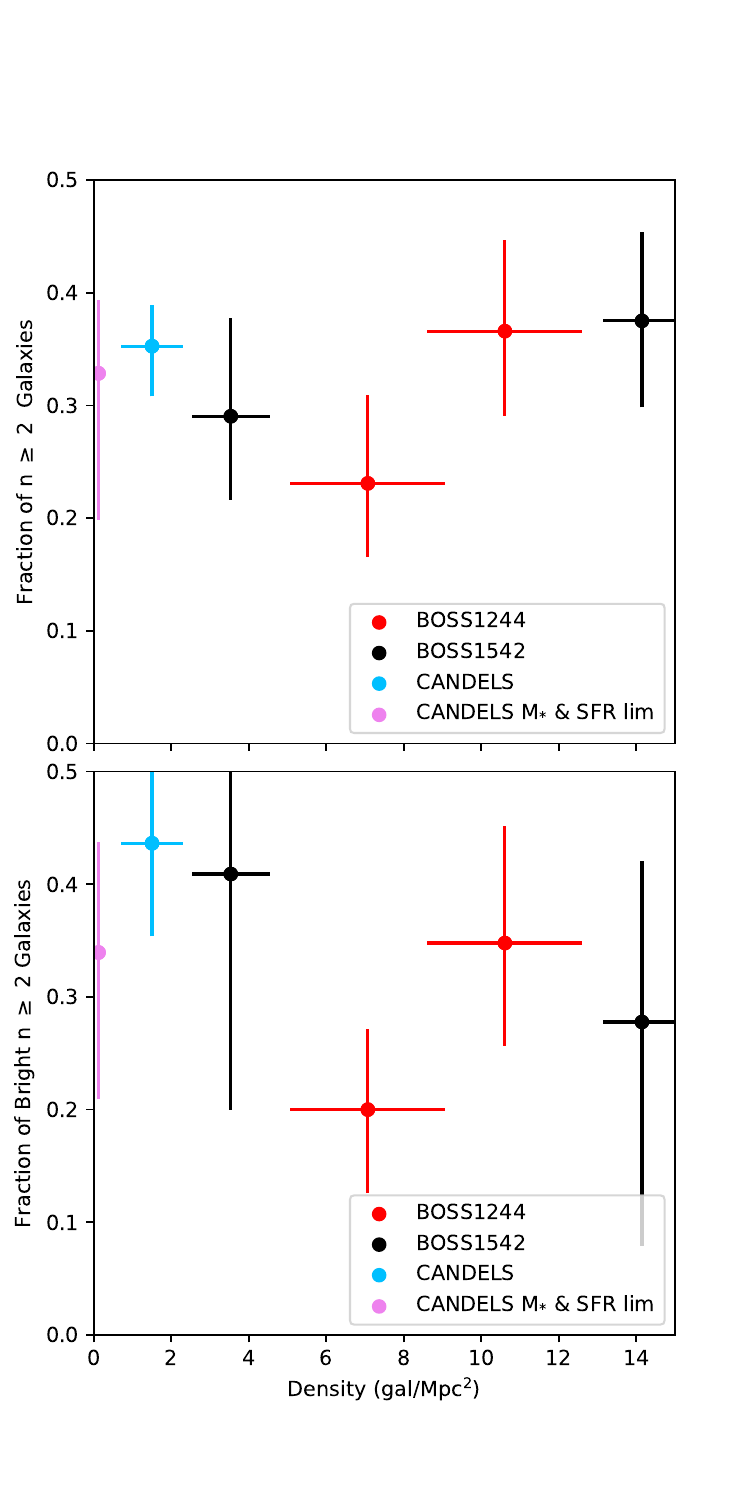}

\caption{The morphology-density relation in two MAMMOTH protoclusters, BOSS1244 (in red) and BOSS1542 (in black).  The weighted CANDELS field is shown in blue and the CANDELS stellar mass and SFR sample (CANDELS M$_{*}$ \& SFR lim) is shown in magenta.  The top panel shows all detected galaxies brighter than our magnitude threshold (m$_{F160W}$ $\leq$ 24.5), while the bottom panel shows all detected galaxies brighter than m*+1, where m* is the modeled magnitude of an L$^{*}$ galaxy at $z$ = 2.246 estimated using EzGal \citep{Mancone2012}  (i.e., m$_{F160W}$ $\leq$ 23.36\,magnitudes).  In each plot, the density measurement is for the entire sample.  The value of the fraction of galaxies with $n$ $\geq$ 2 for each sample is measured by bootstrapping over the measured errors in S\'ersic index over 1000 iterations.  We find evidence for an internal morphology-density trend in both protoclusters among all galaxies, but no evidence for an enhancement of the early-type fraction relative to the field.  However, we do find a difference among the brighter galaxies in each protocluster, where the denser regions of BOSS1542 have fewer early-type massive galaxies.}

\label{Fig:Morph-density}
\end{center}
\end{figure}

The top panel of Figure~\ref{Fig:Morph-density} highlights that both protocluster systems show an increase in the number of early-type galaxies in the denser bin.  However, because these values differ by less than 1$\sigma$, we cannot statistically confirm the morphology-density relation in these two protoclusters.  Since we do not have HST coverage of the entire field and only observe the denser regions of each protocluster, we cannot probe the outskirts of the protoclusters, which limits our ability to look at the build-up of this relation from the outskirts of the protocluster, where the early-type fraction should be closer to the field, to the densest regions.  However, relative to the co-eval field samples we construct from CANDELS, we do not find a statistical enhancement in the fraction of early-type galaxies.  In fact, we find that the early-type fraction in the low-density bin is below field levels in BOSS1244 and that all of the values in the higher density bins are within 1$\sigma$ of the field values.  

To further quantify if we see any differences among the field and protocluster populations, we perform a KS test to compare the distribution of S\'ersic indices among the statistically similar stellar mass and SFR limited co-eval field sample to the two protoclusters (see Figure~\ref{Fig:StatsComp}), finding no evidence that they are drawn from different populations (P$_{1244 - Field}$ = 0.83, P$_{1542 - Field}$ = 0.98).  Ultimately, we find little evidence that the morphology of these protocluster HAEs differs greatly from their field counterparts.  However, given that our sample only includes primarily massive and relatively high star-forming HAEs, it is possible that any morphological transformations due to the protocluster environment occur among populations of galaxies with either lower SFRs that we are not sensitive to or lower masses that we are not probing, which would be similar to the dominance of mass quenching among bright cluster galaxies seen out to $z$ $\sim$ 1.5 \citep[e.g.,][]{YPeng2010,Hewitt2025} (see Section~\ref{Sect:Discussion} for further discussions on the morphological transformation of protocluster galaxies).  Additionally, as seen in Figure~\ref{Fig:Mergers}, we identify a number of multi-peak galaxies, which may not be well characterized by a S\'ersic index and could impact our lack of evidence of a strong morphological difference between protocluster and field galaxies (see Section~\ref{sect:mergers-M}).    

As discussed in \citet{Shi2024}, BOSS1244 contains a population of massive and bright protocluster galaxies.  Because high-$z$ massive cluster galaxies have been shown to be more quenched than their lower mass counterparts \citep[e.g.,][]{Rudnick2012,Kawinwanichakij2017,Lee-Brown2017} and in simulations, protocluster populations are found to be older than their field counterparts \citep[e.g.,][]{Hatch2014,Overzier2016}, we compare the fraction of massive, bright early-type galaxies in our protoclusters to their field counterparts in the bottom panel of Figure~\ref{Fig:Morph-density} (we use m$_{F160W}$ $<$ 23.36, 1 magnitude fainter than the magnitude of an L* galaxy at z = 2.246 as estimated in the HST WFC3 F160W band using EzGal; \citealp{Mancone2012}).  While the overall results for both BOSS1244 and the two co-eval field samples remain relatively unchanged, we see a stark difference between the internal distributions of early-type galaxies in BOSS1244 and BOSS1542, with BOSS1542 hosting more massive early-type galaxies in the low-density region than the high-density region, though the difference is not statistically robust.

Within the full sample, we note a number of massive, strongly bulge-dominated early-type galaxies, with S\'ersic indices $\geq$ 3, similar to what was seen in \citet{Shi2024}.  To determine if this population exists beyond the smaller spectroscopic sample in \citet{Shi2024} and in the population of emitters as a whole, we look at the fraction of strongly bulge-dominated early-type galaxies (n $\geq$ 3) in each protocluster system (see Figure~\ref{Fig:Morph-density-v2}; we note the density measurements in this figure represent the density of all galaxies and are the same as in Figure~\ref{Fig:Morph-density}).  As clearly shown, the trends between co-eval field samples and distributions in BOSS1244 are not changed in any statistically significant way.  Interestingly, for both the entire sample and the sample of only bright n $\geq$ 3 early-type galaxies, the distribution in BOSS1542 is again inverted.  Given that BOSS1244 shows a stronger local morphology-density trend among bright galaxies than BOSS1542, though neither trend is statistically robust, this may represent additional evidence, along with the previously reported quiescent galaxies in BOSS1244 and the difference in protocluster morphology, that these protoclusters are at different evolutionary states.

\begin{figure}[hbt!]
\begin{center}
\includegraphics[scale=0.7,trim={0.0in 0.65in 0.0in 1.0in},clip=true]{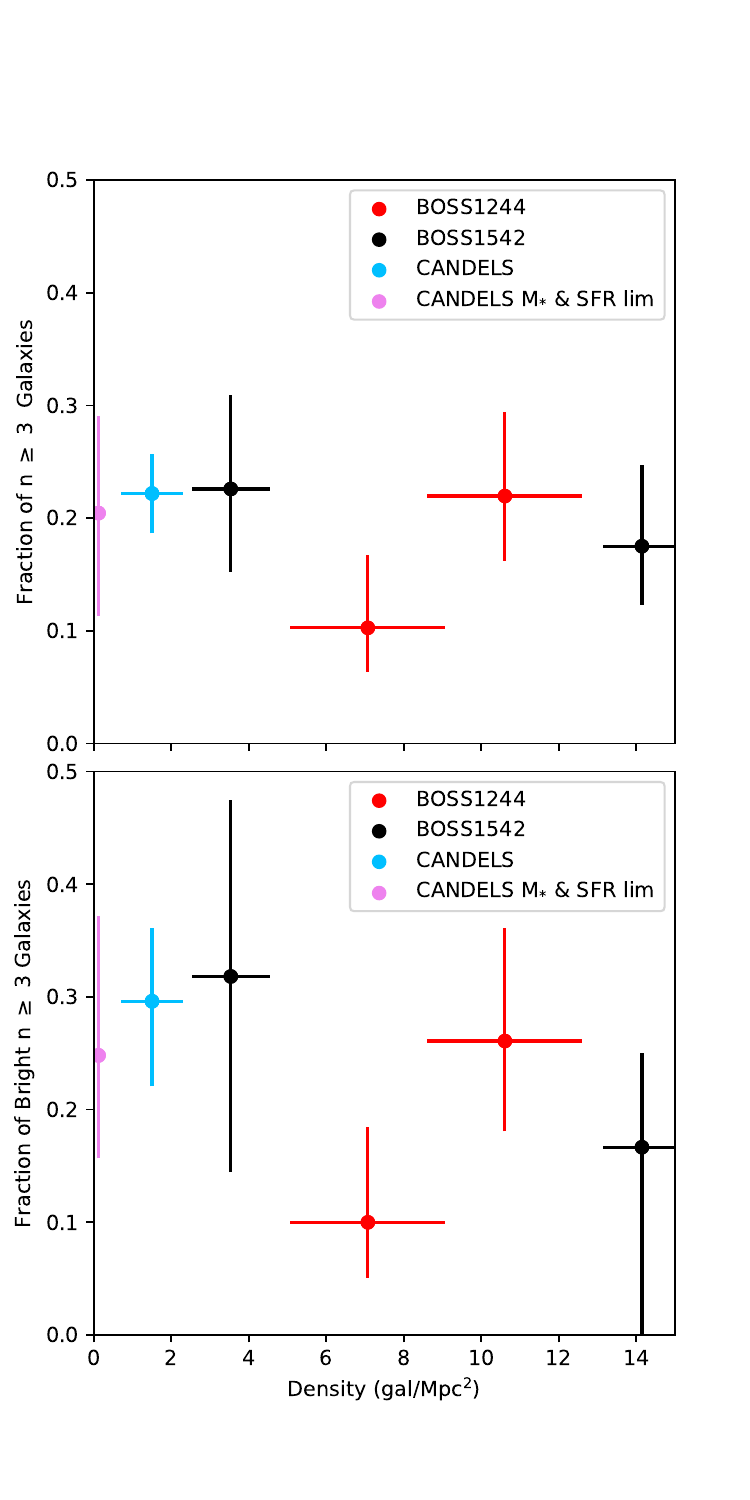}

\caption{The morphology-density relation in BOSS1244 and BOSS1542 for strongly bulge dominated galaxies (n $\geq$ 3).  The same legend is used as in Figure~\ref{Fig:Morph-density}.  The top panel shows all detected galaxies brighter than our magnitude limit (m$_{F160W}$ $< $ 24.5), while the bottom panel shows all detected galaxies brighter than an m$^{*}$+1 galaxy at $z$ = 2.246.  In each plot, the density measurement is for the entire sample.  As in Figure~\ref{Fig:Morph-density}, the value of the fraction of galaxies with $n$ $\geq$ 3 for each sample is measured by bootstrapping over the measured errors in S\'ersic index over 1000 iterations.  Here, we see a local morphology-density relation in BOSS1244, but no such relation in BOSS1542, pointing to differences in the evolutionary states of these protoclusters. }

\label{Fig:Morph-density-v2}
\end{center}
\end{figure}

\subsection{Galaxy mergers and clumpy galaxies}\label{sect:mergers-M}
\begin{figure}[hbt!]
\begin{center}
\includegraphics[scale=0.7,trim={0.0in 0.65in 0.0in 1.0in},clip=true]{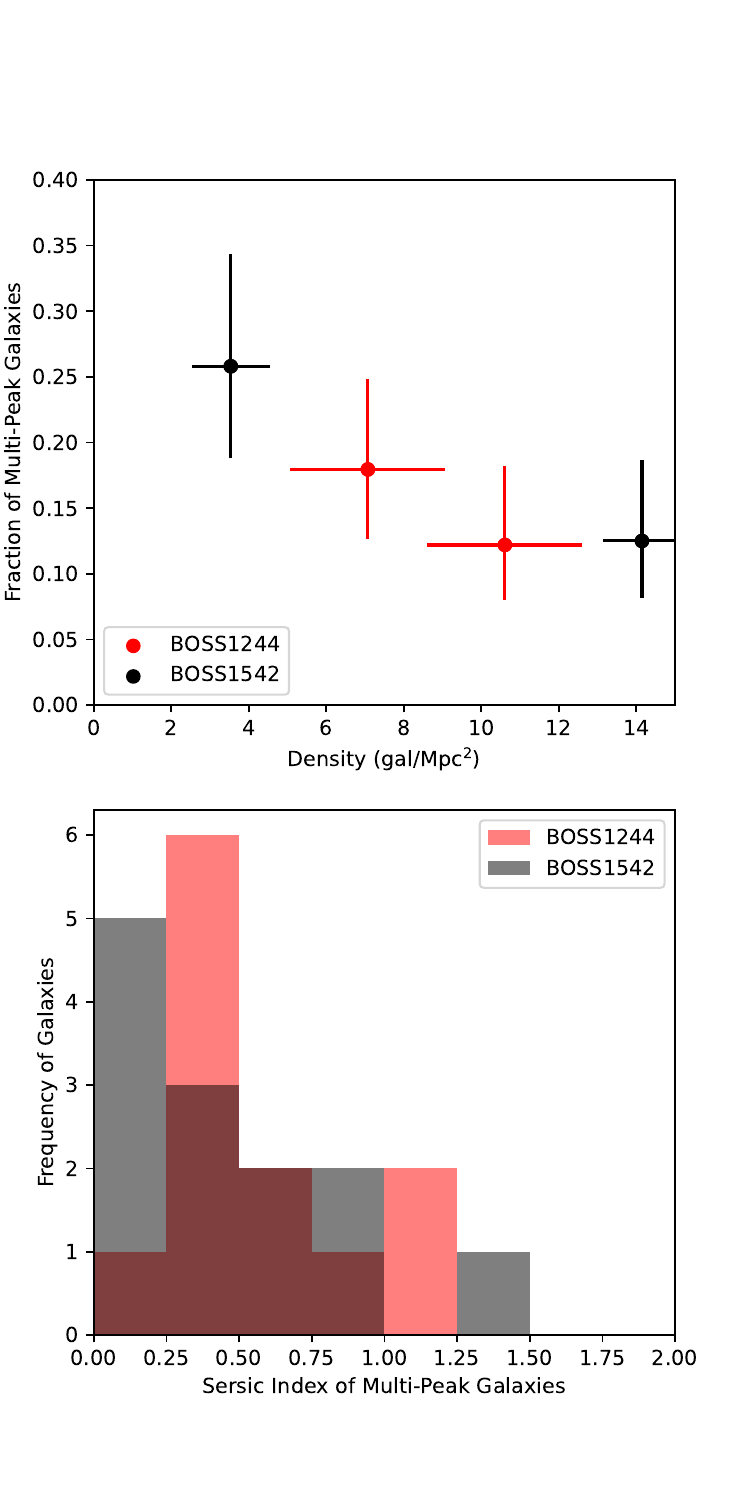}

\caption{Top Panel: The merger-density relation for the multi-peak galaxies in two MAMMOTH protoclusters.  Bottom Panel: A histogram of the S\'ersic Indices of multi-peak galaxies.  As previously noted, BOSS1244 is shown in red and BOSS1542 is shown in black.  We find fewer multi-peak galaxies in the denser regions of both structures and classify all of these systems as late-type galaxies.}

\label{Fig:Merger-Density}
\end{center}
\end{figure}

To further probe galaxy evolution within MAMMOTH protoclusters, we revisit the population of merging/clumpy galaxies.  As highlighted in \citet{Liu2023} and \citet{Liu2025}, both BOSS1244 and BOSS1542 include large populations of close pair galaxies that are merger candidates.  Such mergers could be key to determining the differences between Figures~\ref{Fig:Morph-density} and \ref{Fig:Morph-density-v2} and the previously discussed stronger morphology-density relations found at 1 $<$ $z$ $<$ 2 \citep[e.g.,][]{Sazonova2020,Noordeh2021,Mei2023} that include quiescent galaxies.  Beyond the population of close pairs and mergers identified in \citet{Liu2023}, the similarity of the morphology of protocluster galaxies relative to the co-eval field may be caused by the clumpy, multi-peak galaxies that populate our protoclusters.  
Using the same density measurements as in Figure~\ref{Fig:Morph-density}, we plot the fraction of multi-peak merging galaxy candidates among the total number of emitting galaxies as a function of density (see the top panel of Figure~\ref{Fig:Merger-Density}).  

The majority of our multi-peak merging galaxies are found in the least dense protocluster environments.  This is intriguing because in BOSS1244, the close-pair galaxies from \citet{Liu2023} are in the densest protocluster environment, while BOSS1542 has a closer to uniform fraction of close-pair galaxies in our density bins.  Despite the differences in the location of the merging galaxies, if some fraction of the $\sim$ 20$\%$ of HAEs that are multi-peak galaxies are true mergers, this will increase the overall number of merging galaxies in these systems from the previously measured values 22$\pm$5 (BOSS1244) and 33$\pm$6 $\%$ (BOSS1542) \citep{Liu2023} and indicate an overall elevated level of merging galaxies across the entire protocluster.  This elevated merger fraction would be closer to the fraction of close kinematic pairs ($\sim$ 49$^{+7.4}_{-7.8}$$\%$) in the Hyperion proto-supercluster (z $=$ 2.5) \citep{Giddings2025}, as well as similar merger fractions (20$\%$ - 48$\%$) in protoclusters and field galaxies at 2 $<$ $z$ $<$ 3 \citep[e.g.,][]{Hine2016,Monson2021,Shibuya2025}. 

More important to the evolution of star-forming protocluster galaxies into their eventual quiescent, early-type counterparts is the morphology of these clumpy galaxies.  We find that all multi-peak protocluster galaxies have n $<$ 1.5 when fit with a single S\'ersic function using \textsc{Galapagos}, with $\sim$ 88$\%$ of these systems having n $<$ 1 (see the lower panel of Figure~\ref{Fig:Merger-Density}).  However, that these galaxies do not appear to be classical late-type galaxies is important for our characterization of the similarities between the fraction of early-type galaxies in the protocluster and the field.  Specifically, despite these multi-peak galaxies having \textsc{Galapagos} fits with reduced-$\chi^{2}$ $<$ 2, if these sources are not true late-type galaxies, then our fractions of early-type galaxies may be lower limits on the number of early-type galaxies as these sources have ambiguous morphologies.  However, if the multiple bright peaks are indicative of a recent merger, then these multi-peak galaxies may be an important sign for the build-up of the future morphology-density relation as they could indicate that the morphology of these systems is in flux and will soon transform, increasing the population of early-type galaxies in these systems \citep[e.g.,][]{Toomre1977,Naab2006,Rodriguez-Gomez2017}.

\section{Discussion}\label{Sect:Discussion}
As discussed in Section~\ref{sect:results-M}, using HST WFC3 F160W imaging of two MAMMOTH protoclusters at $z$ $\sim$ 2.23, we probed the morphology of HAEs as a function of density for these star-forming protocluster galaxies.  Although we see a slight increase in the fraction of early-type galaxies in denser regions in BOSS1244, it is not at a statistically significant level relative to the co-eval field and we find general agreement between the fraction of early-type galaxies in the field and protoclusters.  As we are examining exclusively star-forming protocluster galaxies, understanding how the morphology of these star-forming protocluster galaxies compares to similar protocluster studies, particularly those including quiescent galaxies, may help to illuminate the morphological transformation and evolution of protocluster galaxies.

\subsection{A comparison to other protocluster studies}\label{Sect:Comp}

Given that we only have two protoclusters, we contextualize our findings among the broader protocluster/high-$z$ cluster results found in the literature to understand the transformation of late-type galaxies in high-$z$ protoclusters into early-type galaxies in low-$z$ clusters.  Importantly, the MAMMOTH sample is unique compared to many other high-$z$ morphology studies because we are exclusively examining HAEs.  However, one similar sample exists in \citet{Naufal2023}, which studies a sample of four protoclusters populated by HAEs, including the Spiderweb protocluster ($z$ = 2.16; \citealp{Miley2006}), one of the most well-studied high-$z$ systems.  While the evolutionary states of our MAMMOTH protoclusters and the Spiderweb protocluster differ (Spiderweb hosts a virialized core, a lower merger fraction, and a detected proto-ICM;  \citealp{DiMascolo2023}), the other protoclusters (2.16 $<$ $z$ $<$ 2.53) appear to have similar to slightly smaller populations of HAEs compared to the MAMMOTH protoclusters \citep[e.g.,][]{Shimakawa2018a,Darvish2020,Koyama2021,Naufal2023}, making this an ideal comparison sample.  Because \citet{Naufal2023} use HST Advanced Camera Survey (ACS) F814W data, we cannot directly compare the morphology measurements of their sample of 122 protocluster HAEs to our sample, but we can compare their environmental trends.  Specifically, \citet{Naufal2023} find a median S\'ersic index of 0.76$^{+0.03}_{-0.05}$ for protocluster galaxies and 0.80$^{+0.22}_{-0.04}$ for their co-eval field sample, which points to similar morphologies between protocluster and field galaxies, in agreement with our findings.  In doing a non-parametric morphology study, \citet{Naufal2023} further affirm the similarities between the protocluster and field galaxies, finding similar values for the concentration, asymmetry, Gini, and M$_{20}$ parameters.  Although they do not explicitly plot a morphology-density relation, \citet{Naufal2023} perform a stacking analysis and find a difference between the S\'ersic index of protocluster galaxies (n $=$ 1.55$^{+0.33}_{-0.29}$) and field galaxies (n $=$ 1.02$^{+0.09}_{-0.08}$), which would imply a slight environmental impact not seen in MAMMOTH protoclusters, although both populations are still dominated by late-type galaxies.  Beyond galaxy morphology, \citet{Naufal2023} also find an abundance of protocluster HAEs with disturbed and clumpy morphologies relative to their field sample.  Although this classification is not identical to our population of "multi-peak" galaxies, both attempt to identify similarly unique galaxies, and both point to the protocluster environment enhancing the fraction of these clumpy and/or merging star-forming HAEs relative to the field.

\citet{Perez-Martin2023} continued the study of HAEs in the Spiderweb protocluster, finding that most of their spectroscopically confirmed HAEs have surface brightness profiles typical of high-$z$ late-type galaxies.  Given that a separate population of quiescent galaxies was recently identified in Spiderweb \citep{Naufal2024}, the similarity in surface brightness and morphology, as well as the results of \citet{Naufal2023}, may indicate that regardless of the overall protocluster evolutionary state, HAEs have not undergone a large degree of morphological pre-processing in these co-eval protoclusters.  Thus, because we only use HAEs and lack a large population of separately identified quiescent galaxies, this may lead to a potential bias in our estimate of the overall morphology of protocluster populations in BOSS1244 and BOSS1542.

Using similar HST ACS F814W observations to \citet{Naufal2023}, \citet{Peter2007} study a protocluster at $z$ $=$ 2.3 protocluster and also compare the morphology of their protocluster galaxies to a co-eval field sample.  Similar to our results and the results from \citet{Naufal2023}, they find no evidence of an environmental influence.  As this sample consists entirely of UV-selected star-forming galaxies, this adds further credence to lack of environmental impact on the morphology of star-forming galaxies in protoclusters. 

To compare our results to a larger sample, we examine the CARLA clusters at 1.4 $<$ $z$ $<$ 2.8 studied in \cite{Mei2023} and \citet{Afanasiev2023}.  Despite the MAMMOTH protoclusters being within the redshift range of the CARLA sample and \citet{Mei2023} using similar HST imaging to probe the morphology-density relation, we do not find strong agreement. This is likely the result of the difference in galaxy populations and morphology measurements.  While \citet{Mei2023} find a strong abundance of bulge-dominated galaxies in all CARLA protoclusters, the CARLA protoclusters, though spectroscopically-confirmed, use color-cut criteria to identify member galaxies, which allow for populations of star-forming and quiescent galaxies to be identified \citep{Mei2023,Afanasiev2023}.  Furthermore, \citet{Mei2023} posit that the low-frequency of color-color selected star-forming galaxies and the high passive galaxy fraction in their clusters is because these systems have already undergone starburst star formation, which would further separate our populations of protoclusters hosting large populations of HAEs.  Given that all but one CARLA cluster is at $z$ $<$ 2.1, this may also indicate that the CARLA clusters are more dynamically evolved systems and that the MAMMOTH systems are progenitors.  However, we also note that \citet{Mei2023} use visual classification to characterize the morphology of galaxies in the CARLA sample.  Although \citet{Afanasiev2023} find general agreement between their \textsc{Galapagos} morphology measurements and the visual classification done in \citet{Mei2023}, the difference in the strength of the morphology-density relation may be due to inherent differences between a visual classification scheme and our \textsc{Galapagos} measurements \citet{Mei2023}.  

In comparing the highest redshift studies of galaxy morphology at cosmic noon (2.1 $<$ $z$ $<$ 3), most report little environmental trends.  From the \citet{Naufal2023}, \citet{Mei2023}, \citet{Peter2007} studies, and our own work, only \citet{Mei2023} show strong evidence of a morphology-density relation in protoclusters.  In fact, the highest redshift source, which is in \citet{Mei2023}, is at $z$ $\sim$ 2.8 and has a bulge-dominated fractions that is only slightly above their estimate of the co-eval field level and below the level of the rest of their sample. As \citet{Mei2023} report a strong population of early-type galaxies and their sample selection allows for the inclusion of quiescent galaxies, this points to quenching playing a more prominent role in the build-up of the morphology-density relation and may be evidence that these systems are found at different evolutionary states.  This may suggest that the dynamical evolution of clusters/protoclusters is key to enhancing the rapid build-up of the morphology-density relation and further highlights the large degree of protocluster variation that hinders these evolutionary studies.  To fully address the role of protocluster evolution, galaxy quenching, and mergers in the build-up of the morphology-density relation, there is a clear need for a large statistical study to map protocluster populations across cosmic time.

Ultimately, because we only probe star-forming HAEs, our sample is biased against large populations of low-SFR galaxies, and/or low-mass galaxies.  Although BOSS1244 has two spectroscopically confirmed quiescent galaxies (\citealp{Shi2024}; see Section~\ref{sect:1244} for our further analysis of the spectroscopic sample), which may indicate that more quenched galaxies exist in these protocluster systems, we cannot exclude the inherent bias in our sample as a reason why we do not see an environmental impact on galaxy morphology.  However, regardless of any missing populations within our protoclusters, contextualizing the morphology of these star-forming galaxies and how it might evolve is important.  Specifically, \citet{Barro2017} note that the morphological transformation of star-forming galaxies at cosmic noon likely occurs via compaction, which would increase star formation before AGN driven quenching can occur.  As compaction occurs over 0.5 - 1.0\,Gyr timescales, while quenching occurs on Gyr timescales \citep{Barro2017}, the absence of a surplus of early-type star-forming galaxies in MAMMOTH systems relative to the CARLA systems is likely due to the evolutionary timescales.  By $z$ $\sim$ 2, if these HAEs undergo compaction, they could increase the population of early-type galaxies and be in the process of quenching.  Within our protocluster sample, we find that, on average, early-type HAEs in these protoclusters are slightly smaller than their late-type counterparts, which could be evidence of compaction in these galaxies.  However, although it is possible that the early-type galaxies in both the protocluster and the field transform via compaction, a robust study of protocluster and co-eval field galaxy size in conjunction with morphology is necessary to better constrain whether compaction is driving any morphological transformations in MAMMOTH protoclusters.  

Furthermore, the most actively star-forming and potentially most compacted galaxies in these systems may be missed in our analysis because \citet{Zhang2022} find evidence of submillimeter galaxies on the outskirts of each protocluster.  Additionally, recent JWST observations comparing the morphology of massive star-forming galaxies at cosmic noon identified a population of optically disk-dominated galaxies that are characterized as bulge dominated in the infrared \citep{Benton2024}, which may indicate that some of the late-type star-forming galaxies we have characterized have indeed begun their morphological transformation.  Thus, despite the lack of environmental trends regarding the morphological transformation of HAEs in MAMMOTH protoclusters, there are still multiple avenues to explore to continue probing the evolution of these protocluster HAEs.

\subsection{Populations of bright and merging protocluster galaxies}
  
Despite the lack of true quiescent galaxies in our sample of HAEs, we do identify bright, early-type protocluster galaxies in each system, although they are more of a dominant population in BOSS1244 (see the lower panels in Figures~\ref{Fig:Morph-density} and \ref{Fig:Morph-density-v2}).  Although many of these bright galaxies have M$_{*}$ $>$ 10$^{11}$\,M$_{\odot}$, we see no enhancement in the fraction of bright early-type galaxies relative to the field.  Although the majority of newly identified high-$z$ quiescent galaxies are massive \citep[e.g.,][]{Carnall2024,Nanayakkara2025,DeGraaf2025}, the lack of an abundance of massive early-type galaxies among our population of HAEs may further indicate that the environment is not the driving factor for the morphological transformation of these massive galaxies.  While we have already discussed that quenching and morphological transformation are not the same process, our results do agree with prior results studying the evolution of quenched galaxies, which found that the passive galaxy fraction is independent of environment at 1 $<$ $z$ $<$ 3 in the COSMOS field and that the quiescent fraction only depends on stellar mass at $z$ $\sim$ 3.0 \citep{Darvish2015, Darvish2016}.  If the morphological transformation of protocluster galaxies is similarly driven by stellar mass and not the environment, then the similar populations of early-type galaxies among the field and protoclusters, as seen in this work, could be further evidence that the timescale for the morphological transformation of galaxies at cosmic noon does not depend on environment. 

Another potential formation mechanism for early-type galaxies seen in protoclusters, are galaxy mergers \citep[e.g.,][]{Hine2016,Watson2019,Monson2021,Giddings2025,Shibuya2025}.  Importantly, the high merger fraction in BOSS1244 and BOSS1542 (measured in \citealp{Liu2023} as 22$\pm$5 $\%$ and 33$\pm$6 $\%$ in each protocluster relative to the merger fraction of 12$\pm$2 $\%$ in their CANDELS co-eval background field), and in particular, the abundance of visually identified multi-peak galaxies (see Figure~\ref{Fig:Merger-Density}), may also explain why we do not identify a strong morphology-density relation relative to the co-eval field.  \citet{Liu2023} estimate a merger timescale for massive galaxies in BOSS1244 and BOSS1542 of 0.63 $\pm$ 0.05\,Gyr.  If we assume that some fraction of our multi-peak galaxies are post-merger systems, this would mean that new early-type and potentially quenched galaxies would appear in these protoclusters by $z$ $\sim$ 2.  Moreover, the close-pairs from \citet{Liu2023} will also increase the population of early-type galaxies by $z$ $\sim$ 2.  When our multi-peak population and the close-pair population are combined with the existing fraction of n $\geq$ 3 galaxies ($\sim$ 25$\%$; see Figure~\ref{Fig:Morph-density-v2}), the early-type fraction increases to the 50 - 60$\%$ seen in \citet{Mei2023}.

\subsection{A closer look at BOSS1244\label{sect:1244}}
Although the goal of this paper is to present an analysis of the impact of the protocluster environment on the morphology of HAEs in two MAMMOTH protoclusters, here we focus only on BOSS1244, which contains two spectroscopically-confirmed quiescent galaxies \citep{Shi2024}, 46 other spectroscopically-confirmed HAEs  \citep{Shi2021}, and an additional 14 star-forming galaxies identified via HST grism observations \citep{Shi2021} (although four HAEs with spectroscopic redshifts fall outside of our redshift range, each is at 2.09 $<$ $z$ $<$ 2.22, so they remain potential protocluster members and we included them in this analysis).  For this reason, along with the additional HST WFC3 F125W photometry that only exists for BOSS1244, we probe the color and distribution of the protocluster galaxies relative to the known quiescent BCG in BOSS1244.  As with the previously discussed HAEs, we use \textsc{Galapagos} to measure the morphology of all additional spectroscopically-confirmed galaxies in an identical manner to the HAEs.  Although not shown,  despite the presence of two quiescent massive galaxies, we find no statistical difference in the morphology-density relation for BOSS1244 when accounting for the new spectroscopic members.

\begin{figure}[hbt!]
\begin{center}
\includegraphics[scale=0.55,trim={0.2in 0.0in 0.5in 0.5in},clip=true]{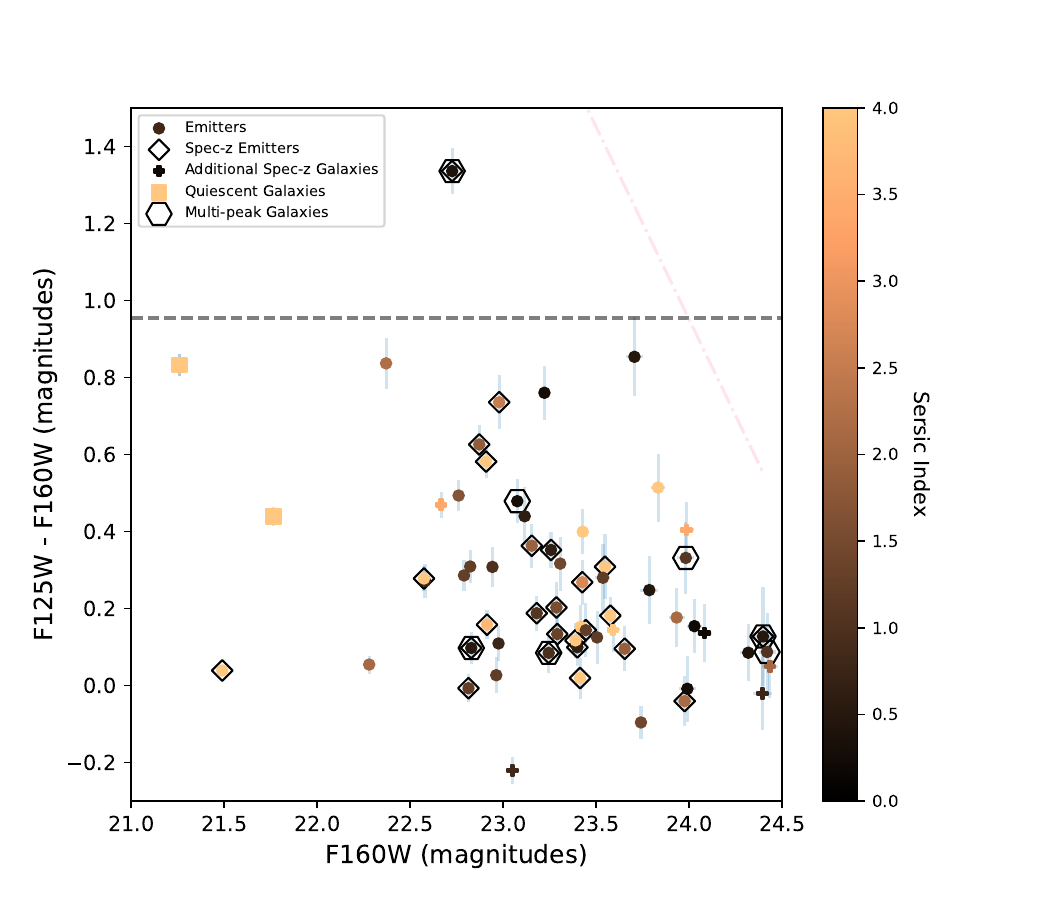}

\caption{The color-magnitude diagram for BOSS1244.  The HAEs are shown in circles.  The spectroscopically-confirmed HAEs have a diamond surrounding them.  The multi-peak galaxies have a hexagon surrounding them.  The additional spectroscopic star-forming galaxies from HST grism are shown by a plus sign.  The quiescent galaxies are shown in squares.  The horizontal dashed line shows the color of a modeled red sequence galaxy based on EzGal models \citep{Mancone2012} for a galaxy at $z$ = 2.246.  The pink dot-dashed line shows the magnitude limit for our analysis.  The color of each point is based on the S\'ersic index; late-type galaxies are darker and early-type galaxies are more orange.}

\label{Fig:CMD}
\end{center}
\end{figure}

The transformation of protocluster galaxies into their low-$z$ counterparts requires both a morphological transformation and quenching.  Thus, despite studying a sample of nearly all HAEs, and given the abundance of recently identified populations of quiescent galaxies in high-$z$ clusters and protoclusters \citep[e.g.,][]{Willis2020,Noordeh2021,McConachie2022,McConachie2025}, we present a color-magnitude diagram of all detected galaxies above our magnitude-limit (m$_{F160W}$ = 24.5\,magnitudes and m$_{F125W}$ = 24.95\,magnitudes) to determine whether any indication of early signs of quenching exists in BOSS1244.  In Figure~\ref{Fig:CMD}, the color of each point is determined by the S\'ersic index.  Although our fitting allows for S\'ersic indices ranging from 0.2 to 8.0, we include a color bar ranging from 0 to 4,  where late-type galaxies are colored black and early-type galaxies are colored orange.  By using the F160W and F125W bands, we straddle the 4000\,\AA\, break at $z$ = 2.246, meaning that we should be able to select red sequence galaxies if any are present.  Also, we note that all spectroscopic galaxies and HAEs are not shown in Figure~\ref{Fig:CMD} as some do not have dual-band imaging and some are not detected at the magnitude limit in both bands (see Appendix~\ref{sect:1244-CMDFaint} and Figure~\ref{Fig:CMD2}). 

Unlike the strong bulge-dominated red sequence found in XLSSC 122 at $z$ $\sim$ 2.0 \citep{Noordeh2021}, we find no evidence for an early forming red sequence, even when including the spectroscopic members.  We also find no correlation between the color and morphology for the total sample.  This is also true among the HST grism spectroscopic sample, which spans a similar range in color and morphology to the HAEs.  Furthermore, of the multi-peak galaxies shown, some have colors similar to the quiescent galaxies, while others are much bluer.  Given the small number of spectroscopic galaxies and the spread in the color and morphology of the emitters as a whole, we cannot claim any strong degree of quenching among this star-forming population, or among our population of HAEs.

Interestingly, despite the differences between BOSS1244 and XLSSC 122 in terms of early-type red sequence populations, \citet{Noordeh2021} find that $\approx$ 33$^{+18}_{-11}$$\%$ of their star-forming galaxies have a bulge-dominated morphology, which agrees with our results.  As \citet{Noordeh2021} posit that the environment does not play a role in the transformation of galaxy morphology at these redshifts, this similarity may further indicate that we are observing structures at different evolutionary points in their lifetimes and that further studies will be necessary to disentangle quenching and morphological transformations.

\begin{figure}[hbt!]
\begin{center}
\includegraphics[scale=0.7,trim={0.0in 0.65in 0.0in 1.0in},clip=true]{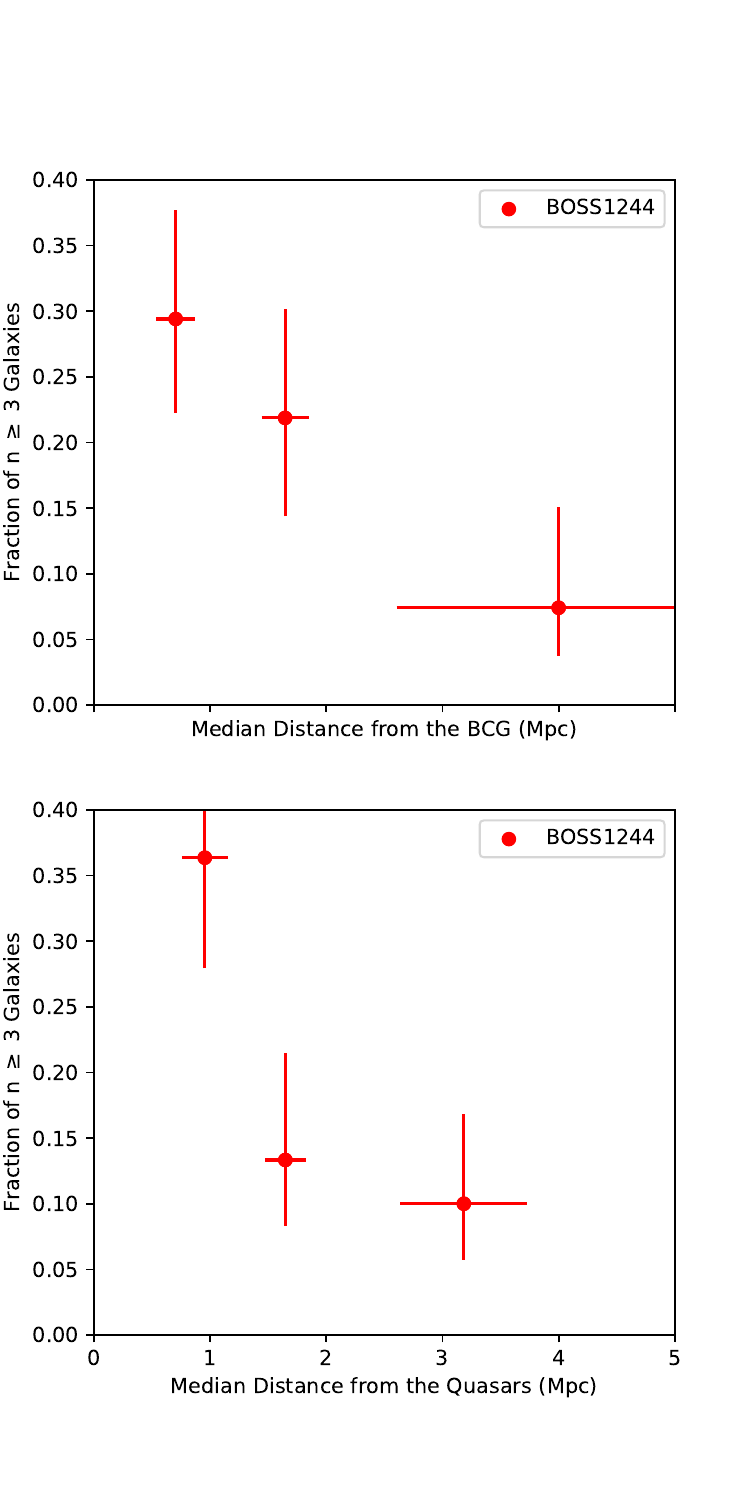}

\caption{The number of strongly bulge-dominated early-type galaxies (n $\geq$ 3) as a function of the distance from the BCG (top panel) and the distance from two quasars (one at $z$ = 2.236 and one at $z$ = 2.229; bottom panel).  The quiescent BCG and the two quasars are $\sim$ 1\,Mpc apart.  In both cases, we see a peak in the distribution of n $\geq$ 3 bulge dominated galaxies near the target galaxy, with the BCG showing a less steep distribution.}

\label{Fig:morph-distance}
\end{center}
\end{figure}

Similar to the morphology-density relation, low-$z$ clusters are commonly characterized by a dense core of red sequence elliptical galaxies (centered on a BCG) and increasing populations of spiral galaxies on the cluster outskirts.  Although BOSS1244 is not virialized, we measure the fraction of early-type galaxies as a function of the distance from the quiescent BCG in the upper panel of Figure~\ref{Fig:morph-distance}.  When we isolate the strongest bulge-dominated early-type galaxies (n $\geq$ 3), we find a peaked distribution with a higher fraction of these galaxies near the BCG.  This finding supports the results of \citet{Shi2024} that the most strongly-bulge dominated spectroscopic galaxies are within 2$\arcmin$ of the BCG ($\sim$ 1\,Mpc) and that the H$\alpha$ equivalent width of their spectroscopic sample increases with distance from the BCG, implying that there is some degree of star formation quenching due to the protocluster environment.  Moreover, \citet{Shi2024} posit that this local overdensity of early-type galaxies may be indicative of an early-forming red sequence core, similar to the results of \citet{Strazzullo2013} in their $z$ $\sim$ 2.0 cluster.  Although \citet{Strazzullo2013} only found strong evidence of environmental morphological quenching within 200\,kpc and our early-type galaxy peak is a factor of 5 times larger, we could be observing BOSS1244 in the early stages of this transformation as the cluster virializes. 

Quiescent BCGs are not the only signpost galaxies that can identify galaxy clusters at high-$z$.  Radio-loud AGNs, and quasars specifically, are excellent tracers for high-$z$ structures \citep[e.g.,][]{Miley2006,Galametz2012,Wylezalek2013,Paterno-Mahler2017,Golden-Marx2019,Shen2019,Moravec2019,Moravec2020,Shen2021,Golden-Marx2021,Golden-Marx2023,CWatson2025}.  While there are multiple SDSS-identified spectroscopic quasars as well as HAEs identified as quasars in \citet{Shi2021} in BOSS1244 (see the left panel of Figure~\ref{Fig:protoclusters} for the locations of the quasars), only two ($z$ = 2.23362 and $z$ = 2.229) fall within our HST coverage (and are thus in the denser part of the protocluster).  Given how close these two quasars are, we characterize their environment jointly.  We see a much sharper peak in the distribution of n $\geq$ 3 bulge-dominated early-type galaxies surrounding the quasars (see the lower panel of Figure~\ref{Fig:morph-distance}).  However, given that the BCG and these quasar are $\sim$ 1\,Mpc apart and thus the inner regions showing the higher fraction of early-type galaxies contains many overlapping galaxies, this may indicate a second potential proto-BCG candidate, similar to the multiple BCG candidates found in \citet{Shen2021}, or be additional evidence of the peaked early forming cluster core.  Since we find a stronger difference in the fraction of early-type galaxies as a function of the distance from the BCG as compared to the fraction of early-type galaxies as a function of local environment in BOSS1244 (see Fig~\ref{Fig:Morph-density}), this may be evidence that the global environment, as estimated by the distance from the BCG, dominates over the local environment when impacting galaxy evolution and the morphological transformation of protocluster galaxies.  However, it is important to note that the two quasars are a confounding factor to this argument as not all protoclusters host central quasars and it may be a confluence of these two characteristics, and not the BCG alone, that yields the peaked distribution of early-type galaxies.   

To determine the potential role quasars play in tracing overdensities of bulge-dominated galaxies in high-$z$ protoclusters, we similarly look at the quasars in BOSS1542.  Unlike BOSS1244, in BOSS1542 (see the right panel of Figure~\ref{Fig:protoclusters} for the locations of the quasars) none of the quasars show a peaked distribution of n $\geq$ 3 early-type galaxies. Therefore, the lack of a central core of early-type galaxies in BOSS1542 and the lack of known quiescent galaxies are further evidence of a difference in the evolutionary state of these two protoclusters despite them being co-eval.  More importantly, this highlights the challenge in comparing high-$z$ protoclusters; despite similar observed redshifts, we cannot observationally constrain formation redshift to adequately determine evolutionary differences between protocluster systems.

\section{Conclusion}\label{sect:Conclusions}
Understanding how galaxy populations evolve within dense environments is key to determining the role the cluster/protocluster environment plays in the transformation of high-$z$ star-forming galaxies into their more ubiquitous low-$z$ quiescent, red, early-type counterparts.  Using high resolution HST WFC3 F160W imaging of BOSS1244 and BOSS1542, we characterized the morphology of a large population of HAEs at $z$ $\approx$ 2.23.  Using \textsc{Galapagos}, we measured the S\'ersic index of 80 HAEs in BOSS1244 and 71 HAEs in BOSS1542.  We summarize our results below.
\begin{itemize}
    \item{In comparing the morphology of our samples of protocluster star-forming HAEs to their co-eval field counterparts, we find strong agreement between the morphologies of these galaxies, with no evidence that they are drawn from separate populations.  This points to the protocluster environment having little impact on galaxy morphology.}
    \item {In BOSS1244 we find a slight internal morphology-density trend, with denser regions hosting more early-type galaxies, regardless of our S\'ersic index threshold or magnitude cut.  However, in all cases, the difference in the population of early-type galaxies is less than 1$\sigma$ and is in agreement with the field values.  Conversely, we do not see such a positive trend in all of the samples for BOSS1542.}
    \item{We identify a large population of multi-peak galaxies that are potential mergers and/or clumpy galaxies.  All of these galaxies have n $<$ 1.5.  As such, we hypothesize that, when combined with the large fraction of close-pair galaxies identified in these protoclusters in \citet{Liu2023}, we are observing the build-up of the future population of early-type galaxies.}  
    \item{In BOSS1244, we examine the color and distribution of protocluster galaxies, including the spectroscopic sample from \citet{Shi2021} and \citet{Shi2024}.  While we find no evidence of large quenched population among the entire system, we find that the strongest bulge-dominated early-type galaxies ($n$ $\geq$ 3) are clustered close to the quiescent BCG, as well as two nearby co-eval quasars.  This core of early-type galaxies could be evidence that the global environment dominates the local environment when impacting the morphological transformation of protocluster galaxies.  The appearance of this clustered early forming core, along with the previously reported differences in the overall protocluster morphology between BOSS1244 and BOSS1542 \citep[e.g.,][]{Zheng2021,Shi2021} point to BOSS1244 being a more evolved protocluster system when compared to BOSS1542.}
\end{itemize}

We successfully measure the fraction of early-type HAEs as a function of density within these two MAMMOTH protoclusters and find little evidence for a local morphology-density relation among HAEs in protocluster.  However, to fully trace the morphology-density relation and the build-up of early-type galaxies, as well as determine the role of virialization and galaxy mergers in this evolution, we require a more complete sampling of the galaxy population.  Although the additional spectroscopically-confirmed galaxies (including the two quiescent galaxies in BOSS1244) do not change our results, to accurately trace the impact of the protocluster environment on galaxy morphology, we must account for all protocluster galaxies across the entire structure and not just within the denser regions, as well as the lower mass galaxy populations.  However, more robust observations of MAMMOTH protoclusters is not enough.  To determine the importance of dynamical evolution and mergers in the build-up of the morphology-density relation, a larger statistical sample of clusters/protoclusters at 1.5 $<$ $z$ $<$ 4.5 across a range of cluster masses is needed.  Given the success of recent observations from CARLA for a small subset of their high-$z$ structures \citep[e.g.,][]{Mei2023,Afanasiev2023}, the plethora of newly identified high-$z$ ($z$ $<$ 2.0) structures from MaDCoWS-2 \citep{Thongkham2024}, new high-$z$ protostructures (2 $<$ $z$ $<$ 5) from the Charting Cluster Cosmology with ORELSE (C3VO) survey \citep[e.g.,][]{Forrest2023,Shah2024,Staab2024,Hung2025}, and soon to be studied high-$z$ clusters and protoclusters observed as part of COSMOS-Web with JWST \citep[e.g.,][]{Gozaliasl2025} and observed with \textit{Euclid}, a true statistical study of the build-up of the morphology-density relation may be possible in the near future.  Furthermore, with the higher angular resolution available with JWST observations, as well as with future telescopes, including the Extremely Large Telescope, it will be possible to determine the true impact of merging/and or clumpy galaxies on the transformation of protocluster populations.

\begin{acknowledgements}
    EGM would like to thank the Anonymous Referee for their useful insights and suggestions during the referee process.  EGM would also like to thank Jesse Golden-Marx for reading drafts of this paper and useful discussions regarding SDSS.  EGM would like to thank members of the C3VO collaboration for useful discussions on protoclusters and galaxy morphology, and in particular Ekta Shah for useful discussions on galaxy mergers.  EGM would also like to thank members of the galaxy evolution group at INAF-OAPd for useful discussions regarding high-$z$ galaxies.  EGM would like to thank the High-$z$ group at Tsinghua University for useful discussions regarding protoclusters.  Additionally, EGM would like to thank Chien Peng, Michael McDonald, and Song Huang for useful discussions regarding \textsc{Galfit} and \textsc{Galapagos}.  EGM would also like to thank the organizers of the First Structures in the Universe Conference for creating useful discussions that aided in this work. EGM would like to thank Richard Grumitt, Tom Binney, and Daniele Spinoso for useful discussions.\\

    EGM and PR acknowledges the support of this work by the Tsinghua Shui Mu Scholarship.  Additionally, EGM, ZC, and PR acknowledge that this work was funded by the National Key R\&D Program of China (grant no.\ 2018YFA0404503), the National Science Foundation of China (grant no.\ 12073014). The science research grants from the China Manned Space Project with No. CMS-CSST2021-A05, and Tsinghua University Initiative Scientific Research Program (No. 20223080023).\\

    DDS acknowledges the support from the National Science Foundation of China (12303015), the National Science Foundation of Jiangsu Province (BK20231106) and the China Manned Space Program with grant No. CMS-CSST- 2025-A20.\\

    X. W. is supported by the National Natural Science Foundation of China (grant 12373009), the CAS Project for Young Scientists in Basic Research Grant No. YSBR-062, the Fundamental Research Funds for the Central Universities, the Xiaomi Young Talents Program, and the science research grant from the China Manned Space Project. X. W. also acknowledges work carried out, in part, at the Swinburne University of Technology, sponsored by the ACAMAR visiting fellowship.\\

    B.C.L. is supported by the international Gemini Observatory, a program of NSF NOIRLab, which is managed by the Association of Universities for Research in Astronomy (AURA) under a cooperative agreement with the U.S. National Science Foundation, on behalf of the Gemini partnership of Argentina, Brazil, Canada, Chile, the Republic of Korea, and the United States of America.\\
    
    Funding for SDSS-III has been provided by the Alfred P. Sloan Foundation, the Participating Institutions, the National Science Foundation, and the U.S. Department of Energy Office of Science. The SDSS-III web site is http://www.sdss3.org/.\\

    SDSS-III is managed by the Astrophysical Research Consortium for the Participating Institutions of the SDSS-III Collaboration including the University of Arizona, the Brazilian Participation Group, Brookhaven National Laboratory, Carnegie Mellon University, University of Florida, the French Participation Group, the German Participation Group, Harvard University, the Instituto de Astrofisica de Canarias, the Michigan State/Notre Dame/JINA Participation Group, Johns Hopkins University, Lawrence Berkeley National Laboratory, Max Planck Institute for Astrophysics, Max Planck Institute for Extraterrestrial Physics, New Mexico State University, New York University, Ohio State University, Pennsylvania State University, University of Portsmouth, Princeton University, the Spanish Participation Group, University of Tokyo, University of Utah, Vanderbilt University, University of Virginia, University of Washington, and Yale University.\\

    This research made use of Astropy (http://www.astropy.org) a community-developed core Python package for Astronomy \citep{astropy:2013, astropy:2018}.\\

\end{acknowledgements}

\bibliography{references}

@ARTICLE{Paterno-Mahler2017,
   author = {{Paterno-Mahler}, R. and {Blanton}, E.~L. and {Brodwin}, M. and 
	{Ashby}, M.~L.~N. and {Golden-Marx}, E. and {Decker}, B. and 
	{Wing}, J.~D. and {Anand}, G.},
    title = "{The High-redshift Clusters Occupied by Bent Radio AGN (COBRA) Survey: The Spitzer  Catalog}",
  journal = {\apj},
archivePrefix = "arXiv",
   eprint = {1611.00746},
     year = 2017,
    month = jul,
   volume = 844,
      eid = {78},
    pages = {78},
      doi = {10.3847/1538-4357/aa7b89},
   adsurl = {http://adsabs.harvard.edu/abs/2017ApJ...844...78P}
}

@ARTICLE{Wing2011,
   author = {{Wing}, J.~D. and {Blanton}, E.~L.},
    title = "{Galaxy Cluster Environments of Radio Sources}",
  journal = {\aj},
archivePrefix = "arXiv",
   eprint = {1008.1099},
     year = 2011,
    month = mar,
   volume = 141,
      eid = {88},
    pages = {88},
      doi = {10.1088/0004-6256/141/3/88},
   adsurl = {http://adsabs.harvard.edu/abs/2011AJ....141...88W}
}

@ARTICLE{Cooke2015,
   author = {{Cooke}, E.~A. and {Hatch}, N.~A. and {Rettura}, A. and {Wylezalek}, D. and 
	{Galametz}, A. and {Stern}, D. and {Brodwin}, M. and {Muldrew}, S.~I. and 
	{Almaini}, O. and {Conselice}, C.~J. and {Eisenhardt}, P.~R. and 
	{Hartley}, W.~G. and {Jarvis}, M. and {Seymour}, N. and {Stanford}, S.~A.
	},
    title = "{The formation history of massive cluster galaxies as revealed by CARLA}",
  journal = {\mnras},
archivePrefix = "arXiv",
   eprint = {1507.00350},
     year = 2015,
    month = sep,
   volume = 452,
    pages = {2318-2336},
      doi = {10.1093/mnras/stv1413},
   adsurl = {http://adsabs.harvard.edu/abs/2015MNRAS.452.2318C}
}

@ARTICLE{Cooke2016,
   author = {{Cooke}, E.~A. and {Hatch}, N.~A. and {Stern}, D. and {Rettura}, A. and 
	{Brodwin}, M. and {Galametz}, A. and {Wylezalek}, D. and {Bridge}, C. and 
	{Conselice}, C.~J. and {De Breuck}, C. and {Gonzalez}, A.~H. and 
	{Jarvis}, M.},
    title = "{A Mature Galaxy Cluster at z=1.58 around the Radio Galaxy 7C1753+6311}",
  journal = {\apj},
archivePrefix = "arXiv",
   eprint = {1511.05150},
     year = 2016,
    month = jan,
   volume = 816,
      eid = {83},
    pages = {83},
      doi = {10.3847/0004-637X/816/2/83},
   adsurl = {http://adsabs.harvard.edu/abs/2016ApJ...816...83C}
}

@ARTICLE{Gladders2000,
   author = {{Gladders}, M.~D. and {Yee}, H.~K.~C.},
    title = "{A New Method For Galaxy Cluster Detection. I. The Algorithm}",
  journal = {\aj},
   eprint = {astro-ph/0004092},
     year = 2000,
    month = oct,
   volume = 120,
    pages = {2148-2162},
      doi = {10.1086/301557},
   adsurl = {http://adsabs.harvard.edu/abs/2000AJ....120.2148G}
}

@ARTICLE{Cerulo2016,
   author = {{Cerulo}, P. and {Couch}, W.~J. and {Lidman}, C. and {Demarco}, R. and 
	{Huertas-Company}, M. and {Mei}, S. and {S{\'a}nchez-Janssen}, R. and 
	{Barrientos}, L.~F. and {Mu{\~n}oz}, R.~P.},
    title = "{The accelerated build-up of the red sequence in high-redshift galaxy clusters}",
  journal = {\mnras},
archivePrefix = "arXiv",
   eprint = {1601.07578},
     year = 2016,
    month = apr,
   volume = 457,
    pages = {2209-2235},
      doi = {10.1093/mnras/stw080},
   adsurl = {http://adsabs.harvard.edu/abs/2016MNRAS.457.2209C}
}

@ARTICLE{Marchesini2014,
   author = {{Marchesini}, D. and {Muzzin}, A. and {Stefanon}, M. and {Franx}, M. and 
	{Brammer}, G.~G. and {Marsan}, C.~Z. and {Vulcani}, B. and {Fynbo}, J.~P.~U. and 
	{Milvang-Jensen}, B. and {Dunlop}, J.~S. and {Buitrago}, F.},
    title = "{The Progenitors of Local Ultra-massive Galaxies Across Cosmic Time: From Dusty Star-bursting to Quiescent Stellar Populations}",
  journal = {\apj},
archivePrefix = "arXiv",
   eprint = {1402.0003},
     year = 2014,
    month = oct,
   volume = 794,
      eid = {65},
    pages = {65},
      doi = {10.1088/0004-637X/794/1/65},
   adsurl = {http://adsabs.harvard.edu/abs/2014ApJ...794...65M}
}

@ARTICLE{Galametz2012,
   author = {{Galametz}, A. and {Stern}, D. and {De Breuck}, C. and {Hatch}, N. and 
	{Mayo}, J. and {Miley}, G. and {Rettura}, A. and {Seymour}, N. and 
	{Stanford}, S.~A. and {Vernet}, J.},
    title = "{The Mid-infrared Environments of High-redshift Radio Galaxies}",
  journal = {\apj},
archivePrefix = "arXiv",
   eprint = {1202.4489},
     year = 2012,
    month = apr,
   volume = 749,
      eid = {169},
    pages = {169},
      doi = {10.1088/0004-637X/749/2/169},
   adsurl = {http://adsabs.harvard.edu/abs/2012ApJ...749..169G}
}

@ARTICLE{Wylezalek2013,
   author = {{Wylezalek}, D. and {Galametz}, A. and {Stern}, D. and {Vernet}, J. and 
	{De Breuck}, C. and {Seymour}, N. and {Brodwin}, M. and {Eisenhardt}, P.~R.~M. and 
	{Gonzalez}, A.~H. and {Hatch}, N. and {Jarvis}, M. and {Rettura}, A. and 
	{Stanford}, S.~A. and {Stevens}, J.~A.},
    title = "{Galaxy Clusters around Radio-loud Active Galactic Nuclei at 1.3 < z < 3.2 as Seen by Spitzer}",
  journal = {\apj},
archivePrefix = "arXiv",
   eprint = {1304.0770},
     year = 2013,
    month = may,
   volume = 769,
      eid = {79},
    pages = {79},
      doi = {10.1088/0004-637X/769/1/79},
   adsurl = {http://adsabs.harvard.edu/abs/2013ApJ...769...79W}
}

@ARTICLE{Hatch2014,
   author = {{Hatch}, N.~A. and {Wylezalek}, D. and {Kurk}, J.~D. and {Stern}, D. and 
	{De Breuck}, C. and {Jarvis}, M.~J. and {Galametz}, A. and {Gonzalez}, A.~H. and 
	{Hartley}, W.~G. and {Mortlock}, A. and {Seymour}, N. and {Stevens}, J.~A.
	},
    title = "{Why z < 1 radio-loud galaxies are commonly located in protoclusters}",
  journal = {\mnras},
archivePrefix = "arXiv",
   eprint = {1409.1218},
     year = 2014,
    month = nov,
   volume = 445,
    pages = {280-289},
      doi = {10.1093/mnras/stu1725},
   adsurl = {http://adsabs.harvard.edu/abs/2014MNRAS.445..280H}
}

@ARTICLE{Mancone2012,
   author = {{Mancone}, C.~L. and {Gonzalez}, A.~H.},
    title = "{EzGal: A Flexible Interface for Stellar Population Synthesis Models}",
  journal = {\pasp},
archivePrefix = "arXiv",
   eprint = {1205.0009},
 primaryClass = "astro-ph.IM",
     year = 2012,
    month = jun,
   volume = 124,
    pages = {606},
      doi = {10.1086/666502},
   adsurl = {http://adsabs.harvard.edu/abs/2012PASP..124..606M}
}

@ARTICLE{Bertin1996,
   author = {{Bertin}, E. and {Arnouts}, S.},
    title = "{SExtractor: Software for source extraction.}",
  journal = {\aaps},
     year = 1996,
    month = jun,
   volume = 117,
    pages = {393-404},
      doi = {10.1051/aas:1996164},
   adsurl = {http://adsabs.harvard.edu/abs/1996A%26AS..117..393B}
}

@ARTICLE{Gonzalez2019,
       author = {{Gonzalez}, Anthony H. and {Gettings}, Daniel P. and {Brodwin}, Mark and
         {Eisenhardt}, Peter R.~M. and {Stanford}, S.~A. and
         {Wylezalek}, Dominika and {Decker}, Bandon and {Marrone}, Daniel P. and
         {Moravec}, Emily and {O'Donnell}, Christine and {Stalder}, Brian and
         {Stern}, Daniel and {Abdulla}, Zubair and {Brown}, Gillen and
         {Carlstrom}, John and {Chambers}, Kenneth C. and {Hayden}, Brian and
         {Lin}, Yen-ting and {Magnier}, Eugene and {Masci}, Frank J. and
         {Mantz}, Adam B. and {McDonald}, Michael and {Mo}, Wenli and
         {Perlmutter}, Saul and {Wright}, Edward L. and {Zeimann}, Gregory R.},
        title = "{The Massive and Distant Clusters of WISE Survey. I. Survey Overview and a Catalog of \&gt;2000 Galaxy Clusters at z ≃ 1}",
      journal = {\apjs},
         year = "2019",
        month = "Feb",
       volume = {240},
       number = {2},
          eid = {33},
        pages = {33},
          doi = {10.3847/1538-4365/aafad2},
archivePrefix = {arXiv},
       eprint = {1809.06820},
 primaryClass = {astro-ph.CO},
       adsurl = {https://ui.adsabs.harvard.edu/abs/2019ApJS..240...33G}
}

@ARTICLE{Rykoff2014,
   author = {{Rykoff}, E.~S. and {Rozo}, E. and {Busha}, M.~T. and {Cunha}, C.~E. and 
	{Finoguenov}, A. and {Evrard}, A. and {Hao}, J. and {Koester}, B.~P. and 
	{Leauthaud}, A. and {Nord}, B. and {Pierre}, M. and {Reddick}, R. and 
	{Sadibekova}, T. and {Sheldon}, E.~S. and {Wechsler}, R.~H.},
    title = "{redMaPPer. I. Algorithm and SDSS DR8 Catalog}",
  journal = {\apj},
archivePrefix = "arXiv",
   eprint = {1303.3562},
     year = 2014,
    month = apr,
   volume = 785,
      eid = {104},
    pages = {104},
      doi = {10.1088/0004-637X/785/2/104},
   adsurl = {http://adsabs.harvard.edu/abs/2014ApJ...785..104R}
}

@ARTICLE{Moravec2019,
       author = {{Moravec}, Emily and {Gonzalez}, Anthony H. and {Stern}, Daniel and
         {Brodwin}, Mark and {Clarke}, Tracy and {Decker}, Bandon and
         {Eisenhardt}, Peter R.~M. and {Mo}, Wenli and {O'Donnell}, Christine and
         {Pope}, Alexandra and {Stanford}, Spencer A. and {Wylezalek}, Dominika},
        title = "{The Massive and Distant Clusters of WISE Survey. V. Extended Radio Sources in Massive Galaxy Clusters at z ̃ 1}",
      journal = {\apj},
         year = "2019",
        month = "Feb",
       volume = {871},
       number = {2},
          eid = {186},
        pages = {186},
          doi = {10.3847/1538-4357/aaf569},
       adsurl = {https://ui.adsabs.harvard.edu/abs/2019ApJ...871..186M}
}

@ARTICLE{Lemaux2019,
       author = {{Lemaux}, B.~C. and {Tomczak}, A.~R. and {Lubin}, L.~M. and
         {Gal}, R.~R. and {Shen}, L. and {Pelliccia}, D. and {Wu}, P. -F. and
         {Hung}, D. and {Mei}, S. and {Le F{\`e}vre}, O. and {Rumbaugh}, N. and
         {Kocevski}, D.~D. and {Squires}, G.~K.},
        title = "{Persistence of the colour-density relation and efficient environmental quenching to z ̃ 1.4}",
      journal = {\mnras},
         year = "2019",
        month = "Nov",
       volume = {490},
       number = {1},
        pages = {1231-1254},
          doi = {10.1093/mnras/stz2661},
archivePrefix = {arXiv},
       eprint = {1812.04624},
 primaryClass = {astro-ph.GA},
       adsurl = {https://ui.adsabs.harvard.edu/abs/2019MNRAS.490.1231L}
}

@ARTICLE{Coogan2018,
       author = {{Coogan}, R.~T. and {Daddi}, E. and {Sargent}, M.~T. and
         {Strazzullo}, V. and {Valentino}, F. and {Gobat}, R. and {Magdis}, G. and
         {Bethermin}, M. and {Pannella}, M. and {Onodera}, M. and {Liu}, D. and
         {Cimatti}, A. and {Dannerbauer}, H. and {Carollo}, M. and
         {Renzini}, A. and {Tremou}, E.},
        title = "{Merger-driven star formation activity in Cl J1449+0856 at z = 1.99 as seen by ALMA and JVLA}",
      journal = {\mnras},
         year = "2018",
        month = "Sep",
       volume = {479},
       number = {1},
        pages = {703-729},
          doi = {10.1093/mnras/sty1446},
archivePrefix = {arXiv},
       eprint = {1805.09789},
 primaryClass = {astro-ph.GA},
       adsurl = {https://ui.adsabs.harvard.edu/abs/2018MNRAS.479..703C}
}

@ARTICLE{Shimakawa2018,
       author = {{Shimakawa}, Rhythm and {Koyama}, Yusei and
         {R{\"o}ttgering}, Huub J.~A. and {Kodama}, Tadayuki and
         {Hayashi}, Masao and {Hatch}, Nina A. and {Dannerbauer}, Helmut and
         {Tanaka}, Ichi and {Tadaki}, Ken-ichi and {Suzuki}, Tomoko L. and
         {Fukagawa}, Nao and {Cai}, Zheng and {Kurk}, Jaron D.},
        title = "{MAHALO Deep Cluster Survey II. Characterizing massive forming galaxies in the Spiderweb protocluster at z = 2.2}",
      journal = {\mnras},
         year = "2018",
        month = "Dec",
       volume = {481},
       number = {4},
        pages = {5630-5650},
          doi = {10.1093/mnras/sty2618},
archivePrefix = {arXiv},
       eprint = {1809.08755},
 primaryClass = {astro-ph.GA},
       adsurl = {https://ui.adsabs.harvard.edu/abs/2018MNRAS.481.5630S}
}

@ARTICLE{Lemaux2012,
       author = {{Lemaux}, B.~C. and {Gal}, R.~R. and {Lubin}, L.~M. and
         {Kocevski}, D.~D. and {Fassnacht}, C.~D. and {McGrath}, E.~J. and
         {Squires}, G.~K. and {Surace}, J.~A. and {Lacy}, M.},
        title = "{The Assembly of the Red Sequence at z \raisebox{-0.5ex}\textasciitilde 1: The Color and Spectral Properties of Galaxies in the Cl1604 Supercluster}",
      journal = {\apj},
         year = "2012",
        month = "Feb",
       volume = {745},
       number = {2},
          eid = {106},
        pages = {106},
          doi = {10.1088/0004-637X/745/2/106},
archivePrefix = {arXiv},
       eprint = {1108.5799},
 primaryClass = {astro-ph.CO},
       adsurl = {https://ui.adsabs.harvard.edu/abs/2012ApJ...745..106L}
}

@ARTICLE{Cucciati2014,
       author = {{Cucciati}, O. and {Zamorani}, G. and {Lemaux}, B.~C. and
         {Bardelli}, S. and {Cimatti}, A. and {Le F{\`e}vre}, O. and
         {Cassata}, P. and {Garilli}, B. and {Le Brun}, V. and {Maccagni}, D. and
         {Pentericci}, L. and {Tasca}, L.~A.~M. and {Thomas}, R. and
         {Vanzella}, E. and {Zucca}, E. and {Amorin}, R. and {Capak}, P. and
         {Cassar{\`a}}, L.~P. and {Castellano}, M. and {Cuby}, J.~G. and
         {de la Torre}, S. and {Durkalec}, A. and {Fontana}, A. and
         {Giavalisco}, M. and {Grazian}, A. and {Hathi}, N.~P. and {Ilbert}, O. and
         {Moreau}, C. and {Paltani}, S. and {Ribeiro}, B. and {Salvato}, M. and
         {Schaerer}, D. and {Scodeggio}, M. and {Sommariva}, V. and {Talia}, M. and
         {Taniguchi}, Y. and {Tresse}, L. and {Vergani}, D. and {Wang}, P.~W. and
         {Charlot}, S. and {Contini}, T. and {Fotopoulou}, S. and
         {L{\'o}pez-Sanjuan}, C. and {Mellier}, Y. and {Scoville}, N.},
        title = "{Discovery of a rich proto-cluster at z = 2.9 and associated diffuse cold gas in the VIMOS Ultra-Deep Survey (VUDS)}",
      journal = {\aap},
         year = "2014",
        month = "Oct",
       volume = {570},
          eid = {A16},
        pages = {A16},
          doi = {10.1051/0004-6361/201423811},
archivePrefix = {arXiv},
       eprint = {1403.3691},
 primaryClass = {astro-ph.CO},
       adsurl = {https://ui.adsabs.harvard.edu/abs/2014A&A...570A..16C}
}

@ARTICLE{Golden-Marx2019,
       author = {{Golden-Marx}, Emmet and {Blanton}, E.~L. and {Paterno-Mahler}, R. and
         {Brodwin}, M. and {Ashby}, M.~L.~N. and {Lemaux}, B.~C. and
         {Lubin}, L.~M. and {Gal}, R.~R. and {Tomczak}, A.~R.},
        title = "{The High-redshift Clusters Occupied by Bent Radio AGN (COBRA) Survey: Follow-up Optical Imaging}",
      journal = {\apj},
         year = "2019",
        month = "Dec",
       volume = {887},
       number = {1},
          eid = {50},
        pages = {50},
          doi = {10.3847/1538-4357/ab5106},
archivePrefix = {arXiv},
       eprint = {1910.11884},
 primaryClass = {astro-ph.GA},
       adsurl = {https://ui.adsabs.harvard.edu/abs/2019ApJ...887...50G}
}

@ARTICLE{Moravec2020,
       author = {{Moravec}, Emily and {Gonzalez}, Anthony H. and {Stern}, Daniel and
         {Clarke}, Tracy and {Brodwin}, Mark and {Decker}, Bandon and
         {Eisenhardt}, Peter R.~M. and {Mo}, Wenli and {Pope}, Alexandra and
         {Stanford}, Spencer A. and {Wylezalek}, Dominika},
        title = "{The Massive and Distant Clusters of WISE Survey. VII. The Environments and Properties of Radio Galaxies in Clusters at z {\ensuremath{\sim}} 1}",
      journal = {\apj},
         year = "2020",
        month = "Jan",
       volume = {888},
       number = {2},
          eid = {74},
        pages = {74},
          doi = {10.3847/1538-4357/ab5af0},
archivePrefix = {arXiv},
       eprint = {1911.09695},
 primaryClass = {astro-ph.GA},
       adsurl = {https://ui.adsabs.harvard.edu/abs/2020ApJ...888...74M}
}

@ARTICLE{Willis2020,
       author = {{Willis}, J.~P. and {Canning}, R.~E.~A. and {Noordeh}, E.~S. and
         {Allen}, S.~W. and {King}, A.~L. and {Mantz}, A. and {Morris}, R.~G. and
         {Stanford}, S.~A. and {Brammer}, G.},
        title = "{Spectroscopic confirmation of a mature galaxy cluster at a redshift of 2}",
      journal = {\nat},
         year = "2020",
        month = "Jan",
       volume = {577},
       number = {7788},
        pages = {39-41},
          doi = {10.1038/s41586-019-1829-4},
archivePrefix = {arXiv},
       eprint = {2001.00549},
 primaryClass = {astro-ph.GA},
       adsurl = {https://ui.adsabs.harvard.edu/abs/2020Natur.577...39W}
}

@ARTICLE{astropy:2018,
       author = {{Astropy Collaboration} and {Price-Whelan}, A.~M. and
         {Sip{\H{o}}cz}, B.~M. and {G{\"u}nther}, H.~M. and {Lim}, P.~L. and
         {Crawford}, S.~M. and {Conseil}, S. and {Shupe}, D.~L. and
         {Craig}, M.~W. and {Dencheva}, N. and {Ginsburg}, A. and {Vand
        erPlas}, J.~T. and {Bradley}, L.~D. and {P{\'e}rez-Su{\'a}rez}, D. and
         {de Val-Borro}, M. and {Aldcroft}, T.~L. and {Cruz}, K.~L. and
         {Robitaille}, T.~P. and {Tollerud}, E.~J. and {Ardelean}, C. and
         {Babej}, T. and {Bach}, Y.~P. and {Bachetti}, M. and {Bakanov}, A.~V. and
         {Bamford}, S.~P. and {Barentsen}, G. and {Barmby}, P. and
         {Baumbach}, A. and {Berry}, K.~L. and {Biscani}, F. and {Boquien}, M. and
         {Bostroem}, K.~A. and {Bouma}, L.~G. and {Brammer}, G.~B. and
         {Bray}, E.~M. and {Breytenbach}, H. and {Buddelmeijer}, H. and
         {Burke}, D.~J. and {Calderone}, G. and {Cano Rodr{\'\i}guez}, J.~L. and
         {Cara}, M. and {Cardoso}, J.~V.~M. and {Cheedella}, S. and {Copin}, Y. and
         {Corrales}, L. and {Crichton}, D. and {D'Avella}, D. and {Deil}, C. and
         {Depagne}, {\'E}. and {Dietrich}, J.~P. and {Donath}, A. and
         {Droettboom}, M. and {Earl}, N. and {Erben}, T. and {Fabbro}, S. and
         {Ferreira}, L.~A. and {Finethy}, T. and {Fox}, R.~T. and
         {Garrison}, L.~H. and {Gibbons}, S.~L.~J. and {Goldstein}, D.~A. and
         {Gommers}, R. and {Greco}, J.~P. and {Greenfield}, P. and
         {Groener}, A.~M. and {Grollier}, F. and {Hagen}, A. and {Hirst}, P. and
         {Homeier}, D. and {Horton}, A.~J. and {Hosseinzadeh}, G. and {Hu}, L. and
         {Hunkeler}, J.~S. and {Ivezi{\'c}}, {\v{Z}}. and {Jain}, A. and
         {Jenness}, T. and {Kanarek}, G. and {Kendrew}, S. and {Kern}, N.~S. and
         {Kerzendorf}, W.~E. and {Khvalko}, A. and {King}, J. and {Kirkby}, D. and
         {Kulkarni}, A.~M. and {Kumar}, A. and {Lee}, A. and {Lenz}, D. and
         {Littlefair}, S.~P. and {Ma}, Z. and {Macleod}, D.~M. and
         {Mastropietro}, M. and {McCully}, C. and {Montagnac}, S. and
         {Morris}, B.~M. and {Mueller}, M. and {Mumford}, S.~J. and {Muna}, D. and
         {Murphy}, N.~A. and {Nelson}, S. and {Nguyen}, G.~H. and
         {Ninan}, J.~P. and {N{\"o}the}, M. and {Ogaz}, S. and {Oh}, S. and
         {Parejko}, J.~K. and {Parley}, N. and {Pascual}, S. and {Patil}, R. and
         {Patil}, A.~A. and {Plunkett}, A.~L. and {Prochaska}, J.~X. and
         {Rastogi}, T. and {Reddy Janga}, V. and {Sabater}, J. and
         {Sakurikar}, P. and {Seifert}, M. and {Sherbert}, L.~E. and
         {Sherwood-Taylor}, H. and {Shih}, A.~Y. and {Sick}, J. and
         {Silbiger}, M.~T. and {Singanamalla}, S. and {Singer}, L.~P. and
         {Sladen}, P.~H. and {Sooley}, K.~A. and {Sornarajah}, S. and
         {Streicher}, O. and {Teuben}, P. and {Thomas}, S.~W. and
         {Tremblay}, G.~R. and {Turner}, J.~E.~H. and {Terr{\'o}n}, V. and
         {van Kerkwijk}, M.~H. and {de la Vega}, A. and {Watkins}, L.~L. and
         {Weaver}, B.~A. and {Whitmore}, J.~B. and {Woillez}, J. and
         {Zabalza}, V. and {Astropy Contributors}},
        title = "{The Astropy Project: Building an Open-science Project and Status of the v2.0 Core Package}",
      journal = {\aj},
         year = 2018,
        month = sep,
       volume = {156},
       number = {3},
          eid = {123},
        pages = {123},
          doi = {10.3847/1538-3881/aabc4f},
archivePrefix = {arXiv},
       eprint = {1801.02634},
 primaryClass = {astro-ph.IM},
       adsurl = {https://ui.adsabs.harvard.edu/abs/2018AJ....156..123A}
}

@ARTICLE{astropy:2013,
       author = {{Astropy Collaboration} and {Robitaille}, Thomas P. and
         {Tollerud}, Erik J. and {Greenfield}, Perry and {Droettboom}, Michael and
         {Bray}, Erik and {Aldcroft}, Tom and {Davis}, Matt and
         {Ginsburg}, Adam and {Price-Whelan}, Adrian M. and
         {Kerzendorf}, Wolfgang E. and {Conley}, Alexander and {Crighton}, Neil and
         {Barbary}, Kyle and {Muna}, Demitri and {Ferguson}, Henry and
         {Grollier}, Fr{\'e}d{\'e}ric and {Parikh}, Madhura M. and
         {Nair}, Prasanth H. and {Unther}, Hans M. and {Deil}, Christoph and
         {Woillez}, Julien and {Conseil}, Simon and {Kramer}, Roban and
         {Turner}, James E.~H. and {Singer}, Leo and {Fox}, Ryan and
         {Weaver}, Benjamin A. and {Zabalza}, Victor and {Edwards}, Zachary I. and
         {Azalee Bostroem}, K. and {Burke}, D.~J. and {Casey}, Andrew R. and
         {Crawford}, Steven M. and {Dencheva}, Nadia and {Ely}, Justin and
         {Jenness}, Tim and {Labrie}, Kathleen and {Lim}, Pey Lian and
         {Pierfederici}, Francesco and {Pontzen}, Andrew and {Ptak}, Andy and
         {Refsdal}, Brian and {Servillat}, Mathieu and {Streicher}, Ole},
        title = "{Astropy: A community Python package for astronomy}",
      journal = {\aap},
         year = 2013,
        month = oct,
       volume = {558},
          eid = {A33},
        pages = {A33},
          doi = {10.1051/0004-6361/201322068},
archivePrefix = {arXiv},
       eprint = {1307.6212},
 primaryClass = {astro-ph.IM},
       adsurl = {https://ui.adsabs.harvard.edu/abs/2013A&A...558A..33A}
}

@ARTICLE{Tomczak2019,
       author = {{Tomczak}, Adam R. and {Lemaux}, Brian C. and {Lubin}, Lori M. and
         {Pelliccia}, Debora and {Shen}, Lu and {Gal}, Roy R. and
         {Hung}, Denise and {Kocevski}, Dale D. and {Le F{\`e}vre}, Olivier and
         {Mei}, Simona and {Rumbaugh}, Nicholas and {Squires}, Gordon K. and
         {Wu}, Po-Feng},
        title = "{Conditional quenching: a detailed look at the SFR-density relation at z {\ensuremath{\sim}} 0.9 from ORELSE}",
      journal = {\mnras},
         year = 2019,
        month = apr,
       volume = {484},
       number = {4},
        pages = {4695-4710},
          doi = {10.1093/mnras/stz342},
archivePrefix = {arXiv},
       eprint = {1812.04633},
 primaryClass = {astro-ph.GA},
       adsurl = {https://ui.adsabs.harvard.edu/abs/2019MNRAS.484.4695T}
}

@ARTICLE{Foltz2018,
       author = {{Foltz}, R. and {Wilson}, G. and {Muzzin}, A. and {Cooper}, M.~C. and
         {Nantais}, J. and {van der Burg}, R.~F.~J. and {Cerulo}, P. and
         {Chan}, J. and {Fillingham}, S.~P. and {Surace}, J. and {Webb}, T. and
         {Noble}, A. and {Lacy}, M. and {McDonald}, M. and {Rudnick}, G. and
         {Lidman}, C. and {Demarco}, R. and {Hlavacek-Larrondo}, J. and
         {Yee}, H.~K.~C. and {Perlmutter}, S. and {Hayden}, B.},
        title = "{The Evolution of Environmental Quenching Timescales to z ̃ 1.6: Evidence for Dynamically Driven Quenching of the Cluster Galaxy Population}",
      journal = {\apj},
         year = 2018,
        month = oct,
       volume = {866},
       number = {2},
          eid = {136},
        pages = {136},
          doi = {10.3847/1538-4357/aad80d},
archivePrefix = {arXiv},
       eprint = {1803.03305},
 primaryClass = {astro-ph.GA},
       adsurl = {https://ui.adsabs.harvard.edu/abs/2018ApJ...866..136F}
}

@ARTICLE{Nantais2017,
       author = {{Nantais}, Julie B. and {Muzzin}, Adam and {van der Burg}, Remco F.~J. and
         {Wilson}, Gillian and {Lidman}, Chris and {Foltz}, Ryan and
         {DeGroot}, Andrew and {Noble}, Allison and {Cooper}, Michael C. and
         {Demarco}, Ricardo},
        title = "{Evidence for strong evolution in galaxy environmental quenching efficiency between z = 1.6 and z = 0.9}",
      journal = {\mnras},
         year = 2017,
        month = feb,
       volume = {465},
       number = {1},
        pages = {L104-L108},
          doi = {10.1093/mnrasl/slw224},
archivePrefix = {arXiv},
       eprint = {1610.08058},
 primaryClass = {astro-ph.GA},
       adsurl = {https://ui.adsabs.harvard.edu/abs/2017MNRAS.465L.104N}
}

@ARTICLE{Nantais2016,
       author = {{Nantais}, Julie B. and {van der Burg}, Remco F.~J. and {Lidman}, Chris and
         {Demarco}, Ricardo and {Noble}, Allison and {Wilson}, Gillian and
         {Muzzin}, Adam and {Foltz}, Ryan and {DeGroot}, Andrew and
         {Cooper}, Michael C.},
        title = "{Stellar mass function of cluster galaxies at z \raisebox{-0.5ex}\textasciitilde 1.5: evidence for reduced quenching efficiency at high redshift}",
      journal = {\aap},
         year = 2016,
        month = aug,
       volume = {592},
          eid = {A161},
        pages = {A161},
          doi = {10.1051/0004-6361/201628663},
archivePrefix = {arXiv},
       eprint = {1606.07832},
 primaryClass = {astro-ph.GA},
       adsurl = {https://ui.adsabs.harvard.edu/abs/2016A&A...592A.161N}
}

@ARTICLE{Shen2019,
       author = {{Shen}, Lu and {Tomczak}, Adam R. and {Lemaux}, Brian C. and
         {Pelliccia}, Debora and {Lubin}, Lori M. and {Miller}, Neal A. and
         {Perrotta}, Serena and {Fassnacht}, Christopher D. and
         {Becker}, Robert H. and {Gal}, Roy R. and {Wu}, Po-Feng and
         {Squires}, Gordon},
        title = "{Possible evidence of the radio AGN quenching of neighbouring galaxies at z {\ensuremath{\sim}} 1}",
      journal = {\mnras},
         year = 2019,
        month = apr,
       volume = {484},
       number = {2},
        pages = {2433-2446},
          doi = {10.1093/mnras/stz152},
archivePrefix = {arXiv},
       eprint = {1812.04667},
 primaryClass = {astro-ph.GA},
       adsurl = {https://ui.adsabs.harvard.edu/abs/2019MNRAS.484.2433S}
}

@ARTICLE{Hatch2011,
       author = {{Hatch}, N.~A. and {De Breuck}, C. and {Galametz}, A. and
         {Miley}, G.~K. and {Overzier}, R.~A. and {R{\"o}ttgering}, H.~J.~A. and
         {Doherty}, M. and {Kodama}, T. and {Kurk}, J.~D. and {Seymour}, N. and
         {Venemans}, B.~P. and {Vernet}, J. and {Zirm}, A.~W.},
        title = "{Galaxy protocluster candidates around z{\ensuremath{\sim}} 2.4 radio galaxies}",
      journal = {\mnras},
         year = 2011,
        month = jan,
       volume = {410},
       number = {3},
        pages = {1537-1549},
          doi = {10.1111/j.1365-2966.2010.17538.x},
archivePrefix = {arXiv},
       eprint = {1008.4588},
 primaryClass = {astro-ph.CO},
       adsurl = {https://ui.adsabs.harvard.edu/abs/2011MNRAS.410.1537H}
}

@ARTICLE{Golden-Marx2021,
       author = {{Golden-Marx}, Emmet and {Blanton}, E.~L. and {Paterno-Mahler}, R. and {Brodwin}, M. and {Ashby}, M.~L.~N. and {Moravec}, E. and {Shen}, L. and {Lemaux}, B.~C. and {Lubin}, L.~M. and {Gal}, R.~R. and {Tomczak}, A.~R.},
        title = "{The High-redshift Clusters Occupied by Bent Radio AGN (COBRA) Survey: Radio Source Properties}",
      journal = {\apj},
         year = 2021,
        month = feb,
       volume = {907},
       number = {2},
          eid = {65},
        pages = {65},
          doi = {10.3847/1538-4357/abcd96},
archivePrefix = {arXiv},
       eprint = {2011.12313},
 primaryClass = {astro-ph.GA},
       adsurl = {https://ui.adsabs.harvard.edu/abs/2021ApJ...907...65G}
}

@ARTICLE{Miller2005,
       author = {{Miller}, Christopher J. and {Nichol}, Robert C. and {Reichart}, Daniel and
         {Wechsler}, Risa H. and {Evrard}, August E. and {Annis}, James and
         {McKay}, Timothy A. and {Bahcall}, Neta A. and {Bernardi}, Mariangela and
         {Boehringer}, Hans and {Connolly}, Andrew J. and {Goto}, Tomotsugu and
         {Kniazev}, Alexie and {Lamb}, Donald and {Postman}, Marc and
         {Schneider}, Donald P. and {Sheth}, Ravi K. and {Voges}, Wolfgang},
        title = "{The C4 Clustering Algorithm: Clusters of Galaxies in the Sloan Digital Sky Survey}",
      journal = {\aj},
         year = 2005,
        month = sep,
       volume = {130},
       number = {3},
        pages = {968-1001},
          doi = {10.1086/431357},
archivePrefix = {arXiv},
       eprint = {astro-ph/0503713},
 primaryClass = {astro-ph},
       adsurl = {https://ui.adsabs.harvard.edu/abs/2005AJ....130..968M}
}

@ARTICLE{Cai2017,
       author = {{Cai}, Zheng and {Fan}, Xiaohui and {Bian}, Fuyan and {Zabludoff}, Ann and {Yang}, Yujin and {Prochaska}, J. Xavier and {McGreer}, Ian and {Zheng}, Zhen-Ya and {Kashikawa}, Nobunari and {Wang}, Ran and {Frye}, Brenda and {Green}, Richard and {Jiang}, Linhua},
        title = "{Mapping the Most Massive Overdensities through Hydrogen (MAMMOTH). II. Discovery of the Extremely Massive Overdensity BOSS1441 at z = 2.32}",
      journal = {\apj},
         year = 2017,
        month = apr,
       volume = {839},
       number = {2},
          eid = {131},
        pages = {131},
          doi = {10.3847/1538-4357/aa6a1a},
archivePrefix = {arXiv},
       eprint = {1609.02913},
 primaryClass = {astro-ph.GA},
       adsurl = {https://ui.adsabs.harvard.edu/abs/2017ApJ...839..131C}
}

@ARTICLE{Shen2021,
       author = {{Shen}, Lu and {Lemaux}, Brian C. and {Lubin}, Lori M. and {Cucciati}, Olga and {Le F{\`e}vre}, Olivier and {Liu}, Guilin and {Fang}, Wenjuan and {Pelliccia}, Debora and {Tomczak}, Adam and {McKean}, John and {Miller}, Neal A. and {Fassnacht}, Christopher D. and {Gal}, Roy and {Hung}, Denise and {Hathi}, Nimish and {Bardelli}, Sandro and {Vergani}, Daniela and {Zucca}, Elena},
        title = "{Implications of the Environments of Radio-detected Active Galactic Nuclei in a Complex Protostructure at z {\ensuremath{\sim}} 3.3}",
      journal = {\apj},
         year = 2021,
        month = may,
       volume = {912},
       number = {1},
          eid = {60},
        pages = {60},
          doi = {10.3847/1538-4357/abee7510.48550/arXiv.2103.03441},
archivePrefix = {arXiv},
       eprint = {2103.03441},
 primaryClass = {astro-ph.GA},
       adsurl = {https://ui.adsabs.harvard.edu/abs/2021ApJ...912...60S}
}

@ARTICLE{Rudnick2012,
       author = {{Rudnick}, Gregory H. and {Tran}, Kim-Vy and {Papovich}, Casey and
         {Momcheva}, Ivelina and {Willmer}, Christopher},
        title = "{A Tale of Dwarfs and Giants: Using a z = 1.62 Cluster to Understand How the Red Sequence Grew over the Last 9.5 Billion Years}",
      journal = {\apj},
         year = 2012,
        month = aug,
       volume = {755},
       number = {1},
          eid = {14},
        pages = {14},
          doi = {10.1088/0004-637X/755/1/14},
archivePrefix = {arXiv},
       eprint = {1203.3541},
 primaryClass = {astro-ph.CO},
       adsurl = {https://ui.adsabs.harvard.edu/abs/2012ApJ...755...14R}
}

@ARTICLE{Lee-Brown2017,
       author = {{Lee-Brown}, Donald B. and {Rudnick}, Gregory H. and
         {Momcheva}, Ivelina G. and {Papovich}, Casey and {Lotz}, Jennifer M. and
         {Tran}, Kim-Vy H. and {Henke}, Brittany and
         {Willmer}, Christopher N.~A. and {Brammer}, Gabriel B. and
         {Brodwin}, Mark and {Dunlop}, James and {Farrah}, Duncan},
        title = "{The Ages of Passive Galaxies in a z = 1.62 Protocluster}",
      journal = {\apj},
         year = 2017,
        month = jul,
       volume = {844},
       number = {1},
          eid = {43},
        pages = {43},
          doi = {10.3847/1538-4357/aa7948},
archivePrefix = {arXiv},
       eprint = {1706.05017},
 primaryClass = {astro-ph.GA},
       adsurl = {https://ui.adsabs.harvard.edu/abs/2017ApJ...844...43L}
}

@ARTICLE{Cerulo2017,
       author = {{Cerulo}, P. and {Couch}, W.~J. and {Lidman}, C. and {Demarco}, R. and
         {Huertas-Company}, M. and {Mei}, S. and {S{\'a}nchez-Janssen}, R. and
         {Barrientos}, L.~F. and {Mu{\~n}oz}, R.},
        title = "{The morphological transformation of red sequence galaxies in clusters since z {\ensuremath{\sim}} 1}",
      journal = {\mnras},
         year = 2017,
        month = nov,
       volume = {472},
       number = {1},
        pages = {254-272},
          doi = {10.1093/mnras/stx1687},
archivePrefix = {arXiv},
       eprint = {1707.00751},
 primaryClass = {astro-ph.GA},
       adsurl = {https://ui.adsabs.harvard.edu/abs/2017MNRAS.472..254C}
}

@ARTICLE{Roberts2019,
       author = {{Roberts}, Ian D. and {Parker}, Laura C. and {Brown}, Toby and
         {Joshi}, Gandhali D. and {Hlavacek-Larrondo}, Julie and
         {Wadsley}, James},
        title = "{Quenching Low-mass Satellite Galaxies: Evidence for a Threshold ICM Density}",
      journal = {\apj},
         year = 2019,
        month = mar,
       volume = {873},
       number = {1},
          eid = {42},
        pages = {42},
          doi = {10.3847/1538-4357/ab04f7},
archivePrefix = {arXiv},
       eprint = {1902.02820},
 primaryClass = {astro-ph.GA},
       adsurl = {https://ui.adsabs.harvard.edu/abs/2019ApJ...873...42R}
}

@ARTICLE{Overzier2016,
       author = {{Overzier}, Roderik A.},
        title = "{The realm of the galaxy protoclusters. A review}",
      journal = {\aapr},
         year = 2016,
        month = nov,
       volume = {24},
       number = {1},
          eid = {14},
        pages = {14},
          doi = {10.1007/s00159-016-0100-3},
archivePrefix = {arXiv},
       eprint = {1610.05201},
 primaryClass = {astro-ph.GA},
       adsurl = {https://ui.adsabs.harvard.edu/abs/2016A&ARv..24...14O}
}

@ARTICLE{Muldrew2015,
       author = {{Muldrew}, Stuart I. and {Hatch}, Nina A. and {Cooke}, Elizabeth A.},
        title = "{What are protoclusters? - Defining high-redshift galaxy clusters and protoclusters}",
      journal = {\mnras},
         year = 2015,
        month = sep,
       volume = {452},
       number = {3},
        pages = {2528-2539},
          doi = {10.1093/mnras/stv1449},
archivePrefix = {arXiv},
       eprint = {1506.08835},
 primaryClass = {astro-ph.CO},
       adsurl = {https://ui.adsabs.harvard.edu/abs/2015MNRAS.452.2528M}
}

@ARTICLE{Hatch2017,
       author = {{Hatch}, N.~A. and {Cooke}, E.~A. and {Muldrew}, S.~I. and
         {Hartley}, W.~G. and {Almaini}, O. and {Conselice}, C.~J. and
         {Simpson}, C.~J.},
        title = "{The impact of protocluster environments at z = 1.6}",
      journal = {\mnras},
         year = 2017,
        month = jan,
       volume = {464},
       number = {1},
        pages = {876-884},
          doi = {10.1093/mnras/stw2359},
archivePrefix = {arXiv},
       eprint = {1609.08629},
 primaryClass = {astro-ph.GA},
       adsurl = {https://ui.adsabs.harvard.edu/abs/2017MNRAS.464..876H}
}

@ARTICLE{Kurk2004,
       author = {{Kurk}, J.~D. and {Pentericci}, L. and {R{\"o}ttgering}, H.~J.~A. and
         {Miley}, G.~K.},
        title = "{A search for clusters at high redshift. III. Candidate H{\ensuremath{\alpha}} emitters and EROs in the PKS 1138-262 proto-cluster at z = 2.16}",
      journal = {\aap},
         year = 2004,
        month = dec,
       volume = {428},
        pages = {793-815},
          doi = {10.1051/0004-6361:20040075},
archivePrefix = {arXiv},
       eprint = {astro-ph/0410202},
 primaryClass = {astro-ph},
       adsurl = {https://ui.adsabs.harvard.edu/abs/2004A&A...428..793K}
}

@ARTICLE{Hill2020,
       author = {{Hill}, Ryley and {Chapman}, Scott and {Scott}, Douglas and
         {Apostolovski}, Yordanka and {Aravena}, Manuel and
         {B{\'e}thermin}, Matthieu and {Bradford}, C.~M. and
         {Canning}, Rebecca E.~A. and {De Breuck}, Carlos and {Dong}, Chenxing and
         {Gonzalez}, Anthony and {Greve}, Thomas R. and
         {Hayward}, Christopher C. and {Hezaveh}, Yashar and {Litke}, Katrina and
         {Malkan}, Matt and {Marrone}, Daniel P. and {Phadke}, Kedar and
         {Reuter}, Cassie and {Rotermund}, Kaja and {Spilker}, Justin and
         {Vieira}, Joaquin D. and {Wei{\ss}}, Axel},
        title = "{Megaparsec-scale structure around the protocluster core SPT2349-56 at z = 4.3}",
      journal = {\mnras},
         year = 2020,
        month = may,
       volume = {495},
       number = {3},
        pages = {3124-3159},
          doi = {10.1093/mnras/staa1275},
       adsurl = {https://ui.adsabs.harvard.edu/abs/2020MNRAS.495.3124H}
}

@ARTICLE{Muldrew2018,
       author = {{Muldrew}, Stuart I. and {Hatch}, Nina A. and {Cooke}, Elizabeth A.},
        title = "{Galaxy evolution in protoclusters}",
      journal = {\mnras},
         year = 2018,
        month = jan,
       volume = {473},
       number = {2},
        pages = {2335-2347},
          doi = {10.1093/mnras/stx2454},
archivePrefix = {arXiv},
       eprint = {1709.07009},
 primaryClass = {astro-ph.GA},
       adsurl = {https://ui.adsabs.harvard.edu/abs/2018MNRAS.473.2335M}
}

@ARTICLE{Contini2016,
       author = {{Contini}, E. and {De Lucia}, G. and {Hatch}, N. and {Borgani}, S. and
         {Kang}, X.},
        title = "{Semi-analytic model predictions of the galaxy population in protoclusters}",
      journal = {\mnras},
         year = 2016,
        month = feb,
       volume = {456},
       number = {2},
        pages = {1924-1935},
          doi = {10.1093/mnras/stv2852},
archivePrefix = {arXiv},
       eprint = {1512.01561},
 primaryClass = {astro-ph.GA},
       adsurl = {https://ui.adsabs.harvard.edu/abs/2016MNRAS.456.1924C}
}

@ARTICLE{Kawinwanichakij2017,
       author = {{Kawinwanichakij}, Lalitwadee and {Papovich}, Casey and
         {Quadri}, Ryan F. and {Glazebrook}, Karl and {Kacprzak}, Glenn G. and
         {Allen}, Rebecca J. and {Bell}, Eric F. and {Croton}, Darren J. and
         {Dekel}, Avishai and {Ferguson}, Henry C. and {Forrest}, Ben and
         {Grogin}, Norman A. and {Guo}, Yicheng and {Kocevski}, Dale D. and
         {Koekemoer}, Anton M. and {Labb{\'e}}, Ivo and {Lucas}, Ray A. and
         {Nanayakkara}, Themiya and {Spitler}, Lee R. and
         {Straatman}, Caroline M.~S. and {Tran}, Kim-Vy H. and {Tomczak}, Adam and
         {van Dokkum}, Pieter},
        title = "{Effect of Local Environment and Stellar Mass on Galaxy Quenching and Morphology at 0.5 \&lt; z \&lt; 2.0}",
      journal = {\apj},
         year = 2017,
        month = oct,
       volume = {847},
       number = {2},
          eid = {134},
        pages = {134},
          doi = {10.3847/1538-4357/aa8b75},
archivePrefix = {arXiv},
       eprint = {1706.03780},
 primaryClass = {astro-ph.GA},
       adsurl = {https://ui.adsabs.harvard.edu/abs/2017ApJ...847..134K}
}

@ARTICLE{Cappellari2016,
       author = {{Cappellari}, Michele},
        title = "{Structure and Kinematics of Early-Type Galaxies from Integral Field Spectroscopy}",
      journal = {\araa},
         year = 2016,
        month = sep,
       volume = {54},
        pages = {597-665},
          doi = {10.1146/annurev-astro-082214-122432},
archivePrefix = {arXiv},
       eprint = {1602.04267},
 primaryClass = {astro-ph.GA},
       adsurl = {https://ui.adsabs.harvard.edu/abs/2016ARA&A..54..597C}
}

@ARTICLE{Balogh2000,
       author = {{Balogh}, Michael L. and {Navarro}, Julio F. and {Morris}, Simon L.},
        title = "{The Origin of Star Formation Gradients in Rich Galaxy Clusters}",
      journal = {\apj},
         year = 2000,
        month = sep,
       volume = {540},
       number = {1},
        pages = {113-121},
          doi = {10.1086/309323},
archivePrefix = {arXiv},
       eprint = {astro-ph/0004078},
 primaryClass = {astro-ph},
       adsurl = {https://ui.adsabs.harvard.edu/abs/2000ApJ...540..113B}
}

@ARTICLE{Larson1980,
       author = {{Larson}, R.~B. and {Tinsley}, B.~M. and {Caldwell}, C.~N.},
        title = "{The evolution of disk galaxies and the origin of S0 galaxies}",
      journal = {\apj},
         year = 1980,
        month = may,
       volume = {237},
        pages = {692-707},
          doi = {10.1086/157917},
       adsurl = {https://ui.adsabs.harvard.edu/abs/1980ApJ...237..692L}
}

@ARTICLE{Gunn1972,
       author = {{Gunn}, James E. and {Gott}, J. Richard, III},
        title = "{On the Infall of Matter Into Clusters of Galaxies and Some Effects on Their Evolution}",
      journal = {\apj},
         year = 1972,
        month = aug,
       volume = {176},
        pages = {1},
          doi = {10.1086/151605},
       adsurl = {https://ui.adsabs.harvard.edu/abs/1972ApJ...176....1G}
}

@ARTICLE{Quilis2000,
       author = {{Quilis}, Vicent and {Moore}, Ben and {Bower}, Richard},
        title = "{Gone with the Wind: The Origin of S0 Galaxies in Clusters}",
      journal = {Science},
         year = 2000,
        month = jun,
       volume = {288},
       number = {5471},
        pages = {1617-1620},
          doi = {10.1126/science.288.5471.1617},
archivePrefix = {arXiv},
       eprint = {astro-ph/0006031},
 primaryClass = {astro-ph},
       adsurl = {https://ui.adsabs.harvard.edu/abs/2000Sci...288.1617Q}
}

@ARTICLE{Dressler1980,
       author = {{Dressler}, A.},
        title = "{Galaxy morphology in rich clusters: implications for the formation and evolution of galaxies.}",
      journal = {\apj},
         year = 1980,
        month = mar,
       volume = {236},
        pages = {351-365},
          doi = {10.1086/157753},
       adsurl = {https://ui.adsabs.harvard.edu/abs/1980ApJ...236..351D}
}

@ARTICLE{Postman2005,
       author = {{Postman}, M. and {Franx}, M. and {Cross}, N.~J.~G. and {Holden}, B. and {Ford}, H.~C. and {Illingworth}, G.~D. and {Goto}, T. and {Demarco}, R. and {Rosati}, P. and {Blakeslee}, J.~P. and {Tran}, K. -V. and {Ben{\'\i}tez}, N. and {Clampin}, M. and {Hartig}, G.~F. and {Homeier}, N. and {Ardila}, D.~R. and {Bartko}, F. and {Bouwens}, R.~J. and {Bradley}, L.~D. and {Broadhurst}, T.~J. and {Brown}, R.~A. and {Burrows}, C.~J. and {Cheng}, E.~S. and {Feldman}, P.~D. and {Golimowski}, D.~A. and {Gronwall}, C. and {Infante}, L. and {Kimble}, R.~A. and {Krist}, J.~E. and {Lesser}, M.~P. and {Martel}, A.~R. and {Mei}, S. and {Menanteau}, F. and {Meurer}, G.~R. and {Miley}, G.~K. and {Motta}, V. and {Sirianni}, M. and {Sparks}, W.~B. and {Tran}, H.~D. and {Tsvetanov}, Z.~I. and {White}, R.~L. and {Zheng}, W.},
        title = "{The Morphology-Density Relation in z \raisebox{-0.5ex}\textasciitilde 1 Clusters}",
      journal = {\apj},
         year = 2005,
        month = apr,
       volume = {623},
       number = {2},
        pages = {721-741},
          doi = {10.1086/428881},
archivePrefix = {arXiv},
       eprint = {astro-ph/0501224},
 primaryClass = {astro-ph},
       adsurl = {https://ui.adsabs.harvard.edu/abs/2005ApJ...623..721P}
}

@ARTICLE{Mei2023,
       author = {{Mei}, Simona and {Hatch}, Nina A. and {Amodeo}, Stefania and {Afanasiev}, Anton V. and {De Breuck}, Carlos and {Stern}, Daniel and {Cooke}, Elizabeth A. and {Gonzalez}, Anthony H. and {Noirot}, Ga{\"e}l and {Rettura}, Alessandro and {Seymour}, Nick and {Stanford}, Spencer A. and {Vernet}, Jo{\"e}l and {Wylezalek}, Dominika},
        title = "{Morphology-density relation, quenching, and mergers in CARLA clusters and protoclusters at 1.4 < z < 2.8}",
      journal = {\aap},
         year = 2023,
        month = feb,
       volume = {670},
          eid = {A58},
        pages = {A58},
          doi = {10.1051/0004-6361/202243551},
archivePrefix = {arXiv},
       eprint = {2209.02078},
 primaryClass = {astro-ph.GA},
       adsurl = {https://ui.adsabs.harvard.edu/abs/2023A&A...670A..58M}
}

@ARTICLE{Afanasiev2023,
       author = {{Afanasiev}, Anton V. and {Mei}, Simona and {Fu}, Hao and {Shankar}, Francesco and {Amodeo}, Stefania and {Stern}, Daniel and {Cooke}, Elizabeth A. and {Gonzalez}, Anthony H. and {Noirot}, Ga{\"e}l and {Rettura}, Alessandro and {Wylezalek}, Dominika and {De Breuck}, Carlos and {Hatch}, Nina A. and {Stanford}, Spencer A. and {Vernet}, Jo{\"e}l},
        title = "{The galaxy mass-size relation in CARLA clusters and proto-clusters at 1.4 < z < 2.8: Larger cluster galaxy sizes}",
      journal = {\aap},
         year = 2023,
        month = feb,
       volume = {670},
          eid = {A95},
        pages = {A95},
          doi = {10.1051/0004-6361/202244634},
archivePrefix = {arXiv},
       eprint = {2212.00031},
 primaryClass = {astro-ph.GA},
       adsurl = {https://ui.adsabs.harvard.edu/abs/2023A&A...670A..95A}
}

@ARTICLE{Shen2024,
       author = {{Shen}, Lu and {Papovich}, Casey and {Matharu}, Jasleen and {Pirzkal}, Nor and {Hu}, Weida and {Backhaus}, Bren E. and {Bagley}, Micaela B. and {Cheng}, Yingjie and {Cleri}, Nikko J. and {Finkelstein}, Steven L. and {Huertas-Company}, Marc and {Giavalisco}, Mauro and {Grogin}, Norman A. and {Jung}, Intae and {Kartaltepe}, Jeyhan S. and {Koekemoer}, Anton M. and {Lotz}, Jennifer M. and {Maseda}, Michael V. and {P{\'e}rez-Gonz{\'a}lez}, Pablo G. and {Rothberg}, Barry and {Simons}, Raymond C. and {Tacchella}, Sandro and {Williams}, Christina C. and {Yung}, L.~Y. Aaron},
        title = "{NGDEEP Epoch 1: Spatially Resolved H{\ensuremath{\alpha}} Observations of Disk and Bulge Growth in Star-forming Galaxies at z {\ensuremath{\sim}} 0.6{\textendash}2.2 from JWST NIRISS Slitless Spectroscopy}",
      journal = {\apjl},
         year = 2024,
        month = mar,
       volume = {963},
       number = {2},
          eid = {L49},
        pages = {L49},
          doi = {10.3847/2041-8213/ad28bd},
archivePrefix = {arXiv},
       eprint = {2310.13745},
 primaryClass = {astro-ph.GA},
       adsurl = {https://ui.adsabs.harvard.edu/abs/2024ApJ...963L..49S}
}

@ARTICLE{Noordeh2021,
       author = {{Noordeh}, E. and {Canning}, R.~E.~A. and {Willis}, J.~P. and {Allen}, S.~W. and {Mantz}, A. and {Stanford}, S.~A. and {Brammer}, G.},
        title = "{Quiescent galaxies in a virialized cluster at redshift 2: evidence for accelerated size growth}",
      journal = {\mnras},
         year = 2021,
        month = nov,
       volume = {507},
       number = {4},
        pages = {5272-5280},
          doi = {10.1093/mnras/stab2459},
archivePrefix = {arXiv},
       eprint = {2109.02200},
 primaryClass = {astro-ph.GA},
       adsurl = {https://ui.adsabs.harvard.edu/abs/2021MNRAS.507.5272N}
}

@ARTICLE{Sazonova2020,
       author = {{Sazonova}, Elizaveta and {Alatalo}, Katherine and {Lotz}, Jennifer and {Rowlands}, Kate and {Snyder}, Gregory F. and {Boone}, Kyle and {Brodwin}, Mark and {Hayden}, Brian and {Lanz}, Lauranne and {Perlmutter}, Saul and {Rodriguez-Gomez}, Vicente},
        title = "{The Morphology-Density Relationship in 1 < z < 2 Clusters}",
      journal = {\apj},
         year = 2020,
        month = aug,
       volume = {899},
       number = {1},
          eid = {85},
        pages = {85},
          doi = {10.3847/1538-4357/aba42f},
archivePrefix = {arXiv},
       eprint = {2007.03698},
 primaryClass = {astro-ph.GA},
       adsurl = {https://ui.adsabs.harvard.edu/abs/2020ApJ...899...85S}
}

@ARTICLE{Lee2024,
       author = {{Lee}, Jeong Hwan and {Park}, Changbom and {Hwang}, Ho Seong and {Kwon}, Minseong},
        title = "{Morphology of Galaxies in JWST Fields: Initial Distribution and Evolution of Galaxy Morphology}",
      journal = {\apj},
         year = 2024,
        month = may,
       volume = {966},
       number = {1},
          eid = {113},
        pages = {113},
          doi = {10.3847/1538-4357/ad3448},
archivePrefix = {arXiv},
       eprint = {2312.04899},
 primaryClass = {astro-ph.GA},
       adsurl = {https://ui.adsabs.harvard.edu/abs/2024ApJ...966..113L}
}

@ARTICLE{Werner2022,
       author = {{Werner}, S.~V. and {Hatch}, N.~A. and {Muzzin}, A. and {van der Burg}, R.~F.~J. and {Balogh}, M.~L. and {Rudnick}, G. and {Wilson}, G.},
        title = "{Satellite quenching was not important for z {\ensuremath{\sim}} 1 clusters: most quenching occurred during infall}",
      journal = {\mnras},
         year = 2022,
        month = feb,
       volume = {510},
       number = {1},
        pages = {674-686},
          doi = {10.1093/mnras/stab3484},
archivePrefix = {arXiv},
       eprint = {2111.14624},
 primaryClass = {astro-ph.GA},
       adsurl = {https://ui.adsabs.harvard.edu/abs/2022MNRAS.510..674W}
}

@ARTICLE{vanderWel2012,
       author = {{van der Wel}, A. and {Bell}, E.~F. and {H{\"a}ussler}, B. and {McGrath}, E.~J. and {Chang}, Yu-Yen and {Guo}, Yicheng and {McIntosh}, D.~H. and {Rix}, H. -W. and {Barden}, M. and {Cheung}, E. and {Faber}, S.~M. and {Ferguson}, H.~C. and {Galametz}, A. and {Grogin}, N.~A. and {Hartley}, W. and {Kartaltepe}, J.~S. and {Kocevski}, D.~D. and {Koekemoer}, A.~M. and {Lotz}, J. and {Mozena}, M. and {Peth}, M.~A. and {Peng}, Chien Y.},
        title = "{Structural Parameters of Galaxies in CANDELS}",
      journal = {\apjs},
         year = 2012,
        month = dec,
       volume = {203},
       number = {2},
          eid = {24},
        pages = {24},
          doi = {10.1088/0067-0049/203/2/24},
archivePrefix = {arXiv},
       eprint = {1211.6954},
 primaryClass = {astro-ph.CO},
       adsurl = {https://ui.adsabs.harvard.edu/abs/2012ApJS..203...24V}
}

@ARTICLE{vanderWel2024,
       author = {{van der Wel}, Arjen and {Martorano}, Marco and {H{\"a}u{\ss}ler}, Boris and {Nedkova}, Kalina V. and {Miller}, Tim B. and {Brammer}, Gabriel B. and {van de Ven}, Glenn and {Leja}, Joel and {Bezanson}, Rachel S. and {Muzzin}, Adam and {Marchesini}, Danilo and {de Graaff}, Anna and {Nelson}, Erica J. and {Kriek}, Mariska and {Bell}, Eric F. and {Franx}, Marijn},
        title = "{Stellar Half-mass Radii of 0.5 z < 2.3 Galaxies: Comparison with JWST/NIRCam Half-light Radii}",
      journal = {\apj},
         year = 2024,
        month = jan,
       volume = {960},
       number = {1},
          eid = {53},
        pages = {53},
          doi = {10.3847/1538-4357/ad02ee},
archivePrefix = {arXiv},
       eprint = {2307.03264},
 primaryClass = {astro-ph.GA},
       adsurl = {https://ui.adsabs.harvard.edu/abs/2024ApJ...960...53V}
}

@ARTICLE{Cai2016,
       author = {{Cai}, Zheng and {Fan}, Xiaohui and {Peirani}, Sebastien and {Bian}, Fuyan and {Frye}, Brenda and {McGreer}, Ian and {Prochaska}, J. Xavier and {Lau}, Marie Wingyee and {Tejos}, Nicolas and {Ho}, Shirley and {Schneider}, Donald P.},
        title = "{Mapping the Most Massive Overdensity Through Hydrogen (MAMMOTH) I: Methodology}",
      journal = {\apj},
         year = 2016,
        month = dec,
       volume = {833},
       number = {2},
          eid = {135},
        pages = {135},
          doi = {10.3847/1538-4357/833/2/135},
archivePrefix = {arXiv},
       eprint = {1512.06859},
 primaryClass = {astro-ph.GA},
       adsurl = {https://ui.adsabs.harvard.edu/abs/2016ApJ...833..135C}
}

@ARTICLE{Zheng2021,
       author = {{Zheng}, Xian Zhong and {Cai}, Zheng and {An}, Fang Xia and {Fan}, Xiaohui and {Shi}, Dong Dong},
        title = "{MAMMOTH: confirmation of two massive galaxy overdensities at z = 2.24 with H{\ensuremath{\alpha}} emitters}",
      journal = {\mnras},
         year = 2021,
        month = jan,
       volume = {500},
       number = {4},
        pages = {4354-4364},
          doi = {10.1093/mnras/staa2882},
archivePrefix = {arXiv},
       eprint = {2009.08068},
 primaryClass = {astro-ph.GA},
       adsurl = {https://ui.adsabs.harvard.edu/abs/2021MNRAS.500.4354Z}
}

@ARTICLE{Shi2024,
       author = {{Shi}, Dong Dong and {Wang}, Xin and {Zheng}, Xian Zhong and {Cai}, Zheng and {Fan}, Xiaohui and {Bian}, Fuyan and {Teplitz}, Harry I.},
        title = "{The Emergence of a Brightest Cluster Galaxy in a Protocluster Core at z = 2.24}",
      journal = {\apj},
         year = 2024,
        month = mar,
       volume = {963},
       number = {1},
          eid = {21},
        pages = {21},
          doi = {10.3847/1538-4357/ad17c3},
archivePrefix = {arXiv},
       eprint = {2303.09726},
 primaryClass = {astro-ph.GA},
       adsurl = {https://ui.adsabs.harvard.edu/abs/2024ApJ...963...21S}
}

@ARTICLE{Shi2021,
       author = {{Shi}, Dong Dong and {Cai}, Zheng and {Fan}, Xiaohui and {Zheng}, Xian Zhong and {Huang}, Yun-Hsin and {Xu}, Jiachuan},
        title = "{Spectroscopic Confirmation of Two Extremely Massive Protoclusters, BOSS1244 and BOSS1542, at z = 2.24}",
      journal = {\apj},
         year = 2021,
        month = jul,
       volume = {915},
       number = {1},
          eid = {32},
        pages = {32},
          doi = {10.3847/1538-4357/abfec0},
archivePrefix = {arXiv},
       eprint = {2105.02248},
 primaryClass = {astro-ph.GA},
       adsurl = {https://ui.adsabs.harvard.edu/abs/2021ApJ...915...32S}
}

@ARTICLE{Liu2023,
       author = {{Liu}, Shuang and {Zheng}, Xian Zhong and {Shi}, Dong Dong and {Cai}, Zheng and {Fan}, Xiaohui and {Wang}, Xin and {Yuan}, Qirong and {Xu}, Haiguang and {Pan}, Zhizheng and {Liu}, Wenhao and {Qin}, Jianbo and {Zhang}, Yuheng and {Wen}, Run},
        title = "{What boost galaxy mergers in two massive galaxy protoclusters at z = 2.24?}",
      journal = {\mnras},
         year = 2023,
        month = aug,
       volume = {523},
       number = {2},
        pages = {2422-2439},
          doi = {10.1093/mnras/stad1543},
archivePrefix = {arXiv},
       eprint = {2305.10932},
 primaryClass = {astro-ph.GA},
       adsurl = {https://ui.adsabs.harvard.edu/abs/2023MNRAS.523.2422L}
}

@ARTICLE{Peng2002,
       author = {{Peng}, Chien Y. and {Ho}, Luis C. and {Impey}, Chris D. and {Rix}, Hans-Walter},
        title = "{Detailed Structural Decomposition of Galaxy Images}",
      journal = {\aj},
         year = 2002,
        month = jul,
       volume = {124},
       number = {1},
        pages = {266-293},
          doi = {10.1086/340952},
archivePrefix = {arXiv},
       eprint = {astro-ph/0204182},
 primaryClass = {astro-ph},
       adsurl = {https://ui.adsabs.harvard.edu/abs/2002AJ....124..266P}
}

@ARTICLE{Peng2010,
       author = {{Peng}, Chien Y. and {Ho}, Luis C. and {Impey}, Chris D. and {Rix}, Hans-Walter},
        title = "{Detailed Decomposition of Galaxy Images. II. Beyond Axisymmetric Models}",
      journal = {\aj},
         year = 2010,
        month = jun,
       volume = {139},
       number = {6},
        pages = {2097-2129},
          doi = {10.1088/0004-6256/139/6/2097},
archivePrefix = {arXiv},
       eprint = {0912.0731},
 primaryClass = {astro-ph.CO},
       adsurl = {https://ui.adsabs.harvard.edu/abs/2010AJ....139.2097P}
}

@ARTICLE{Haussler2013,
       author = {{H{\"a}u{\ss}ler}, Boris and {Bamford}, Steven P. and {Vika}, Marina and {Rojas}, Alex L. and {Barden}, Marco and {Kelvin}, Lee S. and {Alpaslan}, Mehmet and {Robotham}, Aaron S.~G. and {Driver}, Simon P. and {Baldry}, I.~K. and {Brough}, Sarah and {Hopkins}, Andrew M. and {Liske}, Jochen and {Nichol}, Robert C. and {Popescu}, Cristina C. and {Tuffs}, Richard J.},
        title = "{MegaMorph - multiwavelength measurement of galaxy structure: complete S{\'e}rsic profile information from modern surveys}",
      journal = {\mnras},
         year = 2013,
        month = mar,
       volume = {430},
       number = {1},
        pages = {330-369},
          doi = {10.1093/mnras/sts633},
archivePrefix = {arXiv},
       eprint = {1212.3332},
 primaryClass = {astro-ph.CO},
       adsurl = {https://ui.adsabs.harvard.edu/abs/2013MNRAS.430..330H}
}

@ARTICLE{Kartaltepe2015,
       author = {{Kartaltepe}, Jeyhan S. and {Mozena}, Mark and {Kocevski}, Dale and {McIntosh}, Daniel H. and {Lotz}, Jennifer and {Bell}, Eric F. and {Faber}, Sandy and {Ferguson}, Harry and {Koo}, David and {Bassett}, Robert and {Bernyk}, Maksym and {Blancato}, Kirsten and {Bournaud}, Frederic and {Cassata}, Paolo and {Castellano}, Marco and {Cheung}, Edmond and {Conselice}, Christopher J. and {Croton}, Darren and {Dahlen}, Tomas and {de Mello}, Duilia F. and {DeGroot}, Laura and {Donley}, Jennifer and {Guedes}, Javiera and {Grogin}, Norman and {Hathi}, Nimish and {Hilton}, Matt and {Hollon}, Brett and {Koekemoer}, Anton and {Liu}, Nick and {Lucas}, Ray A. and {Martig}, Marie and {McGrath}, Elizabeth and {McPartland}, Conor and {Mobasher}, Bahram and {Morlock}, Alice and {O'Leary}, Erin and {Peth}, Mike and {Pforr}, Janine and {Pillepich}, Annalisa and {Rosario}, David and {Soto}, Emmaris and {Straughn}, Amber and {Telford}, Olivia and {Sunnquist}, Ben and {Trump}, Jonathan and {Weiner}, Benjamin and {Wuyts}, Stijn and {Inami}, Hanae and {Kassin}, Susan and {Lani}, Caterina and {Poole}, Gregory B. and {Rizer}, Zachary},
        title = "{CANDELS Visual Classifications: Scheme, Data Release, and First Results}",
      journal = {\apjs},
         year = 2015,
        month = nov,
       volume = {221},
       number = {1},
          eid = {11},
        pages = {11},
          doi = {10.1088/0067-0049/221/1/11},
archivePrefix = {arXiv},
       eprint = {1401.2455},
 primaryClass = {astro-ph.GA},
       adsurl = {https://ui.adsabs.harvard.edu/abs/2015ApJS..221...11K}
}

@ARTICLE{Kodra2023,
       author = {{Kodra}, Dritan and {Andrews}, Brett H. and {Newman}, Jeffrey A. and {Finkelstein}, Steven L. and {Fontana}, Adriano and {Hathi}, Nimish and {Salvato}, Mara and {Wiklind}, Tommy and {Wuyts}, Stijn and {Broussard}, Adam and {Chartab}, Nima and {Conselice}, Christopher and {Cooper}, M.~C. and {Dekel}, Avishai and {Dickinson}, Mark and {Ferguson}, Henry C. and {Gawiser}, Eric and {Grogin}, Norman A. and {Iyer}, Kartheik and {Kartaltepe}, Jeyhan and {Kassin}, Susan and {Koekemoer}, Anton M. and {Koo}, David C. and {Lucas}, Ray A. and {Mantha}, Kameswara Bharadwaj and {McIntosh}, Daniel H. and {Mobasher}, Bahram and {Pacifici}, Camilla and {P{\'e}rez-Gonz{\'a}lez}, Pablo G. and {Santini}, Paola},
        title = "{Optimized Photometric Redshifts for the Cosmic Assembly Near-infrared Deep Extragalactic Legacy Survey (CANDELS)}",
      journal = {\apj},
         year = 2023,
        month = jan,
       volume = {942},
       number = {1},
          eid = {36},
        pages = {36},
          doi = {10.3847/1538-4357/ac9f12},
archivePrefix = {arXiv},
       eprint = {2210.01140},
 primaryClass = {astro-ph.GA},
       adsurl = {https://ui.adsabs.harvard.edu/abs/2023ApJ...942...36K}
}

@ARTICLE{McConachie2022,
       author = {{McConachie}, Ian and {Wilson}, Gillian and {Forrest}, Ben and {Marsan}, Z. Cemile and {Muzzin}, Adam and {Cooper}, M.~C. and {Annunziatella}, Marianna and {Marchesini}, Danilo and {Chan}, Jeffrey C.~C. and {Gomez}, Percy and {Abdullah}, Mohamed H. and {Saracco}, Paolo and {Nantais}, Julie},
        title = "{Spectroscopic Confirmation of a Protocluster at z = 3.37 with a High Fraction of Quiescent Galaxies}",
      journal = {\apj},
         year = 2022,
        month = feb,
       volume = {926},
       number = {1},
          eid = {37},
        pages = {37},
          doi = {10.3847/1538-4357/ac2b9f},
archivePrefix = {arXiv},
       eprint = {2109.07696},
 primaryClass = {astro-ph.GA},
       adsurl = {https://ui.adsabs.harvard.edu/abs/2022ApJ...926...37M}
}

@ARTICLE{Naufal2024,
       author = {{Naufal}, Abdurrahman and {Koyama}, Yusei and {D'Eugenio}, Chiara and {Dannerbauer}, Helmut and {Shimakawa}, Rhythm and {P{\'e}rez-Mart{\'\i}nez}, Jose Manuel and {Kodama}, Tadayuki and {Zhang}, Yuheng and {Daikuhara}, Kazuki},
        title = "{Revealing the Quiescent Galaxy Population in the Spiderweb Protocluster at z = 2.16 with Deep HST/WFC3 Slitless Spectroscopy}",
      journal = {\apj},
         year = 2024,
        month = dec,
       volume = {977},
       number = {1},
          eid = {58},
        pages = {58},
          doi = {10.3847/1538-4357/ad8dcf},
archivePrefix = {arXiv},
       eprint = {2410.16643},
 primaryClass = {astro-ph.GA},
       adsurl = {https://ui.adsabs.harvard.edu/abs/2024ApJ...977...58N}
}

@ARTICLE{Forrest2024,
       author = {{Forrest}, Ben and {Lemaux}, Brian C. and {Shah}, Ekta A. and {Staab}, Priti and {Gal}, Roy R. and {Lubin}, Lori M. and {Cooper}, M.~C. and {Cucciati}, Olga and {Hung}, Denise and {McConachie}, Ian and {Muzzin}, Adam and {Wilson}, Gillian and {Bardelli}, Sandro and {Cassar{\`a}}, Letizia P. and {Chang}, Wenjun and {Giddings}, Finn and {Golden-Marx}, Emmet and {Hathi}, Nimish and {Urbano Stawinski}, Stephanie M. and {Zucca}, Elena},
        title = "{Environmental Effects on the Stellar Mass Function in a z {\ensuremath{\sim}} 3.3 Overdensity of Galaxies in the COSMOS Field}",
      journal = {\apj},
         year = 2024,
        month = aug,
       volume = {971},
       number = {2},
          eid = {169},
        pages = {169},
          doi = {10.3847/1538-4357/ad5e78},
archivePrefix = {arXiv},
       eprint = {2405.18491},
 primaryClass = {astro-ph.GA},
       adsurl = {https://ui.adsabs.harvard.edu/abs/2024ApJ...971..169F}
}

@ARTICLE{Tanaka2024,
       author = {{Tanaka}, Masayuki and {Onodera}, Masato and {Shimakawa}, Rhythm and {Ito}, Kei and {Kakimoto}, Takumi and {Kubo}, Mariko and {Morishita}, Takahiro and {Toft}, Sune and {Valentino}, Francesco and {Wu}, Po-Feng},
        title = "{A Protocluster of Massive Quiescent Galaxies at z = 4}",
      journal = {\apj},
         year = 2024,
        month = jul,
       volume = {970},
       number = {1},
          eid = {59},
        pages = {59},
          doi = {10.3847/1538-4357/ad5316},
archivePrefix = {arXiv},
       eprint = {2311.11569},
 primaryClass = {astro-ph.GA},
       adsurl = {https://ui.adsabs.harvard.edu/abs/2024ApJ...970...59T}
}

@ARTICLE{Kiyota2025,
       author = {{Kiyota}, Tomokazu and {Ando}, Makoto and {Tanaka}, Masayuki and {Finoguenov}, Alexis and {Ali}, Sadman Shariar and {Coupon}, Jean and {Desprez}, Guillaume and {Gwyn}, Stephen and {Sawicki}, Marcin and {Shimakawa}, Rhythm},
        title = "{Cluster Candidates with Massive Quiescent Galaxies at z {\ensuremath{\sim}} 2}",
      journal = {\apj},
         year = 2025,
        month = feb,
       volume = {980},
       number = {1},
          eid = {104},
        pages = {104},
          doi = {10.3847/1538-4357/ada5f4},
archivePrefix = {arXiv},
       eprint = {2406.02849},
 primaryClass = {astro-ph.GA},
       adsurl = {https://ui.adsabs.harvard.edu/abs/2025ApJ...980..104K}
}

@ARTICLE{Staab2024,
       author = {{Staab}, Priti and {Lemaux}, Brian C. and {Forrest}, Ben and {Shah}, Ekta and {Cucciati}, Olga and {Lubin}, Lori and {Gal}, Roy R. and {Hung}, Denise and {Shen}, Lu and {Giddings}, Finn and {Khusanova}, Yana and {Zamorani}, Giovanni and {Bardelli}, Sandro and {Cassara}, Letizia Pasqua and {Cassata}, Paolo and {Chiang}, Yi-Kuan and {Fudamoto}, Yoshinobu and {Fukushima}, Shuma and {Garilli}, Bianca and {Giavalisco}, Mauro and {Gruppioni}, Carlotta and {Guaita}, Lucia and {Gururajan}, Gayathri and {Hathi}, Nimish and {Kashino}, Daichi and {Scoville}, Nick and {Talia}, Margherita and {Vergani}, Daniela and {Zucca}, Elena},
        title = "{Protoclusters as drivers of stellar mass growth in the early Universe, a case study: Taralay - a massive protocluster at z   4.57}",
      journal = {\mnras},
         year = 2024,
        month = mar,
       volume = {528},
       number = {4},
        pages = {6934-6958},
          doi = {10.1093/mnras/stae301},
archivePrefix = {arXiv},
       eprint = {2312.11465},
 primaryClass = {astro-ph.GA},
       adsurl = {https://ui.adsabs.harvard.edu/abs/2024MNRAS.528.6934S}
}

@ARTICLE{Strazzullo2023,
       author = {{Strazzullo}, V. and {Pannella}, M. and {Mohr}, J.~J. and {Saro}, A. and {Ashby}, M.~L.~N. and {Bayliss}, M.~B. and {Canning}, R.~E.~A. and {Floyd}, B. and {Gonzalez}, A.~H. and {Khullar}, G. and {Kim}, K.~J. and {McDonald}, M. and {Reichardt}, C.~L. and {Sharon}, K. and {Somboonpanyakul}, T.},
        title = "{Galaxy populations in the most distant SPT-SZ clusters. II. Galaxy structural properties in massive clusters at 1.4 {\ensuremath{\lesssim}} z {\ensuremath{\lesssim}} 1.7}",
      journal = {\aap},
         year = 2023,
        month = jan,
       volume = {669},
          eid = {A131},
        pages = {A131},
          doi = {10.1051/0004-6361/202245268},
archivePrefix = {arXiv},
       eprint = {2212.06853},
 primaryClass = {astro-ph.GA},
       adsurl = {https://ui.adsabs.harvard.edu/abs/2023A&A...669A.131S}
}

@ARTICLE{Forrest2023,
       author = {{Forrest}, Ben and {Lemaux}, Brian C. and {Shah}, Ekta and {Staab}, Priti and {McConachie}, Ian and {Cucciati}, Olga and {Gal}, Roy R. and {Hung}, Denise and {Lubin}, Lori M. and {Cassar{\`a}}, Letizia P. and {Cassata}, Paolo and {Chang}, Wenjun and {Cooper}, M.~C. and {Decarli}, Roberto and {Gomez}, Percy and {Gururajan}, Gayathri and {Hathi}, Nimish and {Kashino}, Daichi and {Marchesini}, Danilo and {Marsan}, Z. Cemile and {McDonald}, Michael and {Muzzin}, Adam and {Shen}, Lu and {Urbano Stawinski}, Stephanie and {Talia}, Margherita and {Vergani}, Daniela and {Wilson}, Gillian and {Zamorani}, Giovanni},
        title = "{Elent{\'a}ri:a massive proto-supercluster at z {\ensuremath{\sim}} 3.3 in the COSMOS field}",
      journal = {\mnras},
         year = 2023,
        month = nov,
       volume = {526},
       number = {1},
        pages = {L56-L62},
          doi = {10.1093/mnrasl/slad114},
archivePrefix = {arXiv},
       eprint = {2307.15113},
 primaryClass = {astro-ph.GA},
       adsurl = {https://ui.adsabs.harvard.edu/abs/2023MNRAS.526L..56F}
}

@ARTICLE{Barden2012,
       author = {{Barden}, Marco and {H{\"a}u{\ss}ler}, Boris and {Peng}, Chien Y. and {McIntosh}, Daniel H. and {Guo}, Yicheng},
        title = "{GALAPAGOS: from pixels to parameters}",
      journal = {\mnras},
         year = 2012,
        month = may,
       volume = {422},
       number = {1},
        pages = {449-468},
          doi = {10.1111/j.1365-2966.2012.20619.x},
archivePrefix = {arXiv},
       eprint = {1203.1831},
 primaryClass = {astro-ph.IM},
       adsurl = {https://ui.adsabs.harvard.edu/abs/2012MNRAS.422..449B}
}

@ARTICLE{Peter2007,
       author = {{Peter}, Annika H.~G. and {Shapley}, Alice E. and {Law}, David R. and {Steidel}, Charles C. and {Erb}, Dawn K. and {Reddy}, Naveen A. and {Pettini}, Max},
        title = "{Morphologies of Galaxies in and around a Protocluster at z = 2.300}",
      journal = {\apj},
         year = 2007,
        month = oct,
       volume = {668},
       number = {1},
        pages = {23-44},
          doi = {10.1086/521184},
archivePrefix = {arXiv},
       eprint = {0706.2865},
 primaryClass = {astro-ph},
       adsurl = {https://ui.adsabs.harvard.edu/abs/2007ApJ...668...23P}
}

@ARTICLE{Liu2025,
       author = {{Liu}, Shuang and {Zheng}, Xian Zhong and {Gonzalez}, Valentino and {Yang}, Xiaohu and {Huang}, Jia-Sheng and {Shi}, Dong Dong and {Xu}, Haiguang and {Yuan}, Qirong and {Zhang}, Yuheng and {Wen}, Run and {Qiao}, Man and {Yang}, Chao and {Lyu}, Zongfei},
        title = "{A first measurement of galaxy merger rate increasing in dynamically colder protoclusters at cosmic noon}",
      journal = {\mnras},
         year = 2025,
        month = jan,
       volume = {536},
       number = {2},
        pages = {2000-2010},
          doi = {10.1093/mnras/stae2746},
archivePrefix = {arXiv},
       eprint = {2412.08336},
 primaryClass = {astro-ph.GA},
       adsurl = {https://ui.adsabs.harvard.edu/abs/2025MNRAS.536.2000L}
}

@ARTICLE{Grogin2011,
       author = {{Grogin}, Norman A. and {Kocevski}, Dale D. and {Faber}, S.~M. and {Ferguson}, Henry C. and {Koekemoer}, Anton M. and {Riess}, Adam G. and {Acquaviva}, Viviana and {Alexander}, David M. and {Almaini}, Omar and {Ashby}, Matthew L.~N. and {Barden}, Marco and {Bell}, Eric F. and {Bournaud}, Fr{\'e}d{\'e}ric and {Brown}, Thomas M. and {Caputi}, Karina I. and {Casertano}, Stefano and {Cassata}, Paolo and {Castellano}, Marco and {Challis}, Peter and {Chary}, Ranga-Ram and {Cheung}, Edmond and {Cirasuolo}, Michele and {Conselice}, Christopher J. and {Roshan Cooray}, Asantha and {Croton}, Darren J. and {Daddi}, Emanuele and {Dahlen}, Tomas and {Dav{\'e}}, Romeel and {de Mello}, Du{\'\i}lia F. and {Dekel}, Avishai and {Dickinson}, Mark and {Dolch}, Timothy and {Donley}, Jennifer L. and {Dunlop}, James S. and {Dutton}, Aaron A. and {Elbaz}, David and {Fazio}, Giovanni G. and {Filippenko}, Alexei V. and {Finkelstein}, Steven L. and {Fontana}, Adriano and {Gardner}, Jonathan P. and {Garnavich}, Peter M. and {Gawiser}, Eric and {Giavalisco}, Mauro and {Grazian}, Andrea and {Guo}, Yicheng and {Hathi}, Nimish P. and {H{\"a}ussler}, Boris and {Hopkins}, Philip F. and {Huang}, Jia-Sheng and {Huang}, Kuang-Han and {Jha}, Saurabh W. and {Kartaltepe}, Jeyhan S. and {Kirshner}, Robert P. and {Koo}, David C. and {Lai}, Kamson and {Lee}, Kyoung-Soo and {Li}, Weidong and {Lotz}, Jennifer M. and {Lucas}, Ray A. and {Madau}, Piero and {McCarthy}, Patrick J. and {McGrath}, Elizabeth J. and {McIntosh}, Daniel H. and {McLure}, Ross J. and {Mobasher}, Bahram and {Moustakas}, Leonidas A. and {Mozena}, Mark and {Nandra}, Kirpal and {Newman}, Jeffrey A. and {Niemi}, Sami-Matias and {Noeske}, Kai G. and {Papovich}, Casey J. and {Pentericci}, Laura and {Pope}, Alexandra and {Primack}, Joel R. and {Rajan}, Abhijith and {Ravindranath}, Swara and {Reddy}, Naveen A. and {Renzini}, Alvio and {Rix}, Hans-Walter and {Robaina}, Aday R. and {Rodney}, Steven A. and {Rosario}, David J. and {Rosati}, Piero and {Salimbeni}, Sara and {Scarlata}, Claudia and {Siana}, Brian and {Simard}, Luc and {Smidt}, Joseph and {Somerville}, Rachel S. and {Spinrad}, Hyron and {Straughn}, Amber N. and {Strolger}, Louis-Gregory and {Telford}, Olivia and {Teplitz}, Harry I. and {Trump}, Jonathan R. and {van der Wel}, Arjen and {Villforth}, Carolin and {Wechsler}, Risa H. and {Weiner}, Benjamin J. and {Wiklind}, Tommy and {Wild}, Vivienne and {Wilson}, Grant and {Wuyts}, Stijn and {Yan}, Hao-Jing and {Yun}, Min S.},
        title = "{CANDELS: The Cosmic Assembly Near-infrared Deep Extragalactic Legacy Survey}",
      journal = {\apjs},
         year = 2011,
        month = dec,
       volume = {197},
       number = {2},
          eid = {35},
        pages = {35},
          doi = {10.1088/0067-0049/197/2/35},
archivePrefix = {arXiv},
       eprint = {1105.3753},
 primaryClass = {astro-ph.CO},
       adsurl = {https://ui.adsabs.harvard.edu/abs/2011ApJS..197...35G}
}

@ARTICLE{Koekemoer2011,
       author = {{Koekemoer}, Anton M. and {Faber}, S.~M. and {Ferguson}, Henry C. and {Grogin}, Norman A. and {Kocevski}, Dale D. and {Koo}, David C. and {Lai}, Kamson and {Lotz}, Jennifer M. and {Lucas}, Ray A. and {McGrath}, Elizabeth J. and {Ogaz}, Sara and {Rajan}, Abhijith and {Riess}, Adam G. and {Rodney}, Steve A. and {Strolger}, Louis and {Casertano}, Stefano and {Castellano}, Marco and {Dahlen}, Tomas and {Dickinson}, Mark and {Dolch}, Timothy and {Fontana}, Adriano and {Giavalisco}, Mauro and {Grazian}, Andrea and {Guo}, Yicheng and {Hathi}, Nimish P. and {Huang}, Kuang-Han and {van der Wel}, Arjen and {Yan}, Hao-Jing and {Acquaviva}, Viviana and {Alexander}, David M. and {Almaini}, Omar and {Ashby}, Matthew L.~N. and {Barden}, Marco and {Bell}, Eric F. and {Bournaud}, Fr{\'e}d{\'e}ric and {Brown}, Thomas M. and {Caputi}, Karina I. and {Cassata}, Paolo and {Challis}, Peter J. and {Chary}, Ranga-Ram and {Cheung}, Edmond and {Cirasuolo}, Michele and {Conselice}, Christopher J. and {Roshan Cooray}, Asantha and {Croton}, Darren J. and {Daddi}, Emanuele and {Dav{\'e}}, Romeel and {de Mello}, Duilia F. and {de Ravel}, Loic and {Dekel}, Avishai and {Donley}, Jennifer L. and {Dunlop}, James S. and {Dutton}, Aaron A. and {Elbaz}, David and {Fazio}, Giovanni G. and {Filippenko}, Alexei V. and {Finkelstein}, Steven L. and {Frazer}, Chris and {Gardner}, Jonathan P. and {Garnavich}, Peter M. and {Gawiser}, Eric and {Gruetzbauch}, Ruth and {Hartley}, Will G. and {H{\"a}ussler}, Boris and {Herrington}, Jessica and {Hopkins}, Philip F. and {Huang}, Jia-Sheng and {Jha}, Saurabh W. and {Johnson}, Andrew and {Kartaltepe}, Jeyhan S. and {Khostovan}, Ali A. and {Kirshner}, Robert P. and {Lani}, Caterina and {Lee}, Kyoung-Soo and {Li}, Weidong and {Madau}, Piero and {McCarthy}, Patrick J. and {McIntosh}, Daniel H. and {McLure}, Ross J. and {McPartland}, Conor and {Mobasher}, Bahram and {Moreira}, Heidi and {Mortlock}, Alice and {Moustakas}, Leonidas A. and {Mozena}, Mark and {Nandra}, Kirpal and {Newman}, Jeffrey A. and {Nielsen}, Jennifer L. and {Niemi}, Sami and {Noeske}, Kai G. and {Papovich}, Casey J. and {Pentericci}, Laura and {Pope}, Alexandra and {Primack}, Joel R. and {Ravindranath}, Swara and {Reddy}, Naveen A. and {Renzini}, Alvio and {Rix}, Hans-Walter and {Robaina}, Aday R. and {Rosario}, David J. and {Rosati}, Piero and {Salimbeni}, Sara and {Scarlata}, Claudia and {Siana}, Brian and {Simard}, Luc and {Smidt}, Joseph and {Snyder}, Diana and {Somerville}, Rachel S. and {Spinrad}, Hyron and {Straughn}, Amber N. and {Telford}, Olivia and {Teplitz}, Harry I. and {Trump}, Jonathan R. and {Vargas}, Carlos and {Villforth}, Carolin and {Wagner}, Cory R. and {Wandro}, Pat and {Wechsler}, Risa H. and {Weiner}, Benjamin J. and {Wiklind}, Tommy and {Wild}, Vivienne and {Wilson}, Grant and {Wuyts}, Stijn and {Yun}, Min S.},
        title = "{CANDELS: The Cosmic Assembly Near-infrared Deep Extragalactic Legacy Survey{\textemdash}The Hubble Space Telescope Observations, Imaging Data Products, and Mosaics}",
      journal = {\apjs},
         year = 2011,
        month = dec,
       volume = {197},
       number = {2},
          eid = {36},
        pages = {36},
          doi = {10.1088/0067-0049/197/2/36},
archivePrefix = {arXiv},
       eprint = {1105.3754},
 primaryClass = {astro-ph.CO},
       adsurl = {https://ui.adsabs.harvard.edu/abs/2011ApJS..197...36K}
}

@ARTICLE{Darvish2015,
       author = {{Darvish}, Behnam and {Mobasher}, Bahram and {Sobral}, David and {Scoville}, Nicholas and {Aragon-Calvo}, Miguel},
        title = "{A Comparative Study of Density Field Estimation for Galaxies: New Insights into the Evolution of Galaxies with Environment in COSMOS out to z{\ensuremath{\sim}}3}",
      journal = {\apj},
         year = 2015,
        month = jun,
       volume = {805},
       number = {2},
          eid = {121},
        pages = {121},
          doi = {10.1088/0004-637X/805/2/121},
archivePrefix = {arXiv},
       eprint = {1503.07879},
 primaryClass = {astro-ph.GA},
       adsurl = {https://ui.adsabs.harvard.edu/abs/2015ApJ...805..121D}
}

@ARTICLE{Darvish2016,
       author = {{Darvish}, Behnam and {Mobasher}, Bahram and {Sobral}, David and {Rettura}, Alessandro and {Scoville}, Nick and {Faisst}, Andreas and {Capak}, Peter},
        title = "{The Effects of the Local Environment and Stellar Mass on Galaxy Quenching to z {\ensuremath{\sim}} 3}",
      journal = {\apj},
         year = 2016,
        month = jul,
       volume = {825},
       number = {2},
          eid = {113},
        pages = {113},
          doi = {10.3847/0004-637X/825/2/113},
archivePrefix = {arXiv},
       eprint = {1605.03182},
 primaryClass = {astro-ph.GA},
       adsurl = {https://ui.adsabs.harvard.edu/abs/2016ApJ...825..113D}
}

@ARTICLE{Perez-Martin2023,
       author = {{P{\'e}rez-Mart{\'\i}nez}, J.~M. and {Dannerbauer}, H. and {Kodama}, T. and {Koyama}, Y. and {Shimakawa}, R. and {Suzuki}, T.~L. and {Calvi}, R. and {Chen}, Z. and {Daikuhara}, K. and {Hatch}, N.~A. and {Laza-Ramos}, A. and {Sobral}, D. and {Stott}, J.~P. and {Tanaka}, I.},
        title = "{Signs of environmental effects on star-forming galaxies in the Spiderweb protocluster at z = 2.16}",
      journal = {\mnras},
         year = 2023,
        month = jan,
       volume = {518},
       number = {2},
        pages = {1707-1734},
          doi = {10.1093/mnras/stac2784},
archivePrefix = {arXiv},
       eprint = {2209.13069},
 primaryClass = {astro-ph.GA},
       adsurl = {https://ui.adsabs.harvard.edu/abs/2023MNRAS.518.1707P}
}

@ARTICLE{Thongkham2024,
       author = {{Thongkham}, Khunanon and {Gonzalez}, Anthony H. and {Brodwin}, Mark and {Trudeau}, Ariane and {Eisenhardt}, Peter and {Stanford}, S.~A. and {Moravec}, Emily and {Connor}, Thomas and {Stern}, Daniel and {Spivey}, Ryan and {Garcia}, Karolina},
        title = "{The Massive and Distant Clusters of WISE Survey 2: Second Data Release}",
      journal = {\apj},
         year = 2024,
        month = dec,
       volume = {976},
       number = {2},
          eid = {186},
        pages = {186},
          doi = {10.3847/1538-4357/ad888c},
archivePrefix = {arXiv},
       eprint = {2410.15303},
 primaryClass = {astro-ph.CO},
       adsurl = {https://ui.adsabs.harvard.edu/abs/2024ApJ...976..186T}
}

@ARTICLE{Hung2025,
       author = {{Hung}, Denise and {Lemaux}, Brian C. and {Cucciati}, Olga and {Forrest}, Ben and {Shah}, Ekta A. and {Gal}, Roy R. and {Giddings}, Finn and {Sikorski}, Derek and {Golden-Marx}, Emmet and {Lubin}, Lori M. and {Hathi}, Nimish and {Zamorani}, Giovanni and {Shen}, Lu and {Bardelli}, Sandro and {Cassar{\`a}}, Letizia P. and {De Lucia}, Gabriella and {Fontanot}, Fabio and {Garilli}, Bianca and {Guaita}, Lucia and {Hirschmann}, Michaela Monika and {Lee}, Kyoung-Soo and {Newman}, Andrew B. and {Ramakrishnan}, Vandana and {Vergani}, Daniela and {Xie}, Lizhi and {Zucca}, Elena},
        title = "{Discovering Large-scale Structure at 2 < z < 5 in the C3VO Survey}",
      journal = {\apj},
         year = 2025,
        month = feb,
       volume = {980},
       number = {1},
          eid = {155},
        pages = {155},
          doi = {10.3847/1538-4357/ada616},
archivePrefix = {arXiv},
       eprint = {2410.00237},
 primaryClass = {astro-ph.GA},
       adsurl = {https://ui.adsabs.harvard.edu/abs/2025ApJ...980..155H}
}

@ARTICLE{Shah2024,
       author = {{Shah}, Ekta A. and {Lemaux}, Brian and {Forrest}, Benjamin and {Cucciati}, Olga and {Hung}, Denise and {Staab}, Priti and {Hathi}, Nimish and {Lubin}, Lori and {Gal}, Roy R. and {Shen}, Lu and {Zamorani}, Giovanni and {Giddings}, Finn and {Bardelli}, Sandro and {Pasqua Cassara}, Letizia and {Cassata}, Paolo and {Contini}, Thierry and {Golden-Marx}, Emmet and {Guaita}, Lucia and {Gururajan}, Gayathri and {Koekemoer}, Anton M. and {McLeod}, Derek and {Tasca}, Lidia A.~M. and {Tresse}, Laurence and {Vergani}, Daniela and {Zucca}, Elena},
        title = "{Identification and characterization of six spectroscopically confirmed massive protostructures at 2.5 < z < 4.5}",
      journal = {\mnras},
         year = 2024,
        month = apr,
       volume = {529},
       number = {2},
        pages = {873-882},
          doi = {10.1093/mnras/stae519},
archivePrefix = {arXiv},
       eprint = {2312.04634},
 primaryClass = {astro-ph.GA},
       adsurl = {https://ui.adsabs.harvard.edu/abs/2024MNRAS.529..873S}
}

@ARTICLE{DiMascolo2023,
       author = {{Di Mascolo}, Luca and {Saro}, Alexandro and {Mroczkowski}, Tony and {Borgani}, Stefano and {Churazov}, Eugene and {Rasia}, Elena and {Tozzi}, Paolo and {Dannerbauer}, Helmut and {Basu}, Kaustuv and {Carilli}, Christopher L. and {Ginolfi}, Michele and {Miley}, George and {Nonino}, Mario and {Pannella}, Maurilio and {Pentericci}, Laura and {Rizzo}, Francesca},
        title = "{Forming intracluster gas in a galaxy protocluster at a redshift of 2.16}",
      journal = {\nat},
         year = 2023,
        month = mar,
       volume = {615},
       number = {7954},
        pages = {809-812},
          doi = {10.1038/s41586-023-05761-x},
archivePrefix = {arXiv},
       eprint = {2303.16226},
 primaryClass = {astro-ph.CO},
       adsurl = {https://ui.adsabs.harvard.edu/abs/2023Natur.615..809D}
}

@ARTICLE{Zhang2022,
       author = {{Zhang}, Yuheng and {Zheng}, Xian Zhong and {Shi}, Dong Dong and {Gao}, Yu and {Dannerbauer}, Helmut and {An}, Fang Xia and {Shu}, Xinwen and {Gao}, Zhen-Kai and {Wang}, Wei-Hao and {Wang}, Xin and {Cai}, Zheng and {Fan}, Xiaohui and {Fang}, Min and {Pan}, Zhizheng and {Liu}, Wenhao and {Tan}, Qinghua and {Qin}, Jianbo and {Ren}, Jian and {Qiao}, Man and {Wen}, Run and {Liu}, Shuang},
        title = "{Submillimetre galaxies in two massive protoclusters at z = 2.24: witnessing the enrichment of extreme starbursts in the outskirts of HAE density peaks}",
      journal = {\mnras},
         year = 2022,
        month = jun,
       volume = {512},
       number = {4},
        pages = {4893-4908},
          doi = {10.1093/mnras/stac824},
archivePrefix = {arXiv},
       eprint = {2203.09260},
 primaryClass = {astro-ph.GA},
       adsurl = {https://ui.adsabs.harvard.edu/abs/2022MNRAS.512.4893Z}
}

@ARTICLE{Watson2019,
       author = {{Watson}, Courtney and {Tran}, Kim-Vy and {Tomczak}, Adam and {Alcorn}, Leo and {Salazar}, Irene V. and {Gupta}, Anshu and {Momcheva}, Ivelina and {Papovich}, Casey and {van Dokkum}, Pieter and {Brammer}, Gabriel and {Lotz}, Jennifer and {Willmer}, Christopher N.~A.},
        title = "{Galaxy Merger Fractions in Two Clusters at z\textbackslashsim 2 Using the Hubble Space Telescope}",
      journal = {\apj},
         year = 2019,
        month = mar,
       volume = {874},
       number = {1},
          eid = {63},
        pages = {63},
          doi = {10.3847/1538-4357/ab06ef},
archivePrefix = {arXiv},
       eprint = {1902.07225},
 primaryClass = {astro-ph.GA},
       adsurl = {https://ui.adsabs.harvard.edu/abs/2019ApJ...874...63W}
}

@ARTICLE{DeGraaf2025,
       author = {{de Graaff}, Anna and {Setton}, David J. and {Brammer}, Gabriel and {Cutler}, Sam and {Suess}, Katherine A. and {Labb{\'e}}, Ivo and {Leja}, Joel and {Weibel}, Andrea and {Maseda}, Michael V. and {Whitaker}, Katherine E. and {Bezanson}, Rachel and {Boogaard}, Leindert A. and {Cleri}, Nikko J. and {De Lucia}, Gabriella and {Franx}, Marijn and {Greene}, Jenny E. and {Hirschmann}, Michaela and {Matthee}, Jorryt and {McConachie}, Ian and {Naidu}, Rohan P. and {Oesch}, Pascal A. and {Price}, Sedona H. and {Rix}, Hans-Walter and {Valentino}, Francesco and {Wang}, Bingjie and {Williams}, Christina C.},
        title = "{Efficient formation of a massive quiescent galaxy at redshift 4.9}",
      journal = {Nature Astronomy},
         year = 2025,
        month = feb,
       volume = {9},
        pages = {280-292},
          doi = {10.1038/s41550-024-02424-3},
       adsurl = {https://ui.adsabs.harvard.edu/abs/2025NatAs...9..280D}
}

@ARTICLE{Nanayakkara2025,
       author = {{Nanayakkara}, Themiya and {Glazebrook}, Karl and {Schreiber}, Corentin and {Chittenden}, Harry and {Brammer}, Gabriel and {Esdaile}, James and {Jacobs}, Colin and {Kacprzak}, Glenn G. and {Kawinwanichakij}, Lalitwadee and {Kimmig}, Lucas C. and {Labbe}, Ivo and {Lagos}, Claudia and {Marchesini}, Danilo and {Mart{\`\i}nez-Mar{\`\i}n}, M. and {Marsan}, Z. Cemile and {Oesch}, Pascal A. and {Papovich}, Casey and {Remus}, Rhea-Silvia and {Tran}, Kim-Vy H.},
        title = "{The Formation Histories of Massive and Quiescent Galaxies in the 3 < z < 4.5 Universe}",
      journal = {\apj},
         year = 2025,
        month = mar,
       volume = {981},
       number = {1},
          eid = {78},
        pages = {78},
          doi = {10.3847/1538-4357/ada6ac},
archivePrefix = {arXiv},
       eprint = {2410.02076},
 primaryClass = {astro-ph.GA},
       adsurl = {https://ui.adsabs.harvard.edu/abs/2025ApJ...981...78N}
}

@ARTICLE{Carnall2024,
       author = {{Carnall}, A.~C. and {Cullen}, F. and {McLure}, R.~J. and {McLeod}, D.~J. and {Begley}, R. and {Donnan}, C.~T. and {Dunlop}, J.~S. and {Shapley}, A.~E. and {Rowlands}, K. and {Almaini}, O. and {Arellano-C{\'o}rdova}, K.~Z. and {Barrufet}, L. and {Cimatti}, A. and {Ellis}, R.~S. and {Grogin}, N.~A. and {Hamadouche}, M.~L. and {Illingworth}, G.~D. and {Koekemoer}, A.~M. and {Leung}, H. -H. and {Lovell}, C.~C. and {P{\'e}rez-Gonz{\'a}lez}, P.~G. and {Santini}, P. and {Stanton}, T.~M. and {Wild}, V.},
        title = "{The JWST EXCELS survey: too much, too young, too fast? Ultra-massive quiescent galaxies at 3 < z < 5}",
      journal = {\mnras},
         year = 2024,
        month = oct,
       volume = {534},
       number = {1},
        pages = {325-348},
          doi = {10.1093/mnras/stae2092},
archivePrefix = {arXiv},
       eprint = {2405.02242},
 primaryClass = {astro-ph.GA},
       adsurl = {https://ui.adsabs.harvard.edu/abs/2024MNRAS.534..325C}
}

@ARTICLE{Strazzullo2013,
       author = {{Strazzullo}, V. and {Gobat}, R. and {Daddi}, E. and {Onodera}, M. and {Carollo}, M. and {Dickinson}, M. and {Renzini}, A. and {Arimoto}, N. and {Cimatti}, A. and {Finoguenov}, A. and {Chary}, R. -R.},
        title = "{Galaxy Evolution in Overdense Environments at High Redshift: Passive Early-type Galaxies in a Cluster at z \raisebox{-0.5ex}\textasciitilde 2}",
      journal = {\apj},
         year = 2013,
        month = aug,
       volume = {772},
       number = {2},
          eid = {118},
        pages = {118},
          doi = {10.1088/0004-637X/772/2/118},
archivePrefix = {arXiv},
       eprint = {1305.3577},
 primaryClass = {astro-ph.CO},
       adsurl = {https://ui.adsabs.harvard.edu/abs/2013ApJ...772..118S}
}

@ARTICLE{Miley2006,
       author = {{Miley}, George K. and {Overzier}, Roderik A. and {Zirm}, Andrew W. and {Ford}, Holland C. and {Kurk}, Jaron and {Pentericci}, Laura and {Blakeslee}, John P. and {Franx}, Marijn and {Illingworth}, Garth D. and {Postman}, Marc and {Rosati}, Piero and {R{\"o}ttgering}, Huub J.~A. and {Venemans}, Bram P. and {Helder}, Eveline},
        title = "{The Spiderweb Galaxy: A Forming Massive Cluster Galaxy at z \raisebox{-0.5ex}\textasciitilde 2}",
      journal = {\apjl},
         year = 2006,
        month = oct,
       volume = {650},
       number = {1},
        pages = {L29-L32},
          doi = {10.1086/508534},
archivePrefix = {arXiv},
       eprint = {astro-ph/0610909},
 primaryClass = {astro-ph},
       adsurl = {https://ui.adsabs.harvard.edu/abs/2006ApJ...650L..29M}
}

@ARTICLE{Vulcani+2023,
       author = {{Vulcani}, Benedetta and {Poggianti}, Bianca M. and {Gullieuszik}, Marco and {Moretti}, Alessia and {Fritz}, Jacopo and {Bettoni}, Daniela and {Facciolli}, Beatrice and {Fasano}, Giovanni and {Omizzolo}, Alessandro},
        title = "{Clustercentric Distance or Local Density? It Depends on Galaxy Morphology}",
      journal = {\apj},
         year = 2023,
        month = jun,
       volume = {949},
       number = {2},
          eid = {73},
        pages = {73},
          doi = {10.3847/1538-4357/acc5e2},
archivePrefix = {arXiv},
       eprint = {2302.02376},
 primaryClass = {astro-ph.GA},
       adsurl = {https://ui.adsabs.harvard.edu/abs/2023ApJ...949...73V}
}

@ARTICLE{Chan2021,
       author = {{Chan}, Jeffrey C.~C. and {Wilson}, Gillian and {Balogh}, Michael and {Rudnick}, Gregory and {van der Burg}, Remco F.~J. and {Muzzin}, Adam and {Webb}, Kristi A. and {Biviano}, Andrea and {Cerulo}, Pierluigi and {Cooper}, M.~C. and {De Lucia}, Gabriella and {Demarco}, Ricardo and {Forrest}, Ben and {Jablonka}, Pascale and {Lidman}, Chris and {McGee}, Sean L. and {Nantais}, Julie and {Old}, Lyndsay and {Pintos-Castro}, Irene and {Poggianti}, Bianca and {Reeves}, Andrew M.~M. and {Vulcani}, Benedetta and {Yee}, Howard K.~C. and {Zaritsky}, Dennis},
        title = "{The GOGREEN Survey: Evidence of an Excess of Quiescent Disks in Clusters at 1.0}",
      journal = {\apj},
         year = 2021,
        month = oct,
       volume = {920},
       number = {1},
          eid = {32},
        pages = {32},
          doi = {10.3847/1538-4357/ac1117},
archivePrefix = {arXiv},
       eprint = {2107.03403},
 primaryClass = {astro-ph.GA},
       adsurl = {https://ui.adsabs.harvard.edu/abs/2021ApJ...920...32C}
}

@ARTICLE{Treu2023,
       author = {{Treu}, T. and {Calabr{\`o}}, A. and {Castellano}, M. and {Leethochawalit}, N. and {Merlin}, E. and {Fontana}, A. and {Yang}, L. and {Morishita}, T. and {Trenti}, M. and {Dressler}, A. and {Mason}, C. and {Paris}, D. and {Pentericci}, L. and {Roberts-Borsani}, G. and {Vulcani}, B. and {Boyett}, K. and {Bradac}, M. and {Glazebrook}, K. and {Jones}, T. and {Marchesini}, D. and {Mascia}, S. and {Nanayakkara}, T. and {Santini}, P. and {Strait}, V. and {Vanzella}, E. and {Wang}, X.},
        title = "{Early Results From GLASS-JWST. XII. The Morphology of Galaxies at the Epoch of Reionization}",
      journal = {\apjl},
         year = 2023,
        month = jan,
       volume = {942},
       number = {2},
          eid = {L28},
        pages = {L28},
          doi = {10.3847/2041-8213/ac9283},
archivePrefix = {arXiv},
       eprint = {2207.13527},
 primaryClass = {astro-ph.GA},
       adsurl = {https://ui.adsabs.harvard.edu/abs/2023ApJ...942L..28T}
}

@ARTICLE{Wang2022,
       author = {{Wang}, Xin and {Li}, Zihao and {Cai}, Zheng and {Shi}, Dong Dong and {Fan}, Xiaohui and {Zheng}, Xian Zhong and {Bian}, Fuyan and {Teplitz}, Harry I. and {Alavi}, Anahita and {Colbert}, James and {Henry}, Alaina L. and {Malkan}, Matthew A.},
        title = "{The Mass-Metallicity Relation at Cosmic Noon in Overdense Environments: First Results from the MAMMOTH-Grism HST Slitless Spectroscopic Survey}",
      journal = {\apj},
         year = 2022,
        month = feb,
       volume = {926},
       number = {1},
          eid = {70},
        pages = {70},
          doi = {10.3847/1538-4357/ac3974},
archivePrefix = {arXiv},
       eprint = {2108.06373},
 primaryClass = {astro-ph.GA},
       adsurl = {https://ui.adsabs.harvard.edu/abs/2022ApJ...926...70W}
}

@ARTICLE{Delahaye2017,
       author = {{Delahaye}, A.~G. and {Webb}, T.~M.~A. and {Nantais}, J. and {DeGroot}, A. and {Wilson}, G. and {Muzzin}, A. and {Yee}, H.~K.~C. and {Foltz}, R. and {Noble}, A.~G. and {Demarco}, R. and {Tudorica}, A. and {Cooper}, M.~C. and {Lidman}, C. and {Perlmutter}, S. and {Hayden}, B. and {Boone}, K. and {Surace}, J.},
        title = "{Galaxy Merger Candidates in High-redshift Cluster Environments}",
      journal = {\apj},
         year = 2017,
        month = jul,
       volume = {843},
       number = {2},
          eid = {126},
        pages = {126},
          doi = {10.3847/1538-4357/aa756a},
archivePrefix = {arXiv},
       eprint = {1705.10849},
 primaryClass = {astro-ph.GA},
       adsurl = {https://ui.adsabs.harvard.edu/abs/2017ApJ...843..126D}
}

@ARTICLE{Monson2021,
       author = {{Monson}, Erik B. and {Lehmer}, Bret D. and {Doore}, Keith and {Eufrasio}, Rafael T. and {Bonine}, Brett and {Alexander}, David M. and {Harrison}, Chris M. and {Kubo}, Mariko and {Mantha}, Kameswara B. and {Saez}, Cristian and {Straughn}, Amber and {Umehata}, Hideki},
        title = "{On the Nature of AGN and Star Formation Enhancement in the z = 3.1 SSA22 Protocluster: The HST WFC3 IR View}",
      journal = {\apj},
         year = 2021,
        month = sep,
       volume = {919},
       number = {1},
          eid = {51},
        pages = {51},
          doi = {10.3847/1538-4357/ac0f84},
archivePrefix = {arXiv},
       eprint = {2106.13846},
 primaryClass = {astro-ph.GA},
       adsurl = {https://ui.adsabs.harvard.edu/abs/2021ApJ...919...51M}
}

@ARTICLE{Merlin2019,
       author = {{Merlin}, E. and {Fortuni}, F. and {Torelli}, M. and {Santini}, P. and {Castellano}, M. and {Fontana}, A. and {Grazian}, A. and {Pentericci}, L. and {Pilo}, S. and {Schmidt}, K.~B.},
        title = "{Red and dead CANDELS: massive passive galaxies at the dawn of the Universe}",
      journal = {\mnras},
         year = 2019,
        month = dec,
       volume = {490},
       number = {3},
        pages = {3309-3328},
          doi = {10.1093/mnras/stz2615},
archivePrefix = {arXiv},
       eprint = {1909.07996},
 primaryClass = {astro-ph.GA},
       adsurl = {https://ui.adsabs.harvard.edu/abs/2019MNRAS.490.3309M}
}

@ARTICLE{Lee2012,
       author = {{Lee}, Janice C. and {Ly}, Chun and {Spitler}, Lee and {Labb{\'e}}, Ivo and {Salim}, Samir and {Persson}, S. Eric and {Ouchi}, Masami and {Dale}, Daniel A. and {Monson}, Andy and {Murphy}, David},
        title = "{A Dual-Narrowband Survey for H{\ensuremath{\alpha}} Emitters at Redshift of 2.2: Demonstration of the Technique and Constraints on the H{\ensuremath{\alpha}} Luminosity Function}",
      journal = {\pasp},
         year = 2012,
        month = jul,
       volume = {124},
       number = {917},
        pages = {782},
          doi = {10.1086/666528},
archivePrefix = {arXiv},
       eprint = {1205.0017},
 primaryClass = {astro-ph.CO},
       adsurl = {https://ui.adsabs.harvard.edu/abs/2012PASP..124..782L}
}

@ARTICLE{Cassata2015,
       author = {{Cassata}, P. and {Tasca}, L.~A.~M. and {Le F{\`e}vre}, O. and {Lemaux}, B.~C. and {Garilli}, B. and {Le Brun}, V. and {Maccagni}, D. and {Pentericci}, L. and {Thomas}, R. and {Vanzella}, E. and {Zamorani}, G. and {Zucca}, E. and {Amorin}, R. and {Bardelli}, S. and {Capak}, P. and {Cassar{\`a}}, L.~P. and {Castellano}, M. and {Cimatti}, A. and {Cuby}, J.~G. and {Cucciati}, O. and {de la Torre}, S. and {Durkalec}, A. and {Fontana}, A. and {Giavalisco}, M. and {Grazian}, A. and {Hathi}, N.~P. and {Ilbert}, O. and {Moreau}, C. and {Paltani}, S. and {Ribeiro}, B. and {Salvato}, M. and {Schaerer}, D. and {Scodeggio}, M. and {Sommariva}, V. and {Talia}, M. and {Taniguchi}, Y. and {Tresse}, L. and {Vergani}, D. and {Wang}, P.~W. and {Charlot}, S. and {Contini}, T. and {Fotopoulou}, S. and {Koekemoer}, A.~M. and {L{\'o}pez-Sanjuan}, C. and {Mellier}, Y. and {Scoville}, N.},
        title = "{The VIMOS Ultra-Deep Survey (VUDS): fast increase in the fraction of strong Lyman-{\ensuremath{\alpha}} emitters from z = 2 to z = 6}",
      journal = {\aap},
         year = 2015,
        month = jan,
       volume = {573},
          eid = {A24},
        pages = {A24},
          doi = {10.1051/0004-6361/201423824},
archivePrefix = {arXiv},
       eprint = {1403.3693},
 primaryClass = {astro-ph.GA},
       adsurl = {https://ui.adsabs.harvard.edu/abs/2015A&A...573A..24C}
}

@ARTICLE{Oteo2015,
       author = {{Oteo}, I. and {Sobral}, D. and {Ivison}, R.~J. and {Smail}, I. and {Best}, P.~N. and {Cepa}, J. and {P{\'e}rez-Garc{\'\i}a}, A.~M.},
        title = "{On the nature of H{\ensuremath{\alpha}} emitters at z {\ensuremath{\sim}} 2 from the HiZELS survey: physical properties, Ly{\ensuremath{\alpha}} escape fraction and main sequence}",
      journal = {\mnras},
         year = 2015,
        month = sep,
       volume = {452},
       number = {2},
        pages = {2018-2033},
          doi = {10.1093/mnras/stv1284},
archivePrefix = {arXiv},
       eprint = {1506.02670},
 primaryClass = {astro-ph.GA},
       adsurl = {https://ui.adsabs.harvard.edu/abs/2015MNRAS.452.2018O}
}

@ARTICLE{Giddings2025,
       author = {{Giddings}, F. and {Lemaux}, B.~C. and {Forrest}, B. and {Shen}, L. and {Sikorski}, D. and {Gal}, R. and {Cucciati}, O. and {Golden-Marx}, E. and {Ronayne}, W. Hu. K. and {Shah}, E. and {Amor{\'\i}n}, R.~O. and {Bardelli}, S. and {Baxter}, D.~C. and {Cassar{\`a}}, L.~P. and {De Lucia}, G. and {Fontanot}, F. and {Gururajan}, G. and {Hathi}, N. and {Hirschmann}, M. and {Hung}, D. and {Lubin}, L. and {Sanders}, D.~B. and {Vergani}, D. and {Xie}, L. and {Zucca}, E.},
        title = "{The HST-Hyperion Survey: Companion Fraction and Overdensity in a z \raisebox{-0.5ex}\textasciitilde 2.5 Proto-supercluster}",
      journal = {arXiv e-prints},
         year = 2025,
        month = mar,
          eid = {arXiv:2503.04913},
        pages = {arXiv:2503.04913},
          doi = {10.48550/arXiv.2503.04913},
archivePrefix = {arXiv},
       eprint = {2503.04913},
 primaryClass = {astro-ph.GA},
       adsurl = {https://ui.adsabs.harvard.edu/abs/2025arXiv250304913G}
}

@ARTICLE{Toshikawa2020,
       author = {{Toshikawa}, Jun and {Malkan}, Matthew A. and {Kashikawa}, Nobunari and {Overzier}, Roderik and {Uchiyama}, Hisakazu and {Ota}, Kazuaki and {Ishikawa}, Shogo and {Ito}, Kei},
        title = "{Discovery of Protoclusters at z {\ensuremath{\sim}} 3.7 and 4.9: Embedded in Primordial Superclusters}",
      journal = {\apj},
         year = 2020,
        month = jan,
       volume = {888},
       number = {2},
          eid = {89},
        pages = {89},
          doi = {10.3847/1538-4357/ab5e85},
archivePrefix = {arXiv},
       eprint = {1912.01625},
 primaryClass = {astro-ph.GA},
       adsurl = {https://ui.adsabs.harvard.edu/abs/2020ApJ...888...89T}
}

@INPROCEEDINGS{Toomre1977,
       author = {{Toomre}, Alar},
        title = "{Mergers and Some Consequences}",
    booktitle = {Evolution of Galaxies and Stellar Populations},
         year = 1977,
       editor = {{Tinsley}, Beatrice M. and {Larson}, D. Campbell, Richard B. Gehret},
        month = jan,
        pages = {401},
       adsurl = {https://ui.adsabs.harvard.edu/abs/1977egsp.conf..401T}
}

@ARTICLE{Naab2006,
       author = {{Naab}, Thorsten and {Jesseit}, Roland and {Burkert}, Andreas},
        title = "{The influence of gas on the structure of merger remnants}",
      journal = {\mnras},
         year = 2006,
        month = oct,
       volume = {372},
       number = {2},
        pages = {839-852},
          doi = {10.1111/j.1365-2966.2006.10902.x},
archivePrefix = {arXiv},
       eprint = {astro-ph/0605155},
 primaryClass = {astro-ph},
       adsurl = {https://ui.adsabs.harvard.edu/abs/2006MNRAS.372..839N}
}

@ARTICLE{Rodriguez-Gomez2017,
       author = {{Rodriguez-Gomez}, Vicente and {Sales}, Laura V. and {Genel}, Shy and {Pillepich}, Annalisa and {Zjupa}, Jolanta and {Nelson}, Dylan and {Griffen}, Brendan and {Torrey}, Paul and {Snyder}, Gregory F. and {Vogelsberger}, Mark and {Springel}, Volker and {Ma}, Chung-Pei and {Hernquist}, Lars},
        title = "{The role of mergers and halo spin in shaping galaxy morphology}",
      journal = {\mnras},
         year = 2017,
        month = may,
       volume = {467},
       number = {3},
        pages = {3083-3098},
          doi = {10.1093/mnras/stx305},
archivePrefix = {arXiv},
       eprint = {1609.09498},
 primaryClass = {astro-ph.GA},
       adsurl = {https://ui.adsabs.harvard.edu/abs/2017MNRAS.467.3083R}
}

@ARTICLE{McConachie2025,
       author = {{McConachie}, Ian and {Wilson}, Gillian and {Forrest}, Ben and {Marsan}, Z. Cemile and {Muzzin}, Adam and {Cooper}, M.~C. and {Annunziatella}, Marianna and {Marchesini}, Danilo and {Gomez}, Percy and {Chang}, Wenjun and {Urbano Stawinski}, Stephanie M. and {McDonald}, Michael and {Webb}, Tracy and {Noble}, Allison and {Lemaux}, Brian C. and {Shah}, Ekta A. and {Staab}, Priti and {Lubin}, Lori M. and {Gal}, Roy R.},
        title = "{MAGAZ3NE: Evidence for Galactic Conformity in z {\ensuremath{\gtrsim}} 3 Protoclusters}",
      journal = {\apj},
         year = 2025,
        month = jan,
       volume = {978},
       number = {1},
          eid = {17},
        pages = {17},
          doi = {10.3847/1538-4357/ad8f36},
archivePrefix = {arXiv},
       eprint = {2411.14641},
 primaryClass = {astro-ph.GA},
       adsurl = {https://ui.adsabs.harvard.edu/abs/2025ApJ...978...17M}
}

@ARTICLE{Chiang2017,
       author = {{Chiang}, Yi-Kuan and {Overzier}, Roderik A. and {Gebhardt}, Karl and {Henriques}, Bruno},
        title = "{Galaxy Protoclusters as Drivers of Cosmic Star Formation History in the First 2 Gyr}",
      journal = {\apjl},
         year = 2017,
        month = aug,
       volume = {844},
       number = {2},
          eid = {L23},
        pages = {L23},
          doi = {10.3847/2041-8213/aa7e7b},
archivePrefix = {arXiv},
       eprint = {1705.01634},
 primaryClass = {astro-ph.GA},
       adsurl = {https://ui.adsabs.harvard.edu/abs/2017ApJ...844L..23C}
}

@ARTICLE{Umehata2019,
       author = {{Umehata}, H. and {Fumagalli}, M. and {Smail}, I. and {Matsuda}, Y. and {Swinbank}, A.~M. and {Cantalupo}, S. and {Sykes}, C. and {Ivison}, R.~J. and {Steidel}, C.~C. and {Shapley}, A.~E. and {Vernet}, J. and {Yamada}, T. and {Tamura}, Y. and {Kubo}, M. and {Nakanishi}, K. and {Kajisawa}, M. and {Hatsukade}, B. and {Kohno}, K.},
        title = "{Gas filaments of the cosmic web located around active galaxies in a protocluster}",
      journal = {Science},
         year = 2019,
        month = oct,
       volume = {366},
       number = {6461},
        pages = {97-100},
          doi = {10.1126/science.aaw5949},
archivePrefix = {arXiv},
       eprint = {1910.01324},
 primaryClass = {astro-ph.GA},
       adsurl = {https://ui.adsabs.harvard.edu/abs/2019Sci...366...97U}
}

@ARTICLE{Lemaux2022,
       author = {{Lemaux}, B.~C. and {Cucciati}, O. and {Le F{\`e}vre}, O. and {Zamorani}, G. and {Lubin}, L.~M. and {Hathi}, N. and {Ilbert}, O. and {Pelliccia}, D. and {Amor{\'\i}n}, R. and {Bardelli}, S. and {Cassata}, P. and {Gal}, R.~R. and {Garilli}, B. and {Guaita}, L. and {Giavalisco}, M. and {Hung}, D. and {Koekemoer}, A. and {Maccagni}, D. and {Pentericci}, L. and {Ribeiro}, B. and {Schaerer}, D. and {Shah}, E. and {Shen}, L. and {Staab}, P. and {Talia}, M. and {Thomas}, R. and {Tomczak}, A.~R. and {Tresse}, L. and {Vanzella}, E. and {Vergani}, D. and {Zucca}, E.},
        title = "{The VIMOS Ultra Deep Survey: The reversal of the star-formation rate {\ensuremath{-}} density relation at 2 < z < 5}",
      journal = {\aap},
         year = 2022,
        month = jun,
       volume = {662},
          eid = {A33},
        pages = {A33},
          doi = {10.1051/0004-6361/202039346},
archivePrefix = {arXiv},
       eprint = {2009.03324},
 primaryClass = {astro-ph.GA},
       adsurl = {https://ui.adsabs.harvard.edu/abs/2022A&A...662A..33L}
}

@ARTICLE{Casey2015,
       author = {{Casey}, C.~M. and {Cooray}, A. and {Capak}, P. and {Fu}, H. and {Kovac}, K. and {Lilly}, S. and {Sanders}, D.~B. and {Scoville}, N.~Z. and {Treister}, E.},
        title = "{A Massive, Distant Proto-cluster at z = 2.47 Caught in a Phase of Rapid Formation?}",
      journal = {\apjl},
         year = 2015,
        month = aug,
       volume = {808},
       number = {2},
          eid = {L33},
        pages = {L33},
          doi = {10.1088/2041-8205/808/2/L33},
archivePrefix = {arXiv},
       eprint = {1506.01715},
 primaryClass = {astro-ph.GA},
       adsurl = {https://ui.adsabs.harvard.edu/abs/2015ApJ...808L..33C}
}

@ARTICLE{Balogh2021,
       author = {{Balogh}, Michael L. and {van der Burg}, Remco F.~J. and {Muzzin}, Adam and {Rudnick}, Gregory and {Wilson}, Gillian and {Webb}, Kristi and {Biviano}, Andrea and {Boak}, Kevin and {Cerulo}, Pierluigi and {Chan}, Jeffrey and {Cooper}, M.~C. and {Gilbank}, David G. and {Gwyn}, Stephen and {Lidman}, Chris and {Matharu}, Jasleen and {McGee}, Sean L. and {Old}, Lyndsay and {Pintos-Castro}, Irene and {Reeves}, Andrew M.~M. and {Shipley}, Heath and {Vulcani}, Benedetta and {Yee}, Howard K.~C. and {Alonso}, M. Victoria and {Bellhouse}, Callum and {Cooke}, Kevin C. and {Davidson}, Anna and {De Lucia}, Gabriella and {Demarco}, Ricardo and {Drakos}, Nicole and {Fillingham}, Sean P. and {Finoguenov}, Alexis and {Forrest}, Ben and {Golledge}, Caelan and {Jablonka}, Pascale and {Lambas Garcia}, Diego and {McNab}, Karen and {Muriel}, Hernan and {Nantais}, Julie B. and {Noble}, Allison and {Parker}, Laura C. and {Petter}, Grayson and {Poggianti}, Bianca M. and {Townsend}, Melinda and {Valotto}, Carlos and {Webb}, Tracy and {Zaritsky}, Dennis},
        title = "{The GOGREEN and GCLASS surveys: first data release}",
      journal = {\mnras},
         year = 2021,
        month = jan,
       volume = {500},
       number = {1},
        pages = {358-387},
          doi = {10.1093/mnras/staa3008},
archivePrefix = {arXiv},
       eprint = {2009.13345},
 primaryClass = {astro-ph.GA},
       adsurl = {https://ui.adsabs.harvard.edu/abs/2021MNRAS.500..358B}
}

@ARTICLE{Edward2024,
       author = {{Edward}, Adit H. and {Balogh}, Michael L. and {Bah{\'e}}, Yannick M. and {Cooper}, M.~C. and {Hatch}, Nina A. and {Marchioni}, Justin and {Muzzin}, Adam and {Noble}, Allison and {Rudnick}, Gregory H. and {Vulcani}, Benedetta and {Wilson}, Gillian and {De Lucia}, Gabriella and {Demarco}, Ricardo and {Forrest}, Ben and {Hirschmann}, Michaela and {Castignani}, Gianluca and {Cerulo}, Pierluigi and {Finn}, Rose A. and {Hewitt}, Guillaume and {Jablonka}, Pascale and {Kodama}, Tadayuki and {Maurogordato}, Sophie and {Nantais}, Julie and {Xie}, Lizhi},
        title = "{The stellar mass function of quiescent galaxies in 2 < z < 2.5 protoclusters}",
      journal = {\mnras},
         year = 2024,
        month = jan,
       volume = {527},
       number = {3},
        pages = {8598-8617},
          doi = {10.1093/mnras/stad3751},
archivePrefix = {arXiv},
       eprint = {2312.12380},
 primaryClass = {astro-ph.GA},
       adsurl = {https://ui.adsabs.harvard.edu/abs/2024MNRAS.527.8598E}
}

@ARTICLE{Naufal2023,
       author = {{Naufal}, Abdurrahman and {Koyama}, Yusei and {Shimakawa}, Rhythm and {Kodama}, Tadayuki},
        title = "{Environmental Impacts on the Rest-frame UV Size and Morphology of Star-forming Galaxies at z   2}",
      journal = {\apj},
         year = 2023,
        month = dec,
       volume = {958},
       number = {2},
          eid = {170},
        pages = {170},
          doi = {10.3847/1538-4357/acfb81},
archivePrefix = {arXiv},
       eprint = {2309.15450},
 primaryClass = {astro-ph.GA},
       adsurl = {https://ui.adsabs.harvard.edu/abs/2023ApJ...958..170N}
}

@ARTICLE{Forrest2025,
       author = {{Forrest}, Ben and {Shen}, Lu and {Lemaux}, Brian C. and {Shah}, Ekta and {Cucciati}, Olga and {Gal}, Roy R. and {Giddings}, Finn and {Golden-Marx}, Emmet and {Hu}, Weida and {Ronayne}, Kaila and {Sikorski}, Derek and {Staab}, Priti and {Amor{\'\i}n}, Ricardo O. and {Bardelli}, Sandro and {Garilli}, Bianca and {Hathi}, Nimish and {Hung}, Denise and {Lubin}, Lori and {Pelliccia}, Debora and {Ryan}, Russell E. and {Zamorani}, Gianni and {Zucca}, Elena},
        title = "{The HST-Hyperion Survey: Grism Observations of a z {\ensuremath{\sim}} 2.5 Protosupercluster}",
      journal = {\apj},
         year = 2025,
        month = may,
       volume = {985},
       number = {1},
          eid = {61},
        pages = {61},
          doi = {10.3847/1538-4357/adc252},
archivePrefix = {arXiv},
       eprint = {2503.04884},
 primaryClass = {astro-ph.GA},
       adsurl = {https://ui.adsabs.harvard.edu/abs/2025ApJ...985...61F}
}

@ARTICLE{Watson2025,
       author = {{Watson}, Peter J. and {Vulcani}, Benedetta and {Treu}, Tommaso and {Roberts-Borsani}, Guido and {Dalmasso}, Nicol{\`o} and {He}, Xianlong and {Malkan}, Matthew A. and {Morishita}, Takahiro and {Rojas Ruiz}, Sof{\'\i}a and {Zhang}, Yechi and {Acharyya}, Ayan and {Bergamini}, Pietro and {Brada{\v{c}}}, Maru{\v{s}}a and {Fontana}, Adriano and {Grillo}, Claudio and {Jones}, Tucker and {Marchesini}, Danilo and {Nanayakkara}, Themiya and {Pentericci}, Laura and {Tubthong}, Chanita and {Wang}, Xin},
        title = "{The GLASS-JWST Early Release Science programme: The NIRISS spectroscopic catalogue}",
      journal = {\aap},
         year = 2025,
        month = jul,
       volume = {699},
          eid = {A225},
        pages = {A225},
          doi = {10.1051/0004-6361/202554954},
archivePrefix = {arXiv},
       eprint = {2504.00823},
 primaryClass = {astro-ph.GA},
       adsurl = {https://ui.adsabs.harvard.edu/abs/2025A&A...699A.225W}
}

@ARTICLE{Muzzin2013A,
       author = {{Muzzin}, Adam and {Marchesini}, Danilo and {Stefanon}, Mauro and {Franx}, Marijn and {McCracken}, Henry J. and {Milvang-Jensen}, Bo and {Dunlop}, James S. and {Fynbo}, J.~P.~U. and {Brammer}, Gabriel and {Labb{\'e}}, Ivo and {van Dokkum}, Pieter G.},
        title = "{The Evolution of the Stellar Mass Functions of Star-forming and Quiescent Galaxies to z = 4 from the COSMOS/UltraVISTA Survey}",
      journal = {\apj},
         year = 2013,
        month = nov,
       volume = {777},
       number = {1},
          eid = {18},
        pages = {18},
          doi = {10.1088/0004-637X/777/1/18},
archivePrefix = {arXiv},
       eprint = {1303.4409},
 primaryClass = {astro-ph.CO},
       adsurl = {https://ui.adsabs.harvard.edu/abs/2013ApJ...777...18M}
}

@ARTICLE{Smethurst2025,
       author = {{Smethurst}, R.~J. and {Simmons}, B.~D. and {G{\'e}ron}, T. and {Dickinson}, H. and {Fortson}, L. and {Garland}, I.~L. and {Kruk}, S. and {Jewell}, S.~M. and {Lintott}, C.~J. and {Makechemu}, J.~S. and {Mantha}, K.~B. and {Masters}, K.~L. and {O'Ryan}, D. and {Roberts}, H. and {Thorne}, M.~R. and {Walmsley}, M. and {Calabr{\`o}}, M. and {Holwerda}, B. and {Kartaltepe}, J.~S. and {Koekemoer}, A.~M. and {Lyu}, Y. and {Lucas}, R. and {Pacucci}, F. and {Tarrasse}, M.},
        title = "{Galaxy Zoo JWST: up to 75 per cent of discs are featureless at 3 < z < 7}",
      journal = {\mnras},
         year = 2025,
        month = may,
       volume = {539},
       number = {2},
        pages = {1359-1371},
          doi = {10.1093/mnras/staf506},
       adsurl = {https://ui.adsabs.harvard.edu/abs/2025MNRAS.539.1359S}
}

@ARTICLE{Chen2022,
       author = {{Chen}, Chian-Chou and {Gao}, Zhen-Kai and {Hsu}, Qi-Ning and {Liao}, Cheng-Lin and {Ling}, Yu-Han and {Lo}, Ching-Min and {Smail}, Ian and {Wang}, Wei-Hao and {Wang}, Yu-Jan},
        title = "{JWST Sneaks a Peek at the Stellar Morphology of z 2 Submillimeter Galaxies: Bulge Formation at Cosmic Noon}",
      journal = {\apjl},
         year = 2022,
        month = nov,
       volume = {939},
       number = {1},
          eid = {L7},
        pages = {L7},
          doi = {10.3847/2041-8213/ac98c6},
archivePrefix = {arXiv},
       eprint = {2208.05296},
 primaryClass = {astro-ph.GA},
       adsurl = {https://ui.adsabs.harvard.edu/abs/2022ApJ...939L...7C}
}

\begin{appendix}

    \section{BOSS1441}\label{sect:1441}
        
    Although this paper focuses on our analysis of the morphology of HAEs in BOSS1244 and BOSS1542, we have additional HST imaging of BOSS1441, another MAMMOTH protocluster at z $\sim$ 2.32 \citep{Cai2017}.  Unlike BOSS1244 and BOSS1542, which are characterized by overdensities of HAEs, BOSS1441 was identified via an overdensity of LAEs (10.8$\pm$1.0 LAEs in a 15\,Mpc$^{3}$ region \citealp{Cai2017}).  These LAEs were identified using deep narrowband imaging in the NB403 filter taken with the MOSAIC1.1 Camera on the 4\,m Mayall Telescope at Kitt Peak.  \citet{Cai2017} report 99 LAEs in the BOSS1441 protocluster structure.  Similar to our pointings for BOSS1244 and BOSS1542, we only observed the densest regions of the protocluster, which yields our 9 HST pointings, including 28 LAEs (see Figure~\ref{Fig:boss1441} for the HST coverage of our LAEs). 
    \begin{figure}[h]
    \begin{center}
    \includegraphics[scale=0.55,trim={0.0in 0.0in 0.0in 0.08in},clip=true]{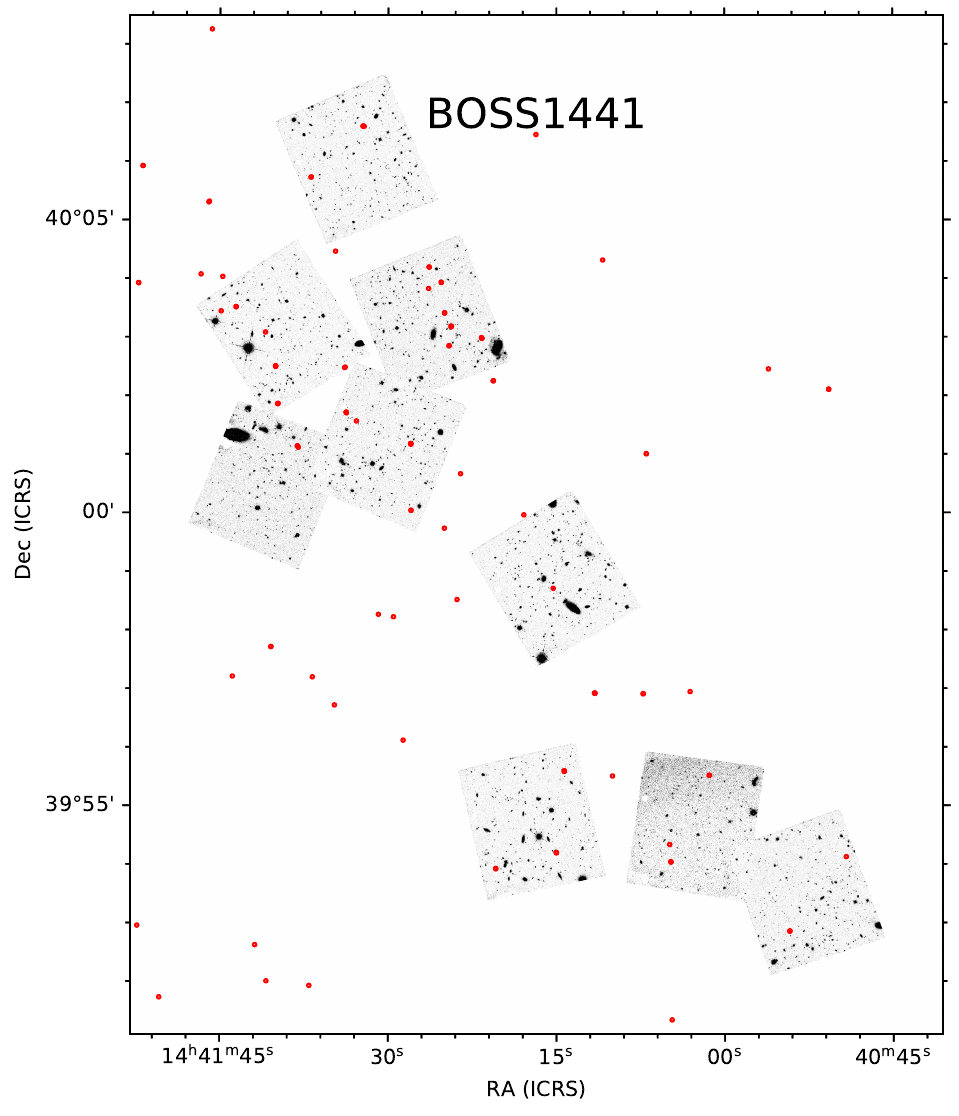}

    \caption{The HST coverage of BOSS1441.  As in Figure~\ref{Fig:protoclusters}, we present our HST WFC3 F160W images and the overlaid red circular regions denote the LAEs (BOSS1441).  The total area shown for BOSS1441 is $\sim$ 13$\farcm$5 $\times$ 17$\arcmin$.}
    \label{Fig:boss1441}
    \end{center}
    \end{figure}
    
    We ultimately focus on BOSS1244 and BOSS1542 in this analysis and not BOSS1441, because of the larger sample of HAEs in the other two protoclusters as well as the representative nature of HAEs as typical star-forming galaxies.  As mentioned previously, at $z$ $>$ 2, star-forming galaxies are the dominant population \citep[e.g.,][]{Muzzin2013A,Marchesini2014,Edward2024} and HAEs span a similar phase space in terms of their stellar mass, SFR, and color to typical star-forming galaxies \citep{Oteo2015}.  However, \citet{Oteo2015} find that only 4.5$\%$ of HAEs are LAEs and that LAEs probe a bluer, lower stellar mass subset of star-forming galaxies.  Additionally, \citet{Cassata2015} find that only $\sim$ 10$\%$ of the star-forming galaxies in the VIMOS Ultra Deep Survey are strong LAEs at $z$ $\sim$ 2.3, further emphasizing their bias. Thus, although a sample of LAEs can obviously be used to identify an overdensity \citep[e.g.,][]{Cai2017,Umehata2019}, it is a biased population, making it difficult to characterize the evolution of protocluster galaxies as a whole.   While we include our analysis of the morphology of LAEs as a function of the environment in BOSS1441 here, we do not include it in the main paper because we cannot place rigorous constraints on the morphology of the protocluster galaxies relative to the field.

    \begin{figure}[h]
        \begin{center}
        \includegraphics[scale=0.7,trim={0.0in 0.6in 0.0in 1.15in},clip=true]{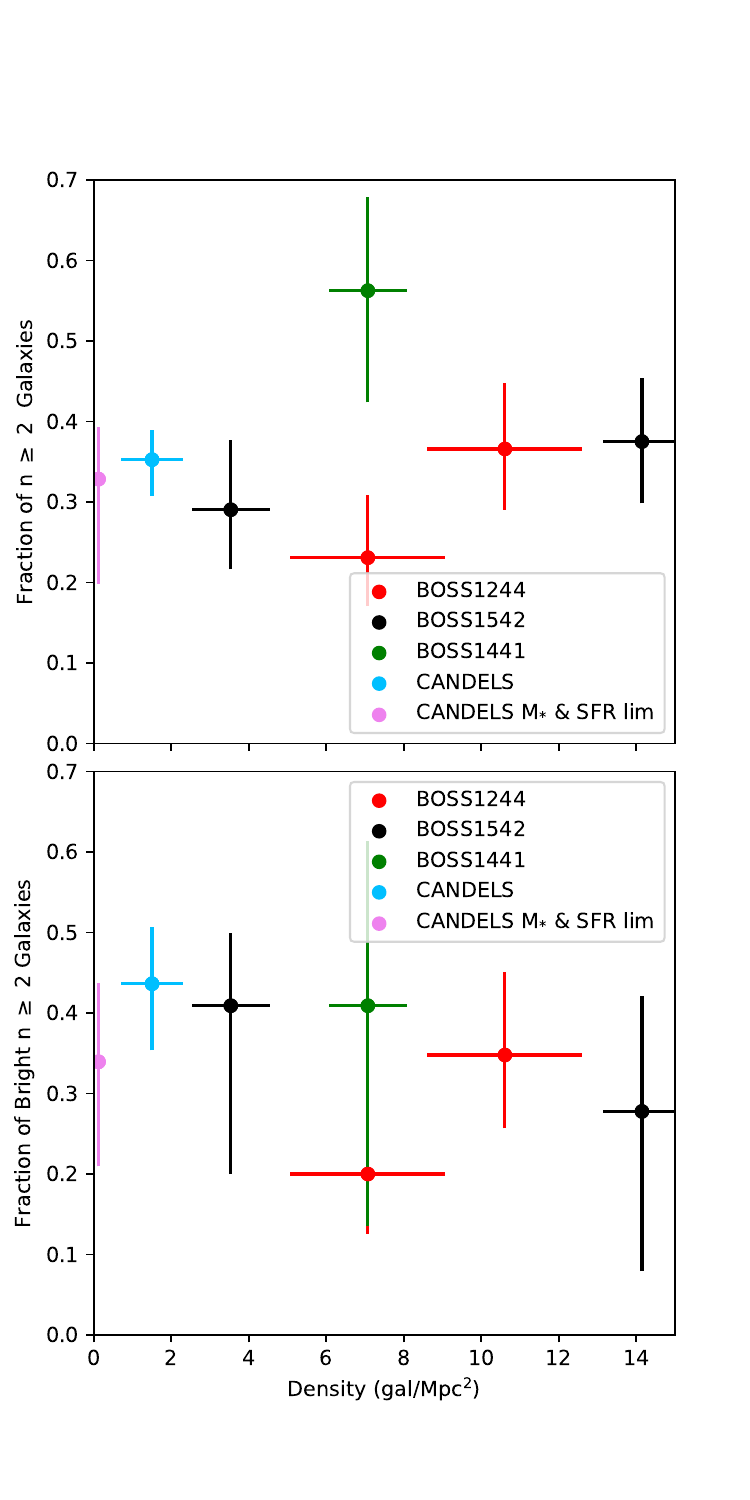}

        \caption{The morphology-density relation in all three MAMMOTH protoclusters.  BOSS1244 is shown in red.  BOSS1542 is shown in black.  BOSS1441 is shown in green.  With the exception of the additional point for BOSS1441, the figure is identical to Figure~\ref{Fig:Morph-density}.  As in Figure~\ref{Fig:Morph-density}, the values of the fractions of galaxies with $n$ $\geq$ 2 for each sample is measured by bootstrapping over the measured errors in S\'ersic index over 1000 iterations.  }
        \label{Fig:Morph-density-ALL3}
    \end{center}    
    \end{figure}

    Unlike the populations of HAEs in BOSS1244 and BOSS1542, we do not have a large population of LAEs.  Rather, only 28 LAEs were covered by our HST coverage and we only had successful \textsc{Galapagos} measurements for 18 of these galaxies.  Thus, we do not have a large enough sample to estimate galaxy density in multiple bins.  However, we do find an abundance of galaxies with $n$ $\geq$ 2 (see Figure~\ref{Fig:Morph-density-ALL3}).  This places 
    \onecolumn
    \noindent the overall fraction of early-type galaxies above the lower density bins for BOSS1244 and BOSS1542 and also approximately 1$\sigma$ above the field (and $\sim$ 1$\sigma$ above the higher density bins in BOSS1244 and BOSS1542).  Interestingly, this seems to be primarily among the fainter galaxies given the similarities to the values in BOSS1441 and the lower density bins in the brighter sample (although the error bars are quite large).  This may also reflect the biases discussed in \citet{Oteo2015} regarding LAEs probing fainter galaxies at $z$ $\sim$ 2.3.  If the abundance of early-type galaxies is real, this could indicate that BOSS1441 is a more evolved protocluster system, at least in regards to its galaxy population.  Interestingly, this result is echoed in the analysis of the spatial distribution of protocluster galaxies in \citet{Shi2021}, who noted that of all three MAMMOTH protoclusters mentioned in this paper, BOSS1441 shows the least substructure and includes only a singular, relatively symmetric distribution of galaxies.  However, given that LAEs are not ubiquitous and we only have a small sample with large error bars, it is difficult to make any strong conclusions.  Additionally, it is interesting to note that while we identify multiple potential interacting galaxies, only one is a multi-peak system, which could suggest that this system is farther along its evolutionary path.

    \section{\textsc{Galapagos} fittings}\label{sect:GalapagosFits}
    
    As part of the measurements of the S\'ersic index, \textsc{Galfit} and \textsc{Galapagos} create models of single-component S\'ersic galaxies, which are then subtracted from the source image.  We include examples of the resulting residuals to show the goodness of these fits (see Figure~\ref{Fig:Residuals}).  For the overwhelming majority of these sources, we find little evidence of substructure among the residuals, pointing to the effectiveness of fitting these galaxies using a single-component S\'ersic index with \textsc{Galapagos}, even for the multi-peak galaxies.  
    
    \begin{figure*}[h]
    \begin{center}
    \includegraphics[scale=0.3,trim={0.0in 0.0in 0.0in 0.0in},clip=true]{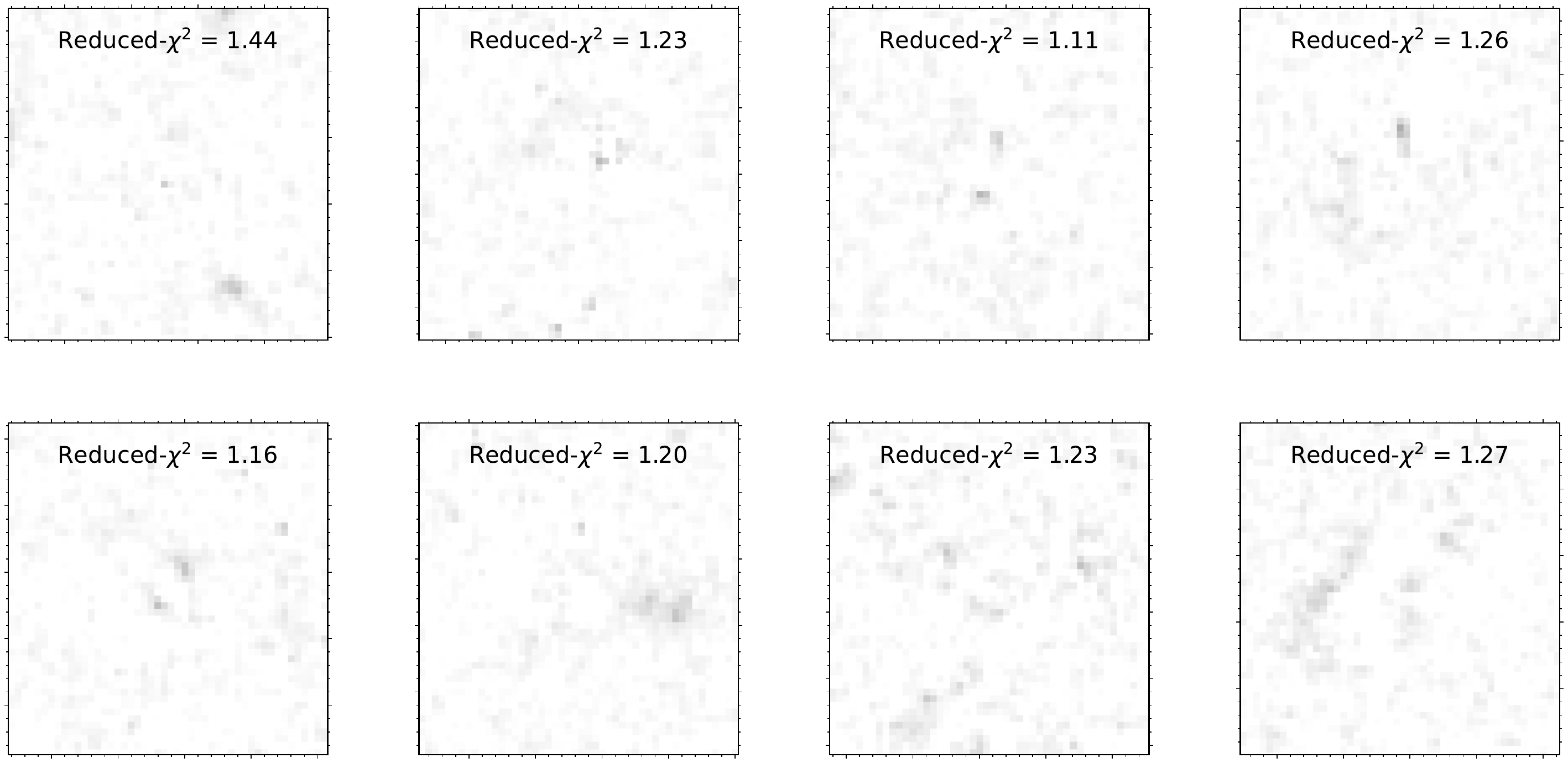}
    \includegraphics[scale=0.3,trim={0.0in 0.0in 0.0in 0.0in},clip=true]{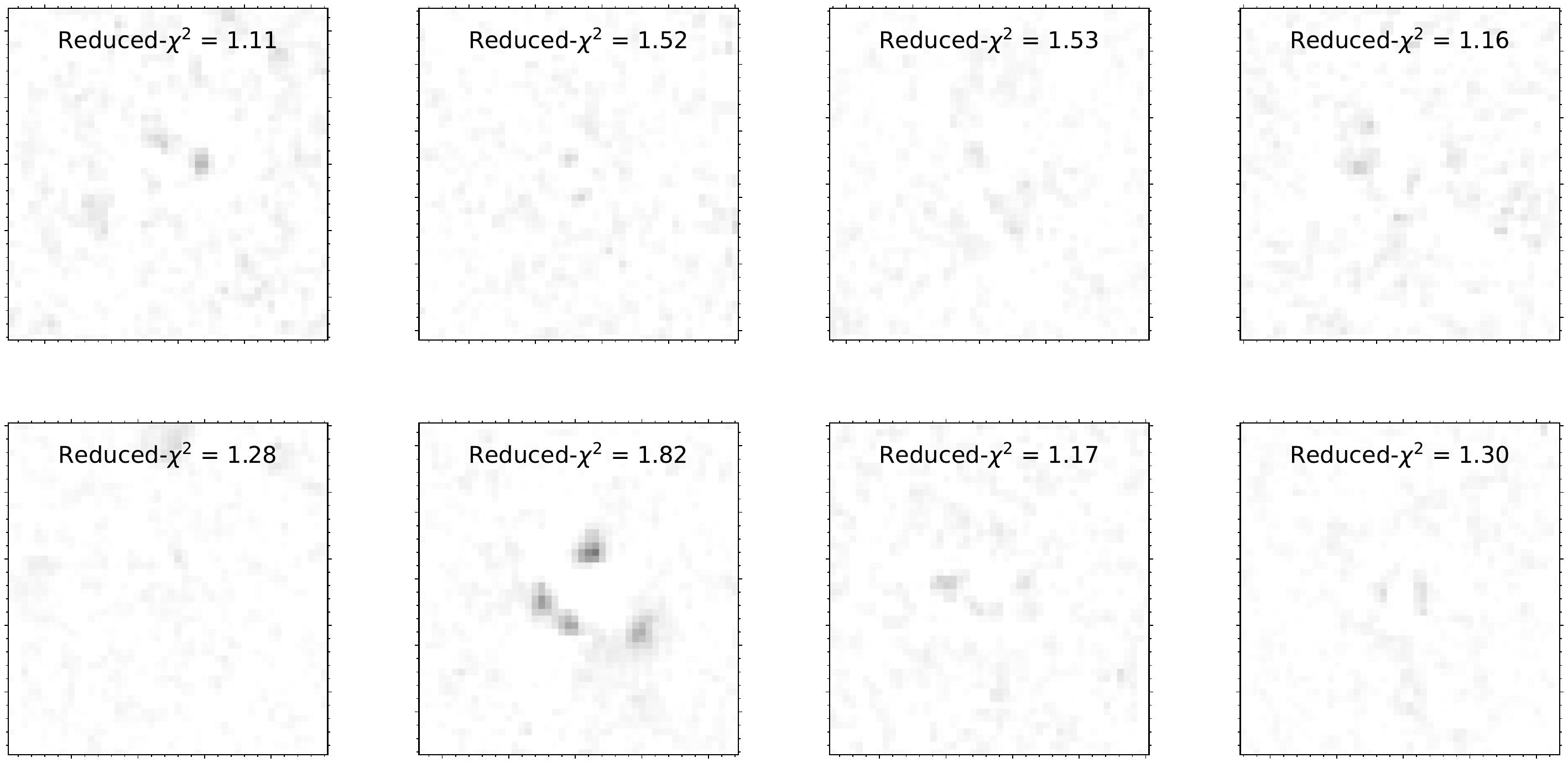}

    \caption{Examples of residuals output from \textsc{Galapagos}.  Each figure shows an $\sim$ 2$\farcs$88 $\times$ 2$\farcs$88 FOV (the same FOV as in Figures~\ref{Fig:Examples} and \ref{Fig:Mergers}) of the \textsc{Galfit} residuals output using \textsc{Galapagos}.  The reduced-$\chi^{2}$ of these fits are included in each image.  The top two rows show the eight galaxies shown in Figure~\ref{Fig:Examples}, while the bottom two rows show the eight multi-peak galaxies shown in Figure~\ref{Fig:Mergers}.  Each residual is scaled identically to the initial image.}

    \label{Fig:Residuals}
    \end{center}
    \end{figure*}
    
    \twocolumn
    \section{The statistical co-eval field sample}\label{sect:Field}

    \begin{figure}[hbt!]
    \begin{center}
    \includegraphics[scale=0.6,trim={0.0in 1.2in 0.5in 1.8in},clip=true]{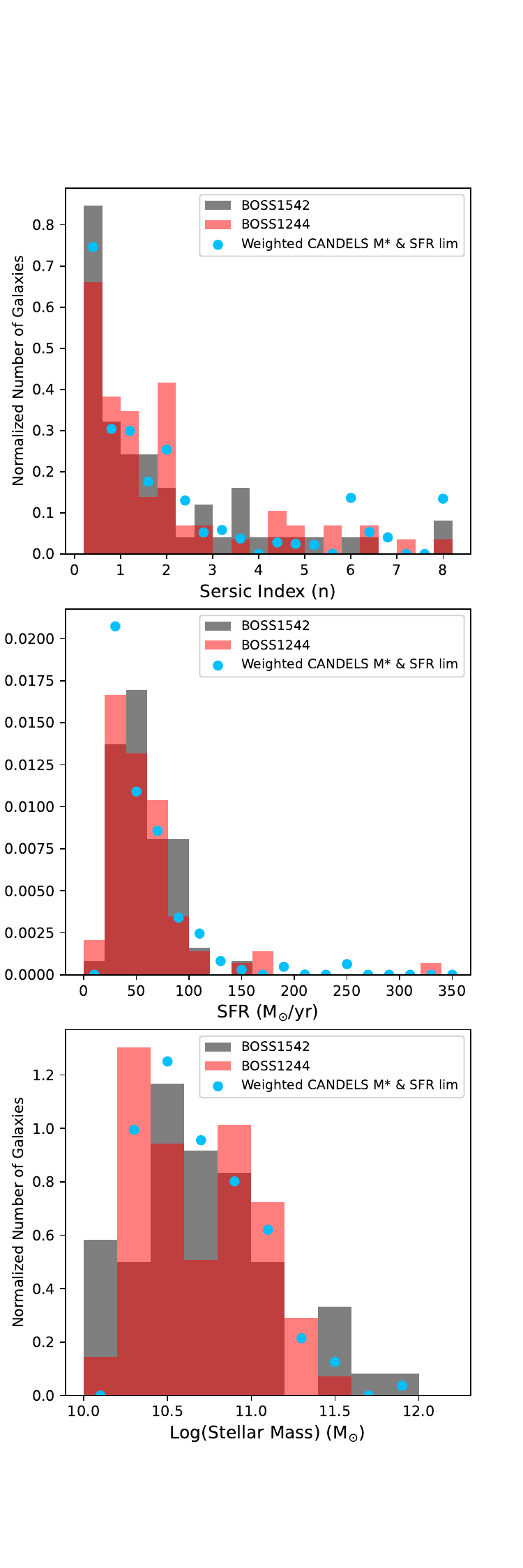}

    \caption{Comparisons of the statistical co-eval field sample to each protocluster.  In each plot, we examine the normalized distribution of various galaxy properties (S\'ersic index [top panel], SFR [middle panel], and Log(Stellar Mass) [bottom panel]) in comparison to the distribution in each protocluster.  Because each co-eval field galaxy has a weighted likelihood of being in our redshift range (2.246 $\pm$ 0.02), the points for the co-eval field sample represent the weighted likelihood of each galaxy with a given S\'ersic index, stellar mass, and SFR being in a given bin.  As seen in each case, we find strong agreement between the co-eval field and protocluster samples, implying that the field is not biasing our results.  }

    \label{Fig:StatsComp}
    \end{center}
    \end{figure}

    As discussed in Section~\ref{sect:background}, we construct a co-eval background field consisting of galaxies in CANDELS with S\'ersic indices measured in \citet{vanderWel2012}.  To statistically show that the co-eval field sample is similar to our protocluster samples in terms of stellar mass and SFR (see Figure~\ref{Fig:StatsComp}), we do a KS-test to compare the S\'ersic indices, stellar masses, and SFRs between the co-eval field sample and each protocluster.  For the co-eval field sample, we created a statistical distribution that weighs the likelihood that each galaxy (with a given S\'ersic index, stellar mass, and SFR) is at the target redshift (2.246 $\pm$ 0.02).  It is the weighted distribution of the background that we compare to the protoclusters in Figure~\ref{Fig:StatsComp}  For the S\'ersic index, we find (P$_{1244 - Field}$ = 0.83, P$_{1542 - Field}$ = 0.98).  For the SFR, we find (P$_{1244 - Field}$ = 0.80, P$_{1542 - Field}$ = 0.79).  For the stellar mass, we find (P$_{1244 - Field}$ = 1.0, P$_{1542 - Field}$ = 0.99).  All of the KS $P$ values indicate that there is no evidence that these sources are drawn from separate samples.

    \section{Examining faint galaxies in the BOSS1244 color-magnitude diagram} \label{sect:1244-CMDFaint}

     As mentioned in Section~\ref{sect:1244}, we examine the population of galaxies in BOSS1244 observed with HST WFC3 F160W and F125W to explore the existence of a red sequence population.  While we find no evidence for a potential red sequence population among the HAEs, we do note a number of sources that are brighter than the magnitude limit in F160W, but below our analysis criterion in F125W.  We include detections as the upward pointing triangles in Figure~\ref{Fig:CMD2}.  These galaxies may represent a sample of extremely dusty, HAEs.  However, their faint detection makes it difficult to fully characterize them at this time and will require additional observations.

    \begin{figure}[hbt!]
    \begin{center}
    \includegraphics[scale=0.6,trim={0.3in 0.0in 0.5in 0.65in},clip=true]{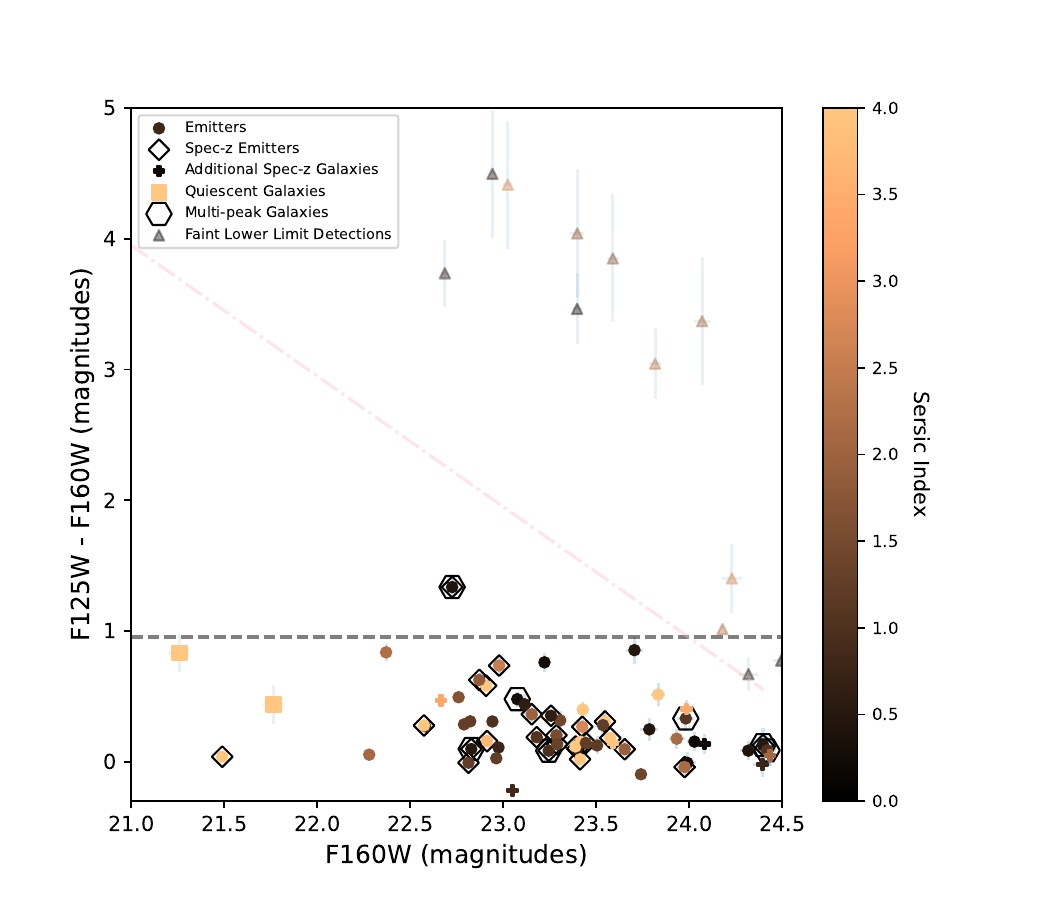}

    \caption{The color-magnitude diagram for BOSS1244 including the lower limit detections.  The legend is identical to Figure~\ref{Fig:CMD}, except we include a subset of galaxies detected in F160W, but not detected above our magnitude limit in F125W.  We have included an estimate of their color shown in upward pointing triangles.  As in Figure~\ref{Fig:CMD}, the pink dot-dashed line marks our detection limit relative to the non-detected F125W galaxies.}

    \label{Fig:CMD2}
    \end{center}
    \end{figure}    
\end{appendix} 

\end{document}